\documentclass[prl,aps,twocolumn,superscriptaddress,showpacs]{revtex4-1}
\usepackage[colorlinks=true,linkcolor=black, citecolor=blue, urlcolor=blue, 
    unicode=true]{hyperref}

\usepackage{graphicx}
\usepackage{epstopdf}

\begin{document}

\title{From Ising resonant fluctuations to static uniaxial order in antiferromagnetic and weakly superconducting CeCo(In$_{1-x}$Hg$_{x}$)$_{5}$ ($x$=0.01)}

\author{C. Stock}
\affiliation{School of Physics and Astronomy, University of Edinburgh, Edinburgh EH9 3JZ, UK}
\author{J. A. Rodriguez-Rivera}
\affiliation{NIST Center for Neutron Research, National Institute of Standards and Technology, 100 Bureau Dr., Gaithersburg, MD 20899}
\affiliation{Department of Materials Science, University of Maryland, College Park, MD  20742}
\author{K. Schmalzl}
\affiliation{Forschungszentrum Juelich GmbH, Juelich Centre for Neutron Science at ILL, 71 avenue des Martyrs, 38000 Grenoble, France}
\author{F. Demmel}
\affiliation{ISIS Facility, Rutherford Appleton Labs, Chilton, Didcot, OX11 0QX}
\author{D. K. Singh}
\affiliation{Department of Physics and Astronomy, University of Missouri, Missouri 65211, USA}
\author{F. Ronning}
\affiliation{Los Alamos National Laboratory, Los Alamos, New Mexico 87545, USA}
\author{J. D. Thompson}
\affiliation{Los Alamos National Laboratory, Los Alamos, New Mexico 87545, USA}
\author{E.D. Bauer}
\affiliation{Los Alamos National Laboratory, Los Alamos, New Mexico 87545, USA}

\date{\today}

\begin{abstract}

CeCo(In$_{0.990}$Hg$_{0.010}$)$_{5}$ is a charge doped variant of the $d$-wave CoCoIn$_{5}$ superconductor with coexistent antiferromagnetic and superconducting transitions occurring at T$_{N}$= 3.4 K and T$_{c}$=1.4 K, respectively.  We use neutron diffraction and spectroscopy to show that the magnetic resonant fluctuations present in the parent superconducting phase are replaced by collinear $c$-axis magnetic order with three-dimensional Ising critical fluctuations.    No low energy transverse spin fluctuations are observable in this doping-induced antiferromagnetic phase and the dynamic resonant spectral weight predominately shifts to the elastic channel.  Static ($\tau$ $>$ 0.2 ns) collinear Ising order is proximate to superconductivity in CeCoIn$_{5}$ and is stabilized through hole doping with Hg.

\end{abstract}


\maketitle

Strong magnetic fluctuations are not compatible with conventional superconductivity~\cite{Miyake86:34},  however are believed to be consistent with a superconducting gap with nodes such as $d$-wave symmetry~\cite{Norman11:332}.  The critical point separating magnetic order and superconductivity is often proximate to new phases.~\cite{Knafo09:5,Norman11:332,Chen17:96,Pfleiderer09:81,Scalapino12:84,Stewart84:56} For example, the cuprate superconducting dome is bracketed by both a pseudogap phase~\cite{Timusk99:62,Varma06:73} and a fermi liquid~\cite{Proust02:89}.  In pnictides, nematic order~\cite{Chu10:329} occurs in the vicinity of superconductivity.   A signature that magnetic fluctuations are important for new superconducting orders is the presence of a magnetic resonance peak observed in many magnetic unconventional superconductors including cuprates~\cite{Fong97:78,Fong00:61,Dai96:77,Dai01:63}, CeCu$_{2}$Si$_{2}$~\cite{Stockert09:7}, UPd$_{2}$Al$_{3}$~\cite{Sato01:410,Bernhoeft98:81,Hess06:18,Hiess07:76}, pnictides~\cite{Christianson08:456,Dai15:87}, and CeCoIn$_{5}$~\cite{Stock08:100,Stock12:109,Panarin09:78} associated with a gap function that undergoes a change in sign~\cite{Norman00:61,Eremin08:101}.   Magnetic resonant excitations also occur when other order parameters are present~\cite{Thalmeier13:86} with the observation of an exciton mode in CeB$_{6}$ an example.~\cite{Friemel12:3}  We investigate the magnetic fluctuations in the CeCoIn$_{5}$ $d$-wave superconductor charge doped to long-range antiferromagnetic (AF) order.  The results illustrate the instability of transverse `spin-waves' in unconventional superconductors in favor of Ising like fluctuations which are condensed via charge doping.

CeCoIn$_{5}$ displays unconventional superconductivity with a transition temperature of T$_c$=2.3 K~\cite{Petrovic01:13} and a $d$-wave superconducting order parameter.~\cite{Thompson12:81}  The crystallographic structure consists of a tetragonal unit cell with layers of magnetic Ce$^{3+}$-In planes stacked along $c$.  Neutron scattering shows the electronic normal state consists of overdamped magnetic excitations peaked near $\vec{Q}$=(1/2, 1/2, 1/2) indicative of antiferromagnetic interactions between the Ce$^{3+}$ ions, both within the $a-b$ plane and along $c$.  The commensurate magnetic spin response differs from non-superconducting  CeRhIn$_{5}$ (at ambient pressure), which displays a magnetic Bragg peak at the incommensurate $\vec{Q}$=(1/2, 1/2, 0.297)~\cite{Bao00:62} characterizing a helical magnetic structure.~\cite{Stock15:114}  The $a-b$ magnetic helix in CeRhIn$_{5}$ contrasts with the commensurate $c$-axis polarized resonant fluctuations which dominate the neutron response in superconducting CeCoIn$_{5}$.~\cite{Stock08:100}

Doping impurities into superconductors has been used to break superconducting Cooper pairs revealing competing phases.~\cite{Nachumi96:77}   Efforts in the `115' system originally were directed to alloying on the Co site in CeCoIn$_{5}$ with either Rh or Ir as a means of tuning from superconducting to magnetic order.~\cite{Zapf01:65,Goh08:101}  However, this phase diagram is complex with CeRhIn$_{5}$ displaying both helical magnetic order~\cite{Stock15:114} and a low temperature superconducting phase under pressure~\cite{Hegger00,Mura01:70,Park08:105,Knebel06:74,Kawasaki03:91,Paglione77:08,Chen06:97}.  Several commensurate magnetic phases are also believed to compete with helical magnetic order~\cite{Bao00:62,Llobet05:95,Fobes17:29} and superconductivity with Rh-Ir alloying. Also, CeIrIn$_{5}$ is a superconductor with a reduced T$_{c}$ of $\sim$ 0.4 K.~\cite{Petrovic01:53}  Replacing Ce by La has been shown to result in a suppression of T$_{c}$.~\cite{Tanatar05:95,Petrovic02:66,Raymond11:80}
 
Another means of electronic tuning CeCoIn$_{5}$ with non-magnetic impurities is through the In site with either electron doping (with Sn) or hole doping (with Cd, Hg, or Ru).~\cite{Bauer05:94,Gofryk12:109,Sakai15:92,Capan10:82,Ou13:88}  Doping magnetic Yb on the Ce site has also been pursued, however is unusual as the suppression of the superconductivity order parameter with doping is very mild~\cite{Xu16:93,Shu11:106} and penetration depth measurements~\cite{Kim15:114} even suggest that nodal $d$-wave superconductivity maybe replaced by a fully gapped order parameter.~\cite{Zhong17:12}  In contrast, hole doping with Cd or Hg on the In site strongly suppresses superconductivity~\cite{Pham06:97,Bauer11:108,Sheo13:10} in favour of a commensurate antiferromagnetic state with a characteristic wavevector of $\vec{Q}$=(1/2, 1/2, 1/2).~\cite{Nicklas07:76,Bao00:62,Bao09:79} We note that this commensurate order contrasts with the incommensurate spin-density-wave reported at large magnetic fields in the superconducting state.~\cite{Kenzelmann08:321,Blackburn10:105,Kim17:95,Michal11:84,Koutroulakis10:104,Raymond14:83}  

\begin{figure}[t]
\includegraphics[angle=0,width=8.7cm] {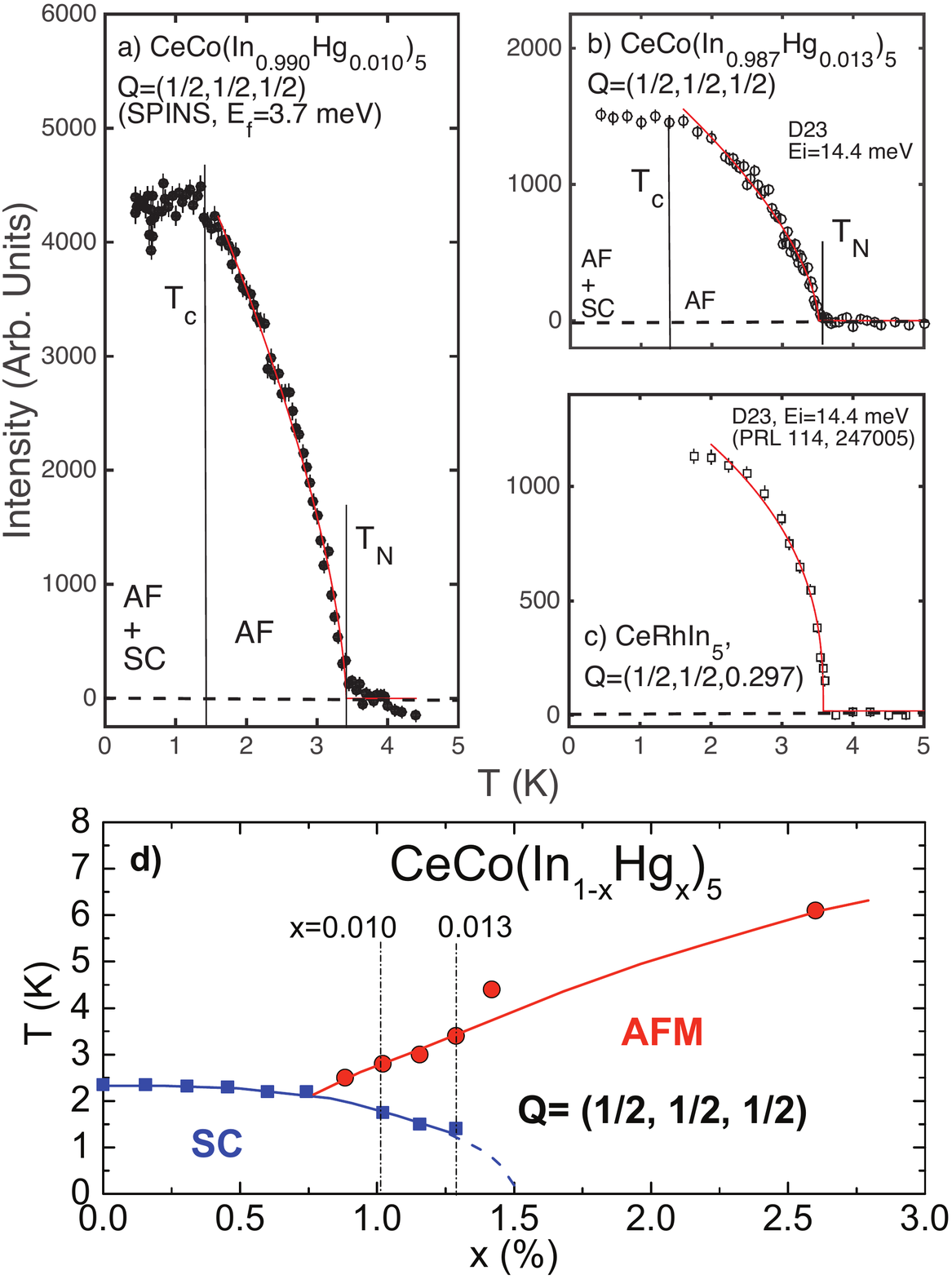}
\caption{\label{order_param}  The magnetic Bragg peak intensity measured as a function of temperature in superconducting and antiferromagnetic $(a)$  CeCo(In$_{0.990}$Hg$_{0.010}$)$_{5}$, $(b)$  CeCo(In$_{0.987}$Hg$_{0.013}$)$_{5}$, and $(c)$ magnetic helically ordered and non-superconducting CeRhIn$_{5}$ (from Ref. \onlinecite{Stock15:114}).  The SC transition T$_{c}$ and N{\'{e}}el T$_N$  temperatures are indicated by vertical lines. $(d)$ Temperature composition $T-x$ phase diagram of CeCo(In$_{1-x}$Hg$_{x}$)$_{5}$ determined from specific heat measurements.}
\end{figure}

We apply neutron diffraction and spectroscopy to study the static and dynamic magnetism in Hg doped CeCoIn$_{5}$.  CeCo(In$_{1-x}$Hg$_{x}$)$_{5}$ samples were grown from In/Hg flux.  Nominal Hg substitution for In of 7\% and 9\% resulted in x=1.0\% and 1.3\% respectively (Fig. \ref{order_param} $d$). Elastic scattering measurements used the D23 diffractometer (ILL, France) and spectroscopic measurements were done on the IN12 (ILL, France) and the MACS (NIST, USA) cold triple-axis spectrometers using a coalignment of $\sim$ 150 crystals (total mass of 4 g).  Experimental details are provided in the Supplementary Information.


The static magnetic properties of antiferromagnetic and superconducting CeCo(In$_{0.990}$Hg$_{0.010}$)$_{5}$ and CeCo(In$_{0.987}$Hg$_{0.013}$)$_{5}$ are shown in Fig. \ref{order_param} and compared to helically magnetically ordered CeRhIn$_{5}$~\cite{Stock15:114}.  The intensity of the magnetic Bragg peak is a measure of the magnetic order parameter and is fit to a power law near T$_{N}$ with $I(T)\propto |M(T)|^{2}\propto (T_{N}-T)^{2\beta}$.  Hg doped CeCoIn$_{5}$ samples in Figs. \ref{order_param} $(a)$ and $(b)$ are shown with a best fit of $\beta$=0.33 $\pm$0.02 and 0.31 $\pm$ 0.02 with magnetic transitions of T$_{N}$=3.41 K and 3.51 K, respectively. CeRhIn$_{5}$ is illustrated in Fig. \ref{order_param}$(c)$ for reference with a similar analysis giving $\beta$=0.19.  The Hg doped samples are are within error of the three dimensional Ising universality class where $\beta$=0.326 \cite{Collins}.  In contrast,  CeRhIn$_{5}$ is consistent with a two dimensional order parameter, a property reflected in the low-energy magnetic excitations~\cite{Das14:113,Stock15:114}, and the anisotropy of the correlation lengths~\cite{Bao02:65}.   Therefore, three dimensional critical dynamics coexist with superconductivity in CeCo(In$_{1-x}$Hg$_{x}$)$_{5}$.  A similar result has been noted for pnictide superconductors near the boundary between antiferromagnetism and superconductivity.~\cite{Wilson10:81,Wilson10:82,Paj13:87}  

The antiferromagnetic order parameter in CeCo(In$_{0.990}$Hg$_{0.010}$)$_{5}$ (Fig. \ref{order_param} $a$) is suggestive of a saturation at the superconducting transition T$_{c}$.  Confirming this is the comparison in Fig. \ref{order_param} $(b)$ which illustrates the magnetic order parameter for a Hg doping of 1.3 \% with a superconducting transition T$_{c}$=1.4 K.    NMR measurements indicate that magnetism forms via localized droplets which are effectively decoupled from superconducting components of the sample~\cite{Urbano07:99}.   Despite this apparent decoupling of magnetic and superconducting orders, the magnetic order parameter for Figs. \ref{order_param} $(a)$ CeCo(In$_{0.990}$Hg$_{0.010}$)$_{5}$ and $(b)$ CeCo(In$_{0.987}$Hg$_{0.013}$)$_{5}$ shows a saturation at the superconducting T$_{c}$, therefore indicating superconductivity interrupts the continuous formation of magnetic order.  We note that a similar low temperature saturation of the magnetic order parameter has been reported in coexistent antiferromagnetic order and superconductivity CeCo(In$_{1-x}$Cd$_{x}$)$_{5}$.~\cite{Nicklas07:76}

\begin{figure}[t]
\includegraphics[angle=0,width=8.7cm] {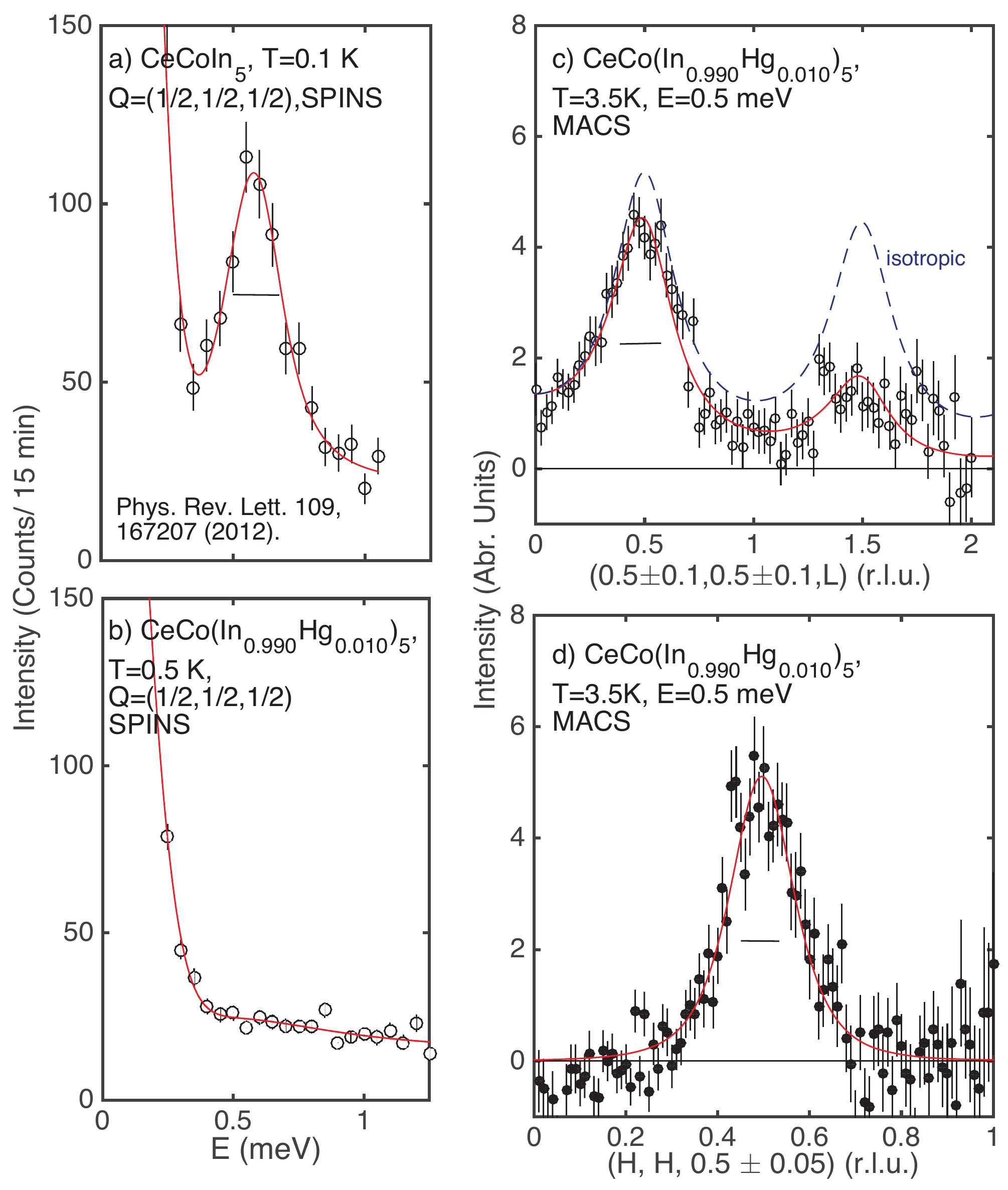}
\caption{\label{compare_115}  Constant $\vec{Q}$=(1/2,1/2,1/2) scans for superconducting and antiferromagnetic $(a)$ CeCo(In$_{0.990}$Hg$_{0.010}$)$_{5}$ and superconducting $(b)$ CeCoIn$_{5}$ with both plots normalized to the same absolute scale.   Panel $(c)$ illustrates the critical magnetic fluctuations showing polarization along $c$ and $(d)$ antiferromagnetic correlations within the $a-b$ plane of CeCo(In$_{0.990}$Hg$_{0.010}$)$_{5}$. The solid lines in $(c)$ and $(d)$ are fits to Eqn. \ref{eq1} corresponding to Ce$^{3+}$ moments polarized along the [001] direction.  The dashed line is the same fit to a model with the isotropic magnetic moments having no preferential direction.}
\end{figure}

The low temperature Ce$^{3+}$ ordered moment in CeCo(In$_{0.987}$Hg$_{0.013}$)$_{5}$ was measured by calibrating against 7 nuclear Bragg peaks to be 0.98 $\pm$ 0.2 $\mu_{B}$, as outlined in the Supplementary Information, while the ordered moment of CeRhIn$_{5}$ and  Cd doped CeCoIn$_{5}$ is $\sim$ 0.3 $\mu_{B}$ and  0.7 $\mu_{B}$, respectively~\cite{Fobes17:29,Nicklas07:76, Urbano07:99}.   Hg doped magnetic order is characterized by a magnetic  moment pointing along the $c$-axis evidenced by a large suppression of intensity at the (1/2, 1/2, 3/2) and (1/2, 1/2, 5/2) magnetic Bragg peaks (see Supplementary Information).  This contrasts with the in-plane helical order of CeRhIn$_{5}$ and also Rh doped CeCoIn$_{5}$~\cite{Llobet05:95}.   


We now discuss the dynamics in superconducting and antiferromagnetic CeCo(In$_{0.990}$Hg$_{0.010}$)$_{5}$ summarized in Fig. \ref{compare_115} and compared to superconducting CeCoIn$_{5}$.~\cite{Stock08:100}   Figs. \ref{compare_115} $(a)$ and $(b)$ illustrates low temperature constant $\vec{Q}$=(1/2,1/2,1/2) scans taken in the superconducting state of both CeCoIn$_{5}$ and CeCo(In$_{0.990}$Hg$_{0.010}$)$_{5}$.  The scans have been normalized to sample mass and confirmed through a comparison of the elastic incoherent scattering.  The superconducting resonance peak in CeCoIn$_{5}$ at $\sim$ 0.5 meV is not observed in antiferromagnetic and superconducting CeCo(In$_{0.990}$Hg$_{0.010}$)$_{5}$ within experimental resolution at T=0.5 K with the solid line in Fig. \ref{compare_115}$(b)$ denoting the measured high temperature background.   Instead, at temperatures near T$_{N}$ (Fig. \ref{order_param} $a$) of 3.5 K, magnetic critical scattering associated with the development of long-range static AF order in CeCo(In$_{0.990}$Hg$_{0.010}$)$_{5}$ is observed with the momentum dependence illustrated in Figs. \ref{compare_115} $(c)$ and $(d)$ from background corrected scans taken using MACS.

\begin{figure}[t]
\includegraphics[angle=0,width=8.7cm] {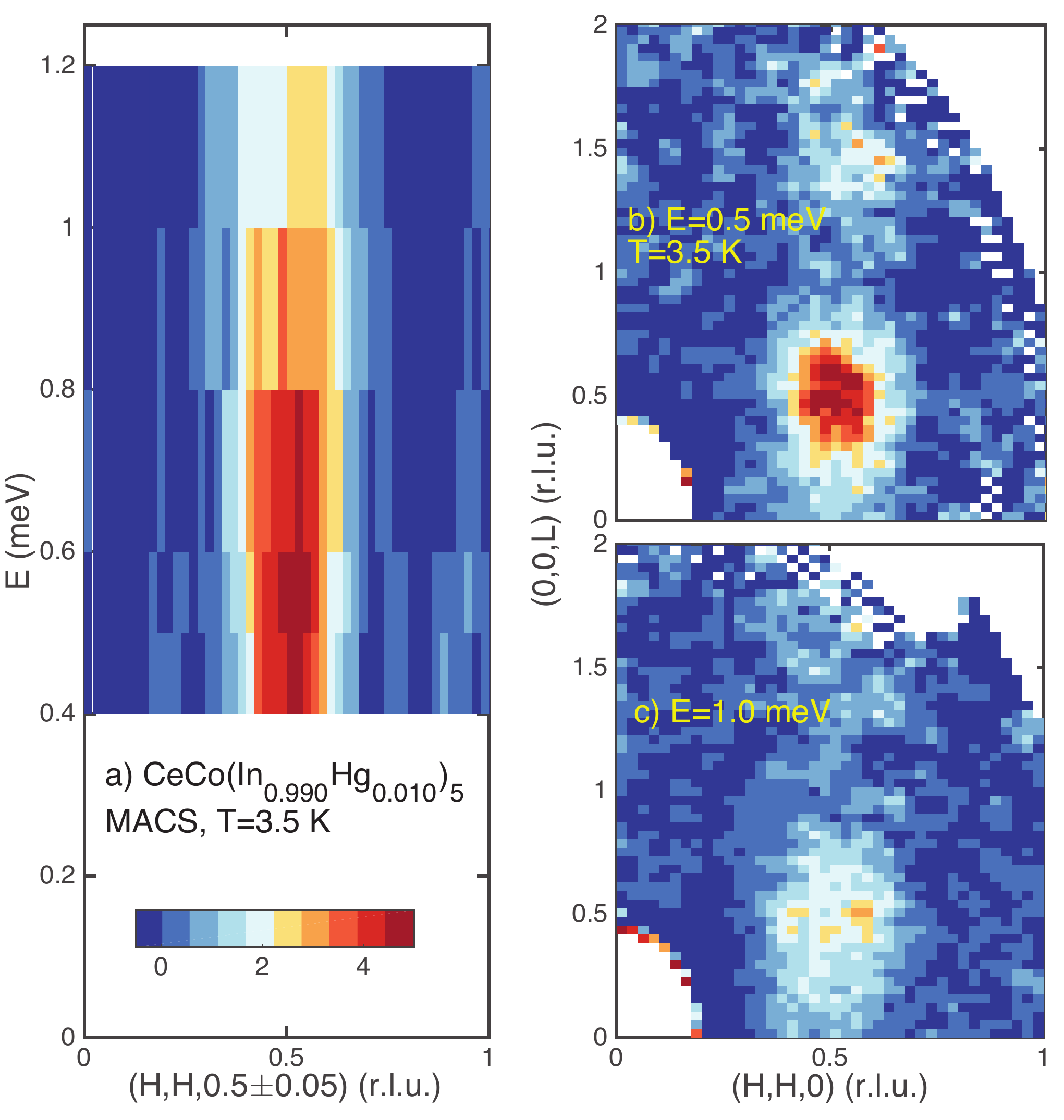}
\caption{\label{macs_cuts}  $(a)$ Constant $\vec{Q}$=(H,H,1/2) slice and constant energy slices at $(b)$ 0.5 meV and $(c)$ 1.0 meV of CeCo(In$_{0.990}$Hg$_{0.010}$)$_{5}$ at 3.5 K.}
\end{figure}

Figs. \ref{compare_115} $(c)$ and $(d)$ show that the magnetic critical dynamics in CeCo(In$_{0.990}$Hg$_{0.010}$)$_{5}$ is highly anisotropic, mimicking the Ising-like critical scattering discussed above.  The solid line in Figs. \ref{compare_115} $(c)$ and $(d)$ is a fit to 

\begin{eqnarray}
I(\vec{Q}) \propto f(Q)^{2}  [1-(\hat{Q} \cdot \hat{c})^{2}] \times \nonumber \\
{\sinh(c/\xi_{c})\over {\cosh(c/\xi_{c})+\cos(\vec{Q}\cdot \hat{c})}} {\xi_{ab}^{2} \over {(1+(\xi_{ab}|{\bf{Q}}_{ab}-{\bf{Q}}|)^{2})^{2}}}\label{eq1}
\end{eqnarray}

\noindent which represents the momentum dependence of short-range antiferromagnetic Ce$^{3+}$ moments polarized along the [001] direction with dynamic correlation lengths of $\xi_{c}$= 6.8 $\pm$ 0.7 \AA\ and $\xi_{ab}$=6.3 $\pm$ 0.5 \AA\ at E=0.5 meV.  $f(Q)^{2}$ is the magnetic form factor~\cite{Blume62:37}.  The dashed line in Fig. \ref{compare_115} $(c)$ is the momentum dependence expected for no preferential Ce$^{3+}$ moment direction.  A Lorentzian squared function was chosen to describe the in-plane momentum dependence as it is normalizable in two dimensions. The ratio of the dynamic correlation lengths is $\xi_{ab}/\xi_{c}\sim$ 1 illustrating a strong three dimensional character. The dynamic correlation length along [001] of CeCo(In$_{0.990}$Hg$_{0.010}$)$_{5}$ is comparable to the value (6.5 $\pm$ 0.9 \AA) for the low temperature resonance peak in superconducting CeCoIn$_{5}$~\cite{Raymond15:115} illustrating that the both the polarization and dynamic correlation lengths have similarities to the parent compound resonant fluctuations.  The uniaxial and three dimensional nature of the fluctuations is consistent with the Ising universality class extracted from the magnetic order parameter discussed above.    

The energy dependence of the critical magnetic fluctuations of CeCo(In$_{0.990}$Hg$_{0.010}$)$_{5}$ is shown in Fig. \ref{macs_cuts}.  Fig. \ref{macs_cuts}$(a)$ displays a constant momentum slice at 3.5 K,  illustrating that these critical fluctuations show little momentum dependence with energy transfer.  Confirming this are constant energy slices at E=0.5 meV (Fig. \ref{macs_cuts} $b$) and 1.0 meV (Fig. \ref{macs_cuts} $c$), which show little change in the lineshape and also the ratio of the dynamic correlation lengths $\xi_{c}$/$\xi_{ab}$ with energy transfer.   The three dimensional character of the critical correlations is robust with energy transfer and there is no observable evidence of fluctuations perpendicular to the [001] direction or transverse to the ordered low temperature magnetic moment direction.

\begin{figure}[t]
\includegraphics[angle=0,width=8.6cm] {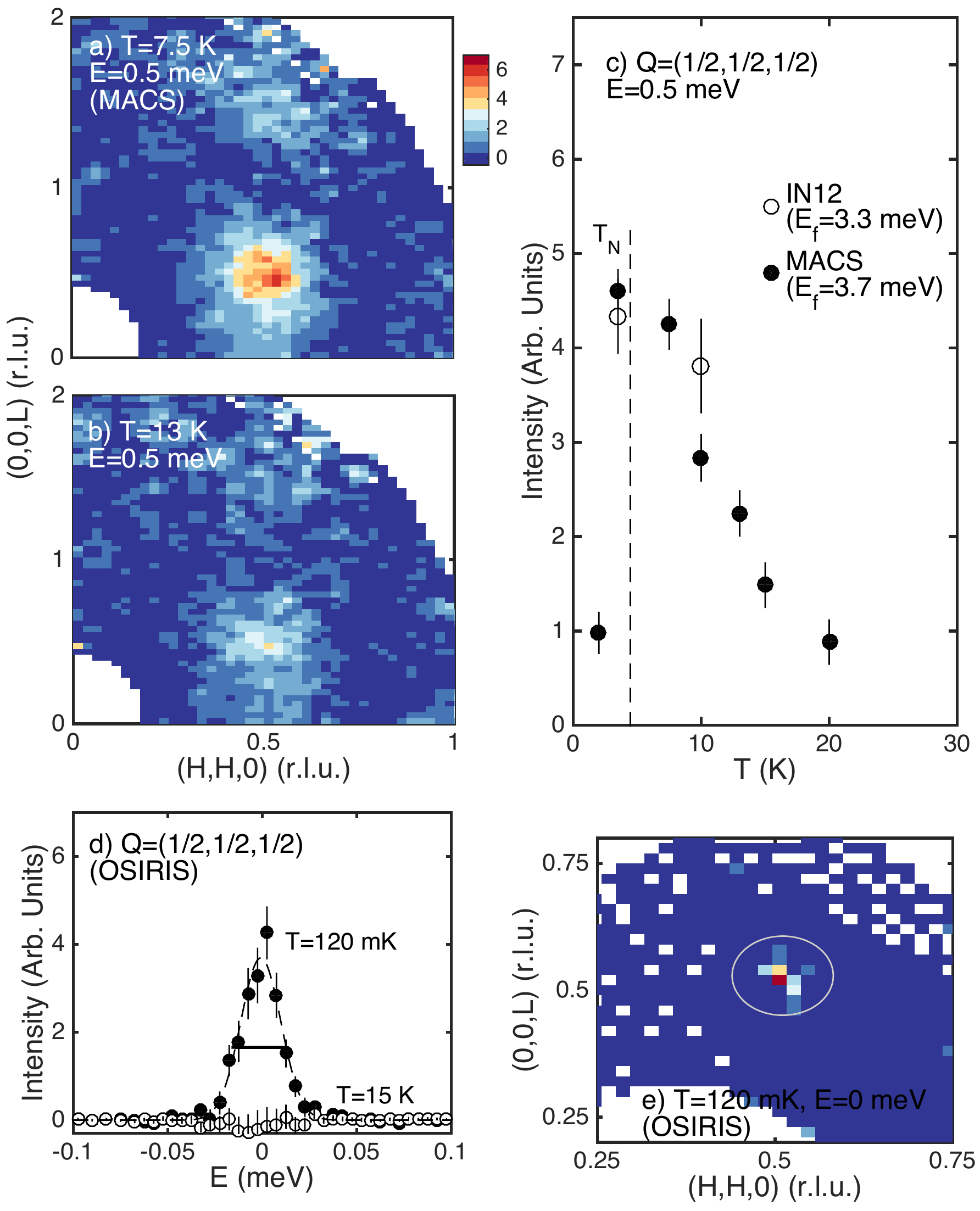}
\caption{\label{T_depend}  The temperature dependence of the magnetic critical scattering in CeCo(In$_{0.990}$Hg$_{0.010}$)$_{5}$.  $(a-b)$ illustrate constant E=0.5 meV slices in the (HHL) scattering plane taken at 7.5 K and 13 K, respectively.  $(c)$ Intensity of the $\vec{Q}$=(1/2,1/2,1/2) magnetic position as a function of temperature. (d-e) illustrate high resolution backscattering measurements probing the temporal nature of the magnetic order.  The horizontal bar is the energy resolution showing that the magnetic order is static to within an energy resolution of 25 $\mu eV$.}
\end{figure}

The temperature dependence of the critical magnetic fluctuations in CeCo(In$_{0.990}$Hg$_{0.010}$)$_{5}$ near $\vec{Q}$=(1/2,1/2,1/2) is displayed in Fig. \ref{T_depend}, measured on IN12 and MACS.  At low temperatures in the superconducting state there is no evidence of any observable magnetic fluctuations confirmed by measurements on IN12 and also momentum space maps on MACS.  Figs. \ref{T_depend}$(a)$ and $(b)$ show further momentum maps illustrating the presence of dynamic magnetic fluctuations above T$_{N}$.  The background corrected intensity as a function of temperature is shown in Fig. \ref{T_depend} $(c)$ displaying a precipitous decrease in magnetic spectral weight below T$_{N}$ and into the superconducting state.  This transfer of spectral weight occurs while static magnetic order is formed (shown in Fig. \ref{order_param}) and results in temporally static order (on order of the experimental resolution of 25 $\mu$ eV shown in Figs. \ref{T_depend} $d$ and $e$).  Any resonant excitation in CeCo(In$_{0.990}$Hg$_{0.010}$)$_{5}$ with comparable spectral weight to that of parent CeCoIn$_{5}$ is either considerably broadened in momentum and energy, or residing within the elastic energy resolution of our measurements.

The results presented here illustrate that collinear $c$-axis polarized magnetic order is parent to superconductivity in the CeCoIn$_{5}$ system.  The magnetic order in CeCo(In$_{1-x}$Hg$_{x}$)$_{5}$ replaces the temporally well defined ``Ising-like"~\cite{Raymond15:115} and longitudinally polarized~\cite{Mazzone17:119} resonant fluctuations reported in Nd doped CeCoIn$_{5}$ which displays both superconductivity and commensurate magnetic order.    The static magnetic structure mimics the predominately [001] polarized resonant magnetic fluctuations present in strongly superconducting samples.   The spectral weight of the static magnetic Bragg peak in  CeCo(In$_{1-x}$Hg$_{x}$)$_{5}$  is similar to $\langle \mu_{eff} ^{2} \rangle\sim$0.4 $\mu_{B}^{2}$ reported to reside in the resonance peak in parent CeCoIn$_{5}$ \cite{Stock08:100}.  Changing the chemical potential through hole doping (Hg) therefore shifts this dynamic spectral weight to E=0 (within the resolution of the measurements discussed here) allowing little spectral weight to form a resonant excitation at finite energy, measured in parent CeCoIn$_{5}$.  This result is in agreement with the significantly reduced superconducting gap, reflected in the specific heat jump $\Delta C/\gamma T_{c} \sim 0.8$ in CeCo(In$_{0.990}$Hg$_{0.015}$)$_{5}$ compared to 4.5 in CeCoIn$_{5}$.~\cite{Mov01:86}

Our work on CeCo(In$_{1-x}$Hg$_{x}$)$_{5}$ shows that a three dimensional Ising phase is parent to superconductivity in CeCoIn$_{5}$.  The lack of evidence of any low energy transverse fluctuations and the presence of a commensurate low temperature structure with [001] ordered moments that mimic the polarization and momentum dependence of the resonance peak in CeCoIn$_{5}$ supports this description of the magnetic order parameter.  This indicates that transverse excitations are likely heavily damped in energy and, hence, are unstable.  While transverse magnetic fluctuations have been reported in CeRhIn$_{5}$~\cite{Das14:113}, we note these fluctuations are also unstable with a momentum and energy broadened continuum of magnetic excitations resulting from spontaneous decay, or multiparticle states~\cite{Stock15:114}.  

Both the resonant peak in CeCoIn$_{5}$ and the $c$-axis polarized three dimensional fluctuations in CeCo(In$_{1-x}$Hg$_{x}$)$_{5}$ differ from the magnetic response in other materials with coexistent magnetism and superconductivity.  Isotropic short-range magnetic order is proximate to unconventional superconductivity in the cuprates~\cite{Fujita02:65,Stock06:73,Stock08:77,Yamani15:91}, while three dimensional order is nearby in pnictides~\cite{Pajerowski13:87,Wilson10:82,Wilson10:81}.  The anisotropy is also reflected in the magnetic dynamics with a predominately $c$-axis polarized resonance present in CeCoIn$_{5}$ compared with the isotropic ``spin-1" response measured in the cuprates.~\cite{Headings10:105,Stock04:69}  The underlying anisotropy has been used to explain the magnetic field dependence of the CeCoIn$_{5}$ resonance peak~\cite{Akbari12:86,Thalmeier13:86} compared to explanations for the cuprates~\cite{Ismer07:99}.  However, similar to the case discussed here, transverse spin-waves are indeed unstable in other unconventional superconductors including iron- and copper-based high-T$_c$ superconductors.~\cite{Headings10:105,Stock07:75,Stock10:82,Stock14:90,Stock17:95}

The work here on CeCo(In$_{1-x}$Hg$_{x}$)$_{5}$ contrasts with the inelastic scattering measurements on Yb doped CeCoIn$_{5}$ that report a robust resonance peak against doping with a dispersing momentum dependence with increasing energy transfer.~\cite{Song16:7}  This observation was used to argue that the resonance peak is associated with spin fluctuations seen in the superconducting state due to a reduction in dampening, similar to magnon quasiparticles in insulating quantum magnets ~\cite{Chubukov08:101}.  In our current work, we report the presence of three dimensional $c$-axis magnetic order and the lack of low-energy transverse magnetic dynamics that reflects the Ising-like order.  The $c$-axis polarized magnetic order in CeCo(In$_{1-x}$Hg$_{x}$)$_{5}$ is similar to the magnetic resonance peak in the superconducting phase of pure CeCoIn$_5$.  We  do not observe a strong resonance peak at $\sim$ 0.5 meV despite superconductivity being present, albeit subdued.  This indicates that such a gapped spin excitation is not present when a superconducting order parameter is suppressed through hole doping in favor of Ising order.  The resonance peak reflects the underlying itinerant response and the superconducting gap symmetry with spectral weight being drawn from competing magnetic orders.  

This work was funded by the Carnegie Trust for the Universities of Scotland,  the STFC, and the EPSRC. Work at Los Alamos National Laboratory (LANL) was supported by the U.S. Department of Energy, Office of Basic Energy Sciences, Division of Materials Sciences and Engineering (material synthesis and characterization).


\begin{thebibliography}{95}%
\makeatletter
\providecommand \@ifxundefined [1]{%
 \@ifx{#1\undefined}
}%
\providecommand \@ifnum [1]{%
 \ifnum #1\expandafter \@firstoftwo
 \else \expandafter \@secondoftwo
 \fi
}%
\providecommand \@ifx [1]{%
 \ifx #1\expandafter \@firstoftwo
 \else \expandafter \@secondoftwo
 \fi
}%
\providecommand \natexlab [1]{#1}%
\providecommand \enquote  [1]{``#1''}%
\providecommand \bibnamefont  [1]{#1}%
\providecommand \bibfnamefont [1]{#1}%
\providecommand \citenamefont [1]{#1}%
\providecommand \href@noop [0]{\@secondoftwo}%
\providecommand \href [0]{\begingroup \@sanitize@url \@href}%
\providecommand \@href[1]{\@@startlink{#1}\@@href}%
\providecommand \@@href[1]{\endgroup#1\@@endlink}%
\providecommand \@sanitize@url [0]{\catcode `\\12\catcode `\$12\catcode
  `\&12\catcode `\#12\catcode `\^12\catcode `\_12\catcode `\%12\relax}%
\providecommand \@@startlink[1]{}%
\providecommand \@@endlink[0]{}%
\providecommand \url  [0]{\begingroup\@sanitize@url \@url }%
\providecommand \@url [1]{\endgroup\@href {#1}{\urlprefix }}%
\providecommand \urlprefix  [0]{URL }%
\providecommand \Eprint [0]{\href }%
\providecommand \doibase [0]{http://dx.doi.org/}%
\providecommand \selectlanguage [0]{\@gobble}%
\providecommand \bibinfo  [0]{\@secondoftwo}%
\providecommand \bibfield  [0]{\@secondoftwo}%
\providecommand \translation [1]{[#1]}%
\providecommand \BibitemOpen [0]{}%
\providecommand \bibitemStop [0]{}%
\providecommand \bibitemNoStop [0]{.\EOS\space}%
\providecommand \EOS [0]{\spacefactor3000\relax}%
\providecommand \BibitemShut  [1]{\csname bibitem#1\endcsname}%
\let\auto@bib@innerbib\@empty
\bibitem [{\citenamefont {Miyake}\ \emph {et~al.}(1986)\citenamefont {Miyake},
  \citenamefont {Schmitt-Rink},\ and\ \citenamefont {Varma}}]{Miyake86:34}%
  \BibitemOpen
  \bibfield  {author} {\bibinfo {author} {\bibfnamefont {K.}~\bibnamefont
  {Miyake}}, \bibinfo {author} {\bibfnamefont {S.}~\bibnamefont
  {Schmitt-Rink}}, \ and\ \bibinfo {author} {\bibfnamefont {C.~M.}\
  \bibnamefont {Varma}},\ }\href@noop {} {\bibfield  {journal} {\bibinfo
  {journal} {Phys. Rev. B}\ }\textbf {\bibinfo {volume} {34}},\ \bibinfo
  {pages} {6554} (\bibinfo {year} {1986})}\BibitemShut {NoStop}%
\bibitem [{\citenamefont {Norman}(2011)}]{Norman11:332}%
  \BibitemOpen
  \bibfield  {author} {\bibinfo {author} {\bibfnamefont {M.~R.}\ \bibnamefont
  {Norman}},\ }\href@noop {} {\bibfield  {journal} {\bibinfo  {journal}
  {Science}\ }\textbf {\bibinfo {volume} {332}},\ \bibinfo {pages} {196}
  (\bibinfo {year} {2011})}\BibitemShut {NoStop}%
\bibitem [{\citenamefont {Knafo}\ \emph {et~al.}(2009)\citenamefont {Knafo},
  \citenamefont {Raymond}, \citenamefont {Lejay},\ and\ \citenamefont
  {Flouquet}}]{Knafo09:5}%
  \BibitemOpen
  \bibfield  {author} {\bibinfo {author} {\bibfnamefont {W.}~\bibnamefont
  {Knafo}}, \bibinfo {author} {\bibfnamefont {S.}~\bibnamefont {Raymond}},
  \bibinfo {author} {\bibfnamefont {P.}~\bibnamefont {Lejay}}, \ and\ \bibinfo
  {author} {\bibfnamefont {J.}~\bibnamefont {Flouquet}},\ }\href@noop {}
  {\bibfield  {journal} {\bibinfo  {journal} {{Nat. Phys.}}\ }\textbf {\bibinfo
  {volume} {{5}}},\ \bibinfo {pages} {{753}} (\bibinfo {year}
  {{2009}})}\BibitemShut {NoStop}%
\bibitem [{\citenamefont {Chen}\ \emph {et~al.}(2017)\citenamefont {Chen},
  \citenamefont {Xu}, \citenamefont {Niu}, \citenamefont {Jiang}, \citenamefont
  {Peng}, \citenamefont {Xu}, \citenamefont {Wen}, \citenamefont {Ding},
  \citenamefont {Huang}, \citenamefont {Shu}, \citenamefont {Zhang},
  \citenamefont {Lee}, \citenamefont {Strocov}, \citenamefont {Shi},
  \citenamefont {Bisti}, \citenamefont {Schmitt}, \citenamefont {Huang},
  \citenamefont {Dudin}, \citenamefont {Lai}, \citenamefont {Kirchner},
  \citenamefont {Yuan},\ and\ \citenamefont {Feng}}]{Chen17:96}%
  \BibitemOpen
  \bibfield  {author} {\bibinfo {author} {\bibfnamefont {Q.~Y.}\ \bibnamefont
  {Chen}}, \bibinfo {author} {\bibfnamefont {D.~F.}\ \bibnamefont {Xu}},
  \bibinfo {author} {\bibfnamefont {X.~H.}\ \bibnamefont {Niu}}, \bibinfo
  {author} {\bibfnamefont {J.}~\bibnamefont {Jiang}}, \bibinfo {author}
  {\bibfnamefont {R.}~\bibnamefont {Peng}}, \bibinfo {author} {\bibfnamefont
  {H.~C.}\ \bibnamefont {Xu}}, \bibinfo {author} {\bibfnamefont {C.~H.~P.}\
  \bibnamefont {Wen}}, \bibinfo {author} {\bibfnamefont {Z.~F.}\ \bibnamefont
  {Ding}}, \bibinfo {author} {\bibfnamefont {K.}~\bibnamefont {Huang}},
  \bibinfo {author} {\bibfnamefont {L.}~\bibnamefont {Shu}}, \bibinfo {author}
  {\bibfnamefont {Y.~J.}\ \bibnamefont {Zhang}}, \bibinfo {author}
  {\bibfnamefont {H.}~\bibnamefont {Lee}}, \bibinfo {author} {\bibfnamefont
  {V.~N.}\ \bibnamefont {Strocov}}, \bibinfo {author} {\bibfnamefont
  {M.}~\bibnamefont {Shi}}, \bibinfo {author} {\bibfnamefont {F.}~\bibnamefont
  {Bisti}}, \bibinfo {author} {\bibfnamefont {T.}~\bibnamefont {Schmitt}},
  \bibinfo {author} {\bibfnamefont {Y.~B.}\ \bibnamefont {Huang}}, \bibinfo
  {author} {\bibfnamefont {P.}~\bibnamefont {Dudin}}, \bibinfo {author}
  {\bibfnamefont {X.~C.}\ \bibnamefont {Lai}}, \bibinfo {author} {\bibfnamefont
  {S.}~\bibnamefont {Kirchner}}, \bibinfo {author} {\bibfnamefont {H.~Q.}\
  \bibnamefont {Yuan}}, \ and\ \bibinfo {author} {\bibfnamefont {D.~L.}\
  \bibnamefont {Feng}},\ }\href@noop {} {\bibfield  {journal} {\bibinfo
  {journal} {Phys. Rev. B}\ }\textbf {\bibinfo {volume} {96}},\ \bibinfo
  {pages} {045107} (\bibinfo {year} {2017})}\BibitemShut {NoStop}%
\bibitem [{\citenamefont {Pfleiderer}(2009)}]{Pfleiderer09:81}%
  \BibitemOpen
  \bibfield  {author} {\bibinfo {author} {\bibfnamefont {C.}~\bibnamefont
  {Pfleiderer}},\ }\href@noop {} {\bibfield  {journal} {\bibinfo  {journal}
  {Rev. Mod. Phys.}\ }\textbf {\bibinfo {volume} {81}},\ \bibinfo {pages}
  {1551} (\bibinfo {year} {2009})}\BibitemShut {NoStop}%
\bibitem [{\citenamefont {Scalapino}(2012)}]{Scalapino12:84}%
  \BibitemOpen
  \bibfield  {author} {\bibinfo {author} {\bibfnamefont {D.~J.}\ \bibnamefont
  {Scalapino}},\ }\href@noop {} {\bibfield  {journal} {\bibinfo  {journal}
  {Rev. Mod. Phys.}\ }\textbf {\bibinfo {volume} {84}},\ \bibinfo {pages}
  {1383} (\bibinfo {year} {2012})}\BibitemShut {NoStop}%
\bibitem [{\citenamefont {Stewart}(1984)}]{Stewart84:56}%
  \BibitemOpen
  \bibfield  {author} {\bibinfo {author} {\bibfnamefont {G.~R.}\ \bibnamefont
  {Stewart}},\ }\href@noop {} {\bibfield  {journal} {\bibinfo  {journal} {Rev.
  Mod. Phys.}\ }\textbf {\bibinfo {volume} {56}},\ \bibinfo {pages} {755}
  (\bibinfo {year} {1984})}\BibitemShut {NoStop}%
\bibitem [{\citenamefont {Timusk}\ and\ \citenamefont
  {Statt}(1999)}]{Timusk99:62}%
  \BibitemOpen
  \bibfield  {author} {\bibinfo {author} {\bibfnamefont {T.}~\bibnamefont
  {Timusk}}\ and\ \bibinfo {author} {\bibfnamefont {B.~W.}\ \bibnamefont
  {Statt}},\ }\href@noop {} {\bibfield  {journal} {\bibinfo  {journal} {Rep.
  Prog. Phys.}\ }\textbf {\bibinfo {volume} {62}},\ \bibinfo {pages} {61}
  (\bibinfo {year} {1999})}\BibitemShut {NoStop}%
\bibitem [{\citenamefont {Varma}(2006)}]{Varma06:73}%
  \BibitemOpen
  \bibfield  {author} {\bibinfo {author} {\bibfnamefont {C.~M.}\ \bibnamefont
  {Varma}},\ }\href@noop {} {\bibfield  {journal} {\bibinfo  {journal} {Phys.
  Rev. B}\ }\textbf {\bibinfo {volume} {73}},\ \bibinfo {pages} {155113}
  (\bibinfo {year} {2006})}\BibitemShut {NoStop}%
\bibitem [{\citenamefont {Proust}\ \emph {et~al.}(2002)\citenamefont {Proust},
  \citenamefont {Boaknin}, \citenamefont {Hill}, \citenamefont {Taillefer},\
  and\ \citenamefont {Mackenzie}}]{Proust02:89}%
  \BibitemOpen
  \bibfield  {author} {\bibinfo {author} {\bibfnamefont {C.}~\bibnamefont
  {Proust}}, \bibinfo {author} {\bibfnamefont {E.}~\bibnamefont {Boaknin}},
  \bibinfo {author} {\bibfnamefont {R.~W.}\ \bibnamefont {Hill}}, \bibinfo
  {author} {\bibfnamefont {L.}~\bibnamefont {Taillefer}}, \ and\ \bibinfo
  {author} {\bibfnamefont {A.~P.}\ \bibnamefont {Mackenzie}},\ }\href@noop {}
  {\bibfield  {journal} {\bibinfo  {journal} {Phys. Rev. Lett.}\ }\textbf
  {\bibinfo {volume} {89}},\ \bibinfo {pages} {147003} (\bibinfo {year}
  {2002})}\BibitemShut {NoStop}%
\bibitem [{\citenamefont {Chu}\ \emph {et~al.}(2010)\citenamefont {Chu},
  \citenamefont {Analytis}, \citenamefont {Greve}, \citenamefont {McMahon},
  \citenamefont {Islam}, \citenamefont {Yamamoto},\ and\ \citenamefont
  {Fisher}}]{Chu10:329}%
  \BibitemOpen
  \bibfield  {author} {\bibinfo {author} {\bibfnamefont {J.~H.}\ \bibnamefont
  {Chu}}, \bibinfo {author} {\bibfnamefont {J.~G.}\ \bibnamefont {Analytis}},
  \bibinfo {author} {\bibfnamefont {K.~D.}\ \bibnamefont {Greve}}, \bibinfo
  {author} {\bibfnamefont {P.~L.}\ \bibnamefont {McMahon}}, \bibinfo {author}
  {\bibfnamefont {Z.}~\bibnamefont {Islam}}, \bibinfo {author} {\bibfnamefont
  {Y.}~\bibnamefont {Yamamoto}}, \ and\ \bibinfo {author} {\bibfnamefont
  {I.~R.}\ \bibnamefont {Fisher}},\ }\href@noop {} {\bibfield  {journal}
  {\bibinfo  {journal} {Science}\ }\textbf {\bibinfo {volume} {329}},\ \bibinfo
  {pages} {824} (\bibinfo {year} {2010})}\BibitemShut {NoStop}%
\bibitem [{\citenamefont {Fong}\ \emph {et~al.}(1997)\citenamefont {Fong},
  \citenamefont {Keimer}, \citenamefont {Milius},\ and\ \citenamefont
  {Aksay}}]{Fong97:78}%
  \BibitemOpen
  \bibfield  {author} {\bibinfo {author} {\bibfnamefont {H.~F.}\ \bibnamefont
  {Fong}}, \bibinfo {author} {\bibfnamefont {B.}~\bibnamefont {Keimer}},
  \bibinfo {author} {\bibfnamefont {D.~L.}\ \bibnamefont {Milius}}, \ and\
  \bibinfo {author} {\bibfnamefont {I.~A.}\ \bibnamefont {Aksay}},\ }\href@noop
  {} {\bibfield  {journal} {\bibinfo  {journal} {Phys. Rev. Lett.}\ }\textbf
  {\bibinfo {volume} {78}},\ \bibinfo {pages} {713} (\bibinfo {year}
  {1997})}\BibitemShut {NoStop}%
\bibitem [{\citenamefont {Fong}\ \emph {et~al.}(2000)\citenamefont {Fong},
  \citenamefont {Bourges}, \citenamefont {Sidis}, \citenamefont {Regnault},
  \citenamefont {Bossy}, \citenamefont {Ivanov}, \citenamefont {Milius},
  \citenamefont {Aksay},\ and\ \citenamefont {Keimer}}]{Fong00:61}%
  \BibitemOpen
  \bibfield  {author} {\bibinfo {author} {\bibfnamefont {H.~F.}\ \bibnamefont
  {Fong}}, \bibinfo {author} {\bibfnamefont {P.}~\bibnamefont {Bourges}},
  \bibinfo {author} {\bibfnamefont {Y.}~\bibnamefont {Sidis}}, \bibinfo
  {author} {\bibfnamefont {L.~P.}\ \bibnamefont {Regnault}}, \bibinfo {author}
  {\bibfnamefont {J.}~\bibnamefont {Bossy}}, \bibinfo {author} {\bibfnamefont
  {A.}~\bibnamefont {Ivanov}}, \bibinfo {author} {\bibfnamefont {D.~L.}\
  \bibnamefont {Milius}}, \bibinfo {author} {\bibfnamefont {I.~A.}\
  \bibnamefont {Aksay}}, \ and\ \bibinfo {author} {\bibfnamefont
  {B.}~\bibnamefont {Keimer}},\ }\href@noop {} {\bibfield  {journal} {\bibinfo
  {journal} {Phys. Rev. B}\ }\textbf {\bibinfo {volume} {61}},\ \bibinfo
  {pages} {14773} (\bibinfo {year} {2000})}\BibitemShut {NoStop}%
\bibitem [{\citenamefont {Dai}\ \emph {et~al.}(1996)\citenamefont {Dai},
  \citenamefont {Yethiraj}, \citenamefont {Mook}, \citenamefont {Lindemer},\
  and\ \citenamefont {Dogan}}]{Dai96:77}%
  \BibitemOpen
  \bibfield  {author} {\bibinfo {author} {\bibfnamefont {P.}~\bibnamefont
  {Dai}}, \bibinfo {author} {\bibfnamefont {M.}~\bibnamefont {Yethiraj}},
  \bibinfo {author} {\bibfnamefont {H.~A.}\ \bibnamefont {Mook}}, \bibinfo
  {author} {\bibfnamefont {T.~B.}\ \bibnamefont {Lindemer}}, \ and\ \bibinfo
  {author} {\bibfnamefont {F.}~\bibnamefont {Dogan}},\ }\href@noop {}
  {\bibfield  {journal} {\bibinfo  {journal} {Phys. Rev. Lett.}\ }\textbf
  {\bibinfo {volume} {77}},\ \bibinfo {pages} {5425} (\bibinfo {year}
  {1996})}\BibitemShut {NoStop}%
\bibitem [{\citenamefont {Dai}\ \emph {et~al.}(2001)\citenamefont {Dai},
  \citenamefont {Mook}, \citenamefont {Hunt},\ and\ \citenamefont
  {Dogan}}]{Dai01:63}%
  \BibitemOpen
  \bibfield  {author} {\bibinfo {author} {\bibfnamefont {P.}~\bibnamefont
  {Dai}}, \bibinfo {author} {\bibfnamefont {H.~A.}\ \bibnamefont {Mook}},
  \bibinfo {author} {\bibfnamefont {R.~D.}\ \bibnamefont {Hunt}}, \ and\
  \bibinfo {author} {\bibfnamefont {F.}~\bibnamefont {Dogan}},\ }\href@noop {}
  {\bibfield  {journal} {\bibinfo  {journal} {Phys. Rev. B}\ }\textbf {\bibinfo
  {volume} {63}},\ \bibinfo {pages} {054525} (\bibinfo {year}
  {2001})}\BibitemShut {NoStop}%
\bibitem [{\citenamefont {Stockert}\ \emph {et~al.}(2011)\citenamefont
  {Stockert}, \citenamefont {Arndt}, \citenamefont {Faulhaber}, \citenamefont
  {Geibel}, \citenamefont {Jeevan}, \citenamefont {Kirchner}, \citenamefont
  {Loewenhaupt}, \citenamefont {Schmalzl}, \citenamefont {Schmidt},
  \citenamefont {Si},\ and\ \citenamefont {Steglich}}]{Stockert09:7}%
  \BibitemOpen
  \bibfield  {author} {\bibinfo {author} {\bibfnamefont {O.}~\bibnamefont
  {Stockert}}, \bibinfo {author} {\bibfnamefont {J.}~\bibnamefont {Arndt}},
  \bibinfo {author} {\bibfnamefont {E.}~\bibnamefont {Faulhaber}}, \bibinfo
  {author} {\bibfnamefont {C.}~\bibnamefont {Geibel}}, \bibinfo {author}
  {\bibfnamefont {H.~S.}\ \bibnamefont {Jeevan}}, \bibinfo {author}
  {\bibfnamefont {S.}~\bibnamefont {Kirchner}}, \bibinfo {author}
  {\bibfnamefont {M.}~\bibnamefont {Loewenhaupt}}, \bibinfo {author}
  {\bibfnamefont {K.}~\bibnamefont {Schmalzl}}, \bibinfo {author}
  {\bibfnamefont {W.}~\bibnamefont {Schmidt}}, \bibinfo {author} {\bibfnamefont
  {Q.}~\bibnamefont {Si}}, \ and\ \bibinfo {author} {\bibfnamefont
  {F.}~\bibnamefont {Steglich}},\ }\href@noop {} {\bibfield  {journal}
  {\bibinfo  {journal} {Nat. Physics}\ }\textbf {\bibinfo {volume} {7}},\
  \bibinfo {pages} {119} (\bibinfo {year} {2011})}\BibitemShut {NoStop}%
\bibitem [{\citenamefont {Sato}\ \emph {et~al.}(2001)\citenamefont {Sato},
  \citenamefont {Aso}, \citenamefont {Miyake}, \citenamefont {Shiina},
  \citenamefont {Thalmeier}, \citenamefont {Varelogiannis}, \citenamefont
  {Geibel}, \citenamefont {Steglich}, \citenamefont {Fulde},\ and\
  \citenamefont {Komatsubara}}]{Sato01:410}%
  \BibitemOpen
  \bibfield  {author} {\bibinfo {author} {\bibfnamefont {N.~K.}\ \bibnamefont
  {Sato}}, \bibinfo {author} {\bibfnamefont {N.}~\bibnamefont {Aso}}, \bibinfo
  {author} {\bibfnamefont {K.}~\bibnamefont {Miyake}}, \bibinfo {author}
  {\bibfnamefont {R.}~\bibnamefont {Shiina}}, \bibinfo {author} {\bibfnamefont
  {P.}~\bibnamefont {Thalmeier}}, \bibinfo {author} {\bibfnamefont
  {G.}~\bibnamefont {Varelogiannis}}, \bibinfo {author} {\bibfnamefont
  {C.}~\bibnamefont {Geibel}}, \bibinfo {author} {\bibfnamefont
  {F.}~\bibnamefont {Steglich}}, \bibinfo {author} {\bibfnamefont
  {P.}~\bibnamefont {Fulde}}, \ and\ \bibinfo {author} {\bibfnamefont
  {T.}~\bibnamefont {Komatsubara}},\ }\href@noop {} {\bibfield  {journal}
  {\bibinfo  {journal} {Nature}\ }\textbf {\bibinfo {volume} {410}},\ \bibinfo
  {pages} {340} (\bibinfo {year} {2001})}\BibitemShut {NoStop}%
\bibitem [{\citenamefont {Bernhoeft}\ \emph {et~al.}(1998)\citenamefont
  {Bernhoeft}, \citenamefont {Sato}, \citenamefont {Roessli}, \citenamefont
  {Aso}, \citenamefont {Hiess}, \citenamefont {Lander}, \citenamefont {Endoh},\
  and\ \citenamefont {Komatsubara}}]{Bernhoeft98:81}%
  \BibitemOpen
  \bibfield  {author} {\bibinfo {author} {\bibfnamefont {N.}~\bibnamefont
  {Bernhoeft}}, \bibinfo {author} {\bibfnamefont {N.}~\bibnamefont {Sato}},
  \bibinfo {author} {\bibfnamefont {B.}~\bibnamefont {Roessli}}, \bibinfo
  {author} {\bibfnamefont {N.}~\bibnamefont {Aso}}, \bibinfo {author}
  {\bibfnamefont {A.}~\bibnamefont {Hiess}}, \bibinfo {author} {\bibfnamefont
  {G.~H.}\ \bibnamefont {Lander}}, \bibinfo {author} {\bibfnamefont
  {Y.}~\bibnamefont {Endoh}}, \ and\ \bibinfo {author} {\bibfnamefont
  {T.}~\bibnamefont {Komatsubara}},\ }\href@noop {} {\bibfield  {journal}
  {\bibinfo  {journal} {Phys. Rev. Lett.}\ }\textbf {\bibinfo {volume} {81}},\
  \bibinfo {pages} {4244} (\bibinfo {year} {1998})}\BibitemShut {NoStop}%
\bibitem [{\citenamefont {Hiess}\ \emph {et~al.}(2006)\citenamefont {Hiess},
  \citenamefont {Bernhoeft}, \citenamefont {Metoki}, \citenamefont {Lander},
  \citenamefont {Roessli}, \citenamefont {Sato}, \citenamefont {Aso},
  \citenamefont {Haga}, \citenamefont {Koike}, \citenamefont {Komatsubara},\
  and\ \citenamefont {Onuki}}]{Hess06:18}%
  \BibitemOpen
  \bibfield  {author} {\bibinfo {author} {\bibfnamefont {A.}~\bibnamefont
  {Hiess}}, \bibinfo {author} {\bibfnamefont {N.}~\bibnamefont {Bernhoeft}},
  \bibinfo {author} {\bibfnamefont {N.}~\bibnamefont {Metoki}}, \bibinfo
  {author} {\bibfnamefont {G.~H.}\ \bibnamefont {Lander}}, \bibinfo {author}
  {\bibfnamefont {B.}~\bibnamefont {Roessli}}, \bibinfo {author} {\bibfnamefont
  {N.~K.}\ \bibnamefont {Sato}}, \bibinfo {author} {\bibfnamefont
  {N.}~\bibnamefont {Aso}}, \bibinfo {author} {\bibfnamefont {Y.}~\bibnamefont
  {Haga}}, \bibinfo {author} {\bibfnamefont {Y.}~\bibnamefont {Koike}},
  \bibinfo {author} {\bibfnamefont {T.}~\bibnamefont {Komatsubara}}, \ and\
  \bibinfo {author} {\bibfnamefont {Y.}~\bibnamefont {Onuki}},\ }\href@noop {}
  {\bibfield  {journal} {\bibinfo  {journal} {J. Phys. Condens. Matter}\
  }\textbf {\bibinfo {volume} {18}},\ \bibinfo {pages} {R437} (\bibinfo {year}
  {2006})}\BibitemShut {NoStop}%
\bibitem [{\citenamefont {Hiess}\ \emph {et~al.}(2007)\citenamefont {Hiess},
  \citenamefont {Blackburn}, \citenamefont {Bernhoeft},\ and\ \citenamefont
  {Lander}}]{Hiess07:76}%
  \BibitemOpen
  \bibfield  {author} {\bibinfo {author} {\bibfnamefont {A.}~\bibnamefont
  {Hiess}}, \bibinfo {author} {\bibfnamefont {E.}~\bibnamefont {Blackburn}},
  \bibinfo {author} {\bibfnamefont {N.}~\bibnamefont {Bernhoeft}}, \ and\
  \bibinfo {author} {\bibfnamefont {G.~H.}\ \bibnamefont {Lander}},\
  }\href@noop {} {\bibfield  {journal} {\bibinfo  {journal} {Phys. Rev. B}\
  }\textbf {\bibinfo {volume} {76}},\ \bibinfo {pages} {132405} (\bibinfo
  {year} {2007})}\BibitemShut {NoStop}%
\bibitem [{\citenamefont {Christianson}\ \emph {et~al.}(2008)\citenamefont
  {Christianson}, \citenamefont {Goremychkin}, \citenamefont {Osborn},
  \citenamefont {Rosenkranz}, \citenamefont {Lumsden}, \citenamefont
  {Malliakas}, \citenamefont {Todorov}, \citenamefont {Claus}, \citenamefont
  {Chung}, \citenamefont {Kanatzidis}, \citenamefont {Bewley},\ and\
  \citenamefont {Guidi}}]{Christianson08:456}%
  \BibitemOpen
  \bibfield  {author} {\bibinfo {author} {\bibfnamefont {A.~D.}\ \bibnamefont
  {Christianson}}, \bibinfo {author} {\bibfnamefont {E.~A.}\ \bibnamefont
  {Goremychkin}}, \bibinfo {author} {\bibfnamefont {R.}~\bibnamefont {Osborn}},
  \bibinfo {author} {\bibfnamefont {S.}~\bibnamefont {Rosenkranz}}, \bibinfo
  {author} {\bibfnamefont {M.~D.}\ \bibnamefont {Lumsden}}, \bibinfo {author}
  {\bibfnamefont {C.~D.}\ \bibnamefont {Malliakas}}, \bibinfo {author}
  {\bibfnamefont {I.~S.}\ \bibnamefont {Todorov}}, \bibinfo {author}
  {\bibfnamefont {H.}~\bibnamefont {Claus}}, \bibinfo {author} {\bibfnamefont
  {D.~Y.}\ \bibnamefont {Chung}}, \bibinfo {author} {\bibfnamefont {M.~G.}\
  \bibnamefont {Kanatzidis}}, \bibinfo {author} {\bibfnamefont {R.~I.}\
  \bibnamefont {Bewley}}, \ and\ \bibinfo {author} {\bibfnamefont
  {T.}~\bibnamefont {Guidi}},\ }\href@noop {} {\bibfield  {journal} {\bibinfo
  {journal} {Nature}\ }\textbf {\bibinfo {volume} {456}},\ \bibinfo {pages}
  {930} (\bibinfo {year} {2008})}\BibitemShut {NoStop}%
\bibitem [{\citenamefont {Dai}(2015)}]{Dai15:87}%
  \BibitemOpen
  \bibfield  {author} {\bibinfo {author} {\bibfnamefont {P.}~\bibnamefont
  {Dai}},\ }\href@noop {} {\bibfield  {journal} {\bibinfo  {journal} {Rev. Mod.
  Phys.}\ }\textbf {\bibinfo {volume} {87}},\ \bibinfo {pages} {855} (\bibinfo
  {year} {2015})}\BibitemShut {NoStop}%
\bibitem [{\citenamefont {Stock}\ \emph
  {et~al.}(2008{\natexlab{a}})\citenamefont {Stock}, \citenamefont {Broholm},
  \citenamefont {Hudis}, \citenamefont {Kang},\ and\ \citenamefont
  {Petrovic}}]{Stock08:100}%
  \BibitemOpen
  \bibfield  {author} {\bibinfo {author} {\bibfnamefont {C.}~\bibnamefont
  {Stock}}, \bibinfo {author} {\bibfnamefont {C.}~\bibnamefont {Broholm}},
  \bibinfo {author} {\bibfnamefont {J.}~\bibnamefont {Hudis}}, \bibinfo
  {author} {\bibfnamefont {H.~J.}\ \bibnamefont {Kang}}, \ and\ \bibinfo
  {author} {\bibfnamefont {C.}~\bibnamefont {Petrovic}},\ }\href@noop {}
  {\bibfield  {journal} {\bibinfo  {journal} {Phys. Rev. Lett.}\ }\textbf
  {\bibinfo {volume} {100}},\ \bibinfo {pages} {087001} (\bibinfo {year}
  {2008}{\natexlab{a}})}\BibitemShut {NoStop}%
\bibitem [{\citenamefont {Stock}\ \emph {et~al.}(2012)\citenamefont {Stock},
  \citenamefont {Broholm}, \citenamefont {Zhao}, \citenamefont {Demmel},
  \citenamefont {Kang}, \citenamefont {Rule},\ and\ \citenamefont
  {Petrovic}}]{Stock12:109}%
  \BibitemOpen
  \bibfield  {author} {\bibinfo {author} {\bibfnamefont {C.}~\bibnamefont
  {Stock}}, \bibinfo {author} {\bibfnamefont {C.}~\bibnamefont {Broholm}},
  \bibinfo {author} {\bibfnamefont {Y.}~\bibnamefont {Zhao}}, \bibinfo {author}
  {\bibfnamefont {F.}~\bibnamefont {Demmel}}, \bibinfo {author} {\bibfnamefont
  {H.~J.}\ \bibnamefont {Kang}}, \bibinfo {author} {\bibfnamefont {K.~C.}\
  \bibnamefont {Rule}}, \ and\ \bibinfo {author} {\bibfnamefont
  {C.}~\bibnamefont {Petrovic}},\ }\href@noop {} {\bibfield  {journal}
  {\bibinfo  {journal} {Phys. Rev. Lett.}\ }\textbf {\bibinfo {volume} {109}},\
  \bibinfo {pages} {167207} (\bibinfo {year} {2012})}\BibitemShut {NoStop}%
\bibitem [{\citenamefont {Panarin}\ \emph {et~al.}(2009)\citenamefont
  {Panarin}, \citenamefont {Raymond}, \citenamefont {Lapertot},\ and\
  \citenamefont {Flouquet}}]{Panarin09:78}%
  \BibitemOpen
  \bibfield  {author} {\bibinfo {author} {\bibfnamefont {J.}~\bibnamefont
  {Panarin}}, \bibinfo {author} {\bibfnamefont {S.}~\bibnamefont {Raymond}},
  \bibinfo {author} {\bibfnamefont {G.}~\bibnamefont {Lapertot}}, \ and\
  \bibinfo {author} {\bibfnamefont {J.}~\bibnamefont {Flouquet}},\ }\href@noop
  {} {\bibfield  {journal} {\bibinfo  {journal} {J. Phys. Soc. Jpn.}\ }\textbf
  {\bibinfo {volume} {78}},\ \bibinfo {pages} {113706} (\bibinfo {year}
  {2009})}\BibitemShut {NoStop}%
\bibitem [{\citenamefont {Norman}(2000)}]{Norman00:61}%
  \BibitemOpen
  \bibfield  {author} {\bibinfo {author} {\bibfnamefont {M.~R.}\ \bibnamefont
  {Norman}},\ }\href@noop {} {\bibfield  {journal} {\bibinfo  {journal} {Phys.
  Rev. B}\ }\textbf {\bibinfo {volume} {61}},\ \bibinfo {pages} {14751}
  (\bibinfo {year} {2000})}\BibitemShut {NoStop}%
\bibitem [{\citenamefont {Eremin}\ \emph {et~al.}(2008)\citenamefont {Eremin},
  \citenamefont {Zwicknagl}, \citenamefont {Thalmeier},\ and\ \citenamefont
  {Fulde}}]{Eremin08:101}%
  \BibitemOpen
  \bibfield  {author} {\bibinfo {author} {\bibfnamefont {I.}~\bibnamefont
  {Eremin}}, \bibinfo {author} {\bibfnamefont {G.}~\bibnamefont {Zwicknagl}},
  \bibinfo {author} {\bibfnamefont {P.}~\bibnamefont {Thalmeier}}, \ and\
  \bibinfo {author} {\bibfnamefont {P.}~\bibnamefont {Fulde}},\ }\href@noop {}
  {\bibfield  {journal} {\bibinfo  {journal} {Phys. Rev. Lett.}\ }\textbf
  {\bibinfo {volume} {101}},\ \bibinfo {pages} {187001} (\bibinfo {year}
  {2008})}\BibitemShut {NoStop}%
\bibitem [{\citenamefont {Thalmeier}\ and\ \citenamefont
  {Akbari}(2013)}]{Thalmeier13:86}%
  \BibitemOpen
  \bibfield  {author} {\bibinfo {author} {\bibfnamefont {P.}~\bibnamefont
  {Thalmeier}}\ and\ \bibinfo {author} {\bibfnamefont {A.}~\bibnamefont
  {Akbari}},\ }\href@noop {} {\bibfield  {journal} {\bibinfo  {journal} {Eur.
  Phys. J. B}\ }\textbf {\bibinfo {volume} {86}},\ \bibinfo {pages} {82}
  (\bibinfo {year} {2013})}\BibitemShut {NoStop}%
\bibitem [{\citenamefont {Friemel}\ \emph {et~al.}(2012)\citenamefont
  {Friemel}, \citenamefont {Li}, \citenamefont {Dukhnenko}, \citenamefont
  {Shitsevalova}, \citenamefont {Sluchanko}, \citenamefont {Ivanov},
  \citenamefont {Filipov}, \citenamefont {Keimer},\ and\ \citenamefont
  {Inosov}}]{Friemel12:3}%
  \BibitemOpen
  \bibfield  {author} {\bibinfo {author} {\bibfnamefont {G.}~\bibnamefont
  {Friemel}}, \bibinfo {author} {\bibfnamefont {Y.}~\bibnamefont {Li}},
  \bibinfo {author} {\bibfnamefont {A.~V.}\ \bibnamefont {Dukhnenko}}, \bibinfo
  {author} {\bibfnamefont {N.~Y.}\ \bibnamefont {Shitsevalova}}, \bibinfo
  {author} {\bibfnamefont {N.~E.}\ \bibnamefont {Sluchanko}}, \bibinfo {author}
  {\bibfnamefont {A.}~\bibnamefont {Ivanov}}, \bibinfo {author} {\bibfnamefont
  {V.~B.}\ \bibnamefont {Filipov}}, \bibinfo {author} {\bibfnamefont
  {B.}~\bibnamefont {Keimer}}, \ and\ \bibinfo {author} {\bibfnamefont {D.~S.}\
  \bibnamefont {Inosov}},\ }\href@noop {} {\bibfield  {journal} {\bibinfo
  {journal} {Nat. Commun.}\ }\textbf {\bibinfo {volume} {3}},\ \bibinfo {pages}
  {380} (\bibinfo {year} {2012})}\BibitemShut {NoStop}%
\bibitem [{\citenamefont {Petrovic}\ \emph
  {et~al.}(2001{\natexlab{a}})\citenamefont {Petrovic}, \citenamefont
  {Pagliuso}, \citenamefont {Hundley}, \citenamefont {Movshovich},
  \citenamefont {Sarrao}, \citenamefont {Thompson}, \citenamefont {Fisk},\ and\
  \citenamefont {Monthoux}}]{Petrovic01:13}%
  \BibitemOpen
  \bibfield  {author} {\bibinfo {author} {\bibfnamefont {C.}~\bibnamefont
  {Petrovic}}, \bibinfo {author} {\bibfnamefont {P.~G.}\ \bibnamefont
  {Pagliuso}}, \bibinfo {author} {\bibfnamefont {M.~F.}\ \bibnamefont
  {Hundley}}, \bibinfo {author} {\bibfnamefont {R.}~\bibnamefont {Movshovich}},
  \bibinfo {author} {\bibfnamefont {J.~L.}\ \bibnamefont {Sarrao}}, \bibinfo
  {author} {\bibfnamefont {J.~D.}\ \bibnamefont {Thompson}}, \bibinfo {author}
  {\bibfnamefont {Z.}~\bibnamefont {Fisk}}, \ and\ \bibinfo {author}
  {\bibfnamefont {P.}~\bibnamefont {Monthoux}},\ }\href@noop {} {\bibfield
  {journal} {\bibinfo  {journal} {J. Phys.: Condens. Matter}\ }\textbf
  {\bibinfo {volume} {13}},\ \bibinfo {pages} {L337} (\bibinfo {year}
  {2001}{\natexlab{a}})}\BibitemShut {NoStop}%
\bibitem [{\citenamefont {Thompson}\ and\ \citenamefont
  {Fisk}(2012)}]{Thompson12:81}%
  \BibitemOpen
  \bibfield  {author} {\bibinfo {author} {\bibfnamefont {J.~D.}\ \bibnamefont
  {Thompson}}\ and\ \bibinfo {author} {\bibfnamefont {Z.}~\bibnamefont
  {Fisk}},\ }\href@noop {} {\bibfield  {journal} {\bibinfo  {journal} {J. Phys.
  Soc. Jpn.}\ }\textbf {\bibinfo {volume} {81}},\ \bibinfo {pages} {011002}
  (\bibinfo {year} {2012})}\BibitemShut {NoStop}%
\bibitem [{\citenamefont {Bao}\ \emph {et~al.}(2000)\citenamefont {Bao},
  \citenamefont {Pagliuso}, \citenamefont {Sarrao}, \citenamefont {Thompson},
  \citenamefont {Fisk}, \citenamefont {Lynn},\ and\ \citenamefont
  {Erwin}}]{Bao00:62}%
  \BibitemOpen
  \bibfield  {author} {\bibinfo {author} {\bibfnamefont {W.}~\bibnamefont
  {Bao}}, \bibinfo {author} {\bibfnamefont {P.~G.}\ \bibnamefont {Pagliuso}},
  \bibinfo {author} {\bibfnamefont {J.~L.}\ \bibnamefont {Sarrao}}, \bibinfo
  {author} {\bibfnamefont {J.~D.}\ \bibnamefont {Thompson}}, \bibinfo {author}
  {\bibfnamefont {Z.}~\bibnamefont {Fisk}}, \bibinfo {author} {\bibfnamefont
  {J.~W.}\ \bibnamefont {Lynn}}, \ and\ \bibinfo {author} {\bibfnamefont
  {R.~W.}\ \bibnamefont {Erwin}},\ }\href@noop {} {\bibfield  {journal}
  {\bibinfo  {journal} {Phys. Rev. B}\ }\textbf {\bibinfo {volume} {62}},\
  \bibinfo {pages} {R14621} (\bibinfo {year} {2000})}\BibitemShut {NoStop}%
\bibitem [{\citenamefont {Stock}\ \emph {et~al.}(2015)\citenamefont {Stock},
  \citenamefont {Rodriguez-Rivera}, \citenamefont {Schmalzl}, \citenamefont
  {Rodriguez}, \citenamefont {Stunault},\ and\ \citenamefont
  {Petrovic}}]{Stock15:114}%
  \BibitemOpen
  \bibfield  {author} {\bibinfo {author} {\bibfnamefont {C.}~\bibnamefont
  {Stock}}, \bibinfo {author} {\bibfnamefont {J.~A.}\ \bibnamefont
  {Rodriguez-Rivera}}, \bibinfo {author} {\bibfnamefont {K.}~\bibnamefont
  {Schmalzl}}, \bibinfo {author} {\bibfnamefont {E.~E.}\ \bibnamefont
  {Rodriguez}}, \bibinfo {author} {\bibfnamefont {A.}~\bibnamefont {Stunault}},
  \ and\ \bibinfo {author} {\bibfnamefont {C.}~\bibnamefont {Petrovic}},\
  }\href@noop {} {\bibfield  {journal} {\bibinfo  {journal} {Phys. Rev. Lett.}\
  }\textbf {\bibinfo {volume} {114}},\ \bibinfo {pages} {247005} (\bibinfo
  {year} {2015})}\BibitemShut {NoStop}%
\bibitem [{\citenamefont {Nachumi}\ \emph {et~al.}(1996)\citenamefont
  {Nachumi}, \citenamefont {Keren}, \citenamefont {Kojima}, \citenamefont
  {Larkin}, \citenamefont {Luke}, \citenamefont {Merrin}, \citenamefont
  {Tchernyshov}, \citenamefont {Uemura}, \citenamefont {Ichikawa},
  \citenamefont {Goto},\ and\ \citenamefont {Uchida}}]{Nachumi96:77}%
  \BibitemOpen
  \bibfield  {author} {\bibinfo {author} {\bibfnamefont {B.}~\bibnamefont
  {Nachumi}}, \bibinfo {author} {\bibfnamefont {A.}~\bibnamefont {Keren}},
  \bibinfo {author} {\bibfnamefont {K.}~\bibnamefont {Kojima}}, \bibinfo
  {author} {\bibfnamefont {M.}~\bibnamefont {Larkin}}, \bibinfo {author}
  {\bibfnamefont {G.~M.}\ \bibnamefont {Luke}}, \bibinfo {author}
  {\bibfnamefont {J.}~\bibnamefont {Merrin}}, \bibinfo {author} {\bibfnamefont
  {O.}~\bibnamefont {Tchernyshov}}, \bibinfo {author} {\bibfnamefont {Y.~J.}\
  \bibnamefont {Uemura}}, \bibinfo {author} {\bibfnamefont {N.}~\bibnamefont
  {Ichikawa}}, \bibinfo {author} {\bibfnamefont {M.}~\bibnamefont {Goto}}, \
  and\ \bibinfo {author} {\bibfnamefont {S.}~\bibnamefont {Uchida}},\
  }\href@noop {} {\bibfield  {journal} {\bibinfo  {journal} {Phys. Rev. Lett.}\
  }\textbf {\bibinfo {volume} {77}},\ \bibinfo {pages} {5421} (\bibinfo {year}
  {1996})}\BibitemShut {NoStop}%
\bibitem [{\citenamefont {Zapf}\ \emph {et~al.}(2001)\citenamefont {Zapf},
  \citenamefont {Freeman}, \citenamefont {Bauer}, \citenamefont {Petricka},
  \citenamefont {Sirvent}, \citenamefont {Frederick}, \citenamefont {Dickey},\
  and\ \citenamefont {Maple}}]{Zapf01:65}%
  \BibitemOpen
  \bibfield  {author} {\bibinfo {author} {\bibfnamefont {V.~S.}\ \bibnamefont
  {Zapf}}, \bibinfo {author} {\bibfnamefont {E.~J.}\ \bibnamefont {Freeman}},
  \bibinfo {author} {\bibfnamefont {E.~D.}\ \bibnamefont {Bauer}}, \bibinfo
  {author} {\bibfnamefont {J.}~\bibnamefont {Petricka}}, \bibinfo {author}
  {\bibfnamefont {C.}~\bibnamefont {Sirvent}}, \bibinfo {author} {\bibfnamefont
  {N.~A.}\ \bibnamefont {Frederick}}, \bibinfo {author} {\bibfnamefont {R.~P.}\
  \bibnamefont {Dickey}}, \ and\ \bibinfo {author} {\bibfnamefont {M.~B.}\
  \bibnamefont {Maple}},\ }\href@noop {} {\bibfield  {journal} {\bibinfo
  {journal} {Phys. Rev. B}\ }\textbf {\bibinfo {volume} {65}},\ \bibinfo
  {pages} {014506} (\bibinfo {year} {2001})}\BibitemShut {NoStop}%
\bibitem [{\citenamefont {Goh}\ \emph {et~al.}(2008)\citenamefont {Goh},
  \citenamefont {Paglione}, \citenamefont {Sutherland}, \citenamefont
  {O'Farrell}, \citenamefont {Bergemann}, \citenamefont {Sayles},\ and\
  \citenamefont {Maple}}]{Goh08:101}%
  \BibitemOpen
  \bibfield  {author} {\bibinfo {author} {\bibfnamefont {S.~K.}\ \bibnamefont
  {Goh}}, \bibinfo {author} {\bibfnamefont {J.}~\bibnamefont {Paglione}},
  \bibinfo {author} {\bibfnamefont {M.}~\bibnamefont {Sutherland}}, \bibinfo
  {author} {\bibfnamefont {E.~C.~T.}\ \bibnamefont {O'Farrell}}, \bibinfo
  {author} {\bibfnamefont {C.}~\bibnamefont {Bergemann}}, \bibinfo {author}
  {\bibfnamefont {T.~A.}\ \bibnamefont {Sayles}}, \ and\ \bibinfo {author}
  {\bibfnamefont {M.~B.}\ \bibnamefont {Maple}},\ }\href@noop {} {\bibfield
  {journal} {\bibinfo  {journal} {Phys. Rev. Lett.}\ }\textbf {\bibinfo
  {volume} {101}},\ \bibinfo {pages} {056402} (\bibinfo {year}
  {2008})}\BibitemShut {NoStop}%
\bibitem [{\citenamefont {Hegger}\ \emph {et~al.}(2000)\citenamefont {Hegger},
  \citenamefont {Petrovic}, \citenamefont {Moshopoulou}, \citenamefont
  {Hundley}, \citenamefont {Sarrao}, \citenamefont {Fisk},\ and\ \citenamefont
  {Thompson}}]{Hegger00}%
  \BibitemOpen
  \bibfield  {author} {\bibinfo {author} {\bibfnamefont {H.}~\bibnamefont
  {Hegger}}, \bibinfo {author} {\bibfnamefont {C.}~\bibnamefont {Petrovic}},
  \bibinfo {author} {\bibfnamefont {E.~G.}\ \bibnamefont {Moshopoulou}},
  \bibinfo {author} {\bibfnamefont {M.~F.}\ \bibnamefont {Hundley}}, \bibinfo
  {author} {\bibfnamefont {J.~L.}\ \bibnamefont {Sarrao}}, \bibinfo {author}
  {\bibfnamefont {Z.}~\bibnamefont {Fisk}}, \ and\ \bibinfo {author}
  {\bibfnamefont {J.~D.}\ \bibnamefont {Thompson}},\ }\href@noop {} {\bibfield
  {journal} {\bibinfo  {journal} {Phys. Rev. Lett.}\ }\textbf {\bibinfo
  {volume} {84}},\ \bibinfo {pages} {4986} (\bibinfo {year}
  {2000})}\BibitemShut {NoStop}%
\bibitem [{\citenamefont {Muramatsu}\ \emph {et~al.}(2001)\citenamefont
  {Muramatsu}, \citenamefont {Tateiwa}, \citenamefont {Kobayashi},
  \citenamefont {Shimizu}, \citenamefont {Amaya}, \citenamefont {Aoki},
  \citenamefont {Shishido}, \citenamefont {Haga},\ and\ \citenamefont
  {Onuki}}]{Mura01:70}%
  \BibitemOpen
  \bibfield  {author} {\bibinfo {author} {\bibfnamefont {T.}~\bibnamefont
  {Muramatsu}}, \bibinfo {author} {\bibfnamefont {N.}~\bibnamefont {Tateiwa}},
  \bibinfo {author} {\bibfnamefont {T.~C.}\ \bibnamefont {Kobayashi}}, \bibinfo
  {author} {\bibfnamefont {K.}~\bibnamefont {Shimizu}}, \bibinfo {author}
  {\bibfnamefont {K.}~\bibnamefont {Amaya}}, \bibinfo {author} {\bibfnamefont
  {D.}~\bibnamefont {Aoki}}, \bibinfo {author} {\bibfnamefont {H.}~\bibnamefont
  {Shishido}}, \bibinfo {author} {\bibfnamefont {Y.}~\bibnamefont {Haga}}, \
  and\ \bibinfo {author} {\bibfnamefont {Y.}~\bibnamefont {Onuki}},\
  }\href@noop {} {\bibfield  {journal} {\bibinfo  {journal} {J. Phys. Soc.
  Jpn.}\ }\textbf {\bibinfo {volume} {70}},\ \bibinfo {pages} {3362} (\bibinfo
  {year} {2001})}\BibitemShut {NoStop}%
\bibitem [{\citenamefont {Park}\ \emph {et~al.}(2008)\citenamefont {Park},
  \citenamefont {Graf}, \citenamefont {Boulaevskii}, \citenamefont {Sarrao},\
  and\ \citenamefont {Thompson}}]{Park08:105}%
  \BibitemOpen
  \bibfield  {author} {\bibinfo {author} {\bibfnamefont {R.}~\bibnamefont
  {Park}}, \bibinfo {author} {\bibfnamefont {M.~J.}\ \bibnamefont {Graf}},
  \bibinfo {author} {\bibfnamefont {L.}~\bibnamefont {Boulaevskii}}, \bibinfo
  {author} {\bibfnamefont {J.~L.}\ \bibnamefont {Sarrao}}, \ and\ \bibinfo
  {author} {\bibfnamefont {J.~D.}\ \bibnamefont {Thompson}},\ }\href@noop {}
  {\bibfield  {journal} {\bibinfo  {journal} {PNAS}\ }\textbf {\bibinfo
  {volume} {105}},\ \bibinfo {pages} {6825} (\bibinfo {year}
  {2008})}\BibitemShut {NoStop}%
\bibitem [{\citenamefont {Knebel}\ \emph {et~al.}(2006)\citenamefont {Knebel},
  \citenamefont {Aoki}, \citenamefont {Braithwaite}, \citenamefont {Salce},\
  and\ \citenamefont {Flouquet}}]{Knebel06:74}%
  \BibitemOpen
  \bibfield  {author} {\bibinfo {author} {\bibfnamefont {G.}~\bibnamefont
  {Knebel}}, \bibinfo {author} {\bibfnamefont {D.}~\bibnamefont {Aoki}},
  \bibinfo {author} {\bibfnamefont {D.}~\bibnamefont {Braithwaite}}, \bibinfo
  {author} {\bibfnamefont {B.}~\bibnamefont {Salce}}, \ and\ \bibinfo {author}
  {\bibfnamefont {J.}~\bibnamefont {Flouquet}},\ }\href@noop {} {\bibfield
  {journal} {\bibinfo  {journal} {Phys. Rev. B}\ }\textbf {\bibinfo {volume}
  {74}},\ \bibinfo {pages} {020501(R)} (\bibinfo {year} {2006})}\BibitemShut
  {NoStop}%
\bibitem [{\citenamefont {Kawasaki}\ \emph {et~al.}(2003)\citenamefont
  {Kawasaki}, \citenamefont {Mito}, \citenamefont {Kawasaki}, \citenamefont
  {q.~Zheng}, \citenamefont {Kitaoka}, \citenamefont {Aoki}, \citenamefont
  {Haga},\ and\ \citenamefont {Onuki}}]{Kawasaki03:91}%
  \BibitemOpen
  \bibfield  {author} {\bibinfo {author} {\bibfnamefont {S.}~\bibnamefont
  {Kawasaki}}, \bibinfo {author} {\bibfnamefont {T.}~\bibnamefont {Mito}},
  \bibinfo {author} {\bibfnamefont {Y.}~\bibnamefont {Kawasaki}}, \bibinfo
  {author} {\bibfnamefont {G.}~\bibnamefont {q.~Zheng}}, \bibinfo {author}
  {\bibfnamefont {Y.}~\bibnamefont {Kitaoka}}, \bibinfo {author} {\bibfnamefont
  {D.}~\bibnamefont {Aoki}}, \bibinfo {author} {\bibfnamefont {Y.}~\bibnamefont
  {Haga}}, \ and\ \bibinfo {author} {\bibfnamefont {Y.}~\bibnamefont {Onuki}},\
  }\href@noop {} {\bibfield  {journal} {\bibinfo  {journal} {Phys. Rev. Lett.}\
  }\textbf {\bibinfo {volume} {91}},\ \bibinfo {pages} {137001} (\bibinfo
  {year} {2003})}\BibitemShut {NoStop}%
\bibitem [{\citenamefont {Paglione}\ \emph {et~al.}(2008)\citenamefont
  {Paglione}, \citenamefont {Ho}, \citenamefont {Maple}, \citenamefont
  {Tanatar}, \citenamefont {Taillefer}, \citenamefont {Lee},\ and\
  \citenamefont {Petrovic}}]{Paglione77:08}%
  \BibitemOpen
  \bibfield  {author} {\bibinfo {author} {\bibfnamefont {J.}~\bibnamefont
  {Paglione}}, \bibinfo {author} {\bibfnamefont {P.~C.}\ \bibnamefont {Ho}},
  \bibinfo {author} {\bibfnamefont {M.~B.}\ \bibnamefont {Maple}}, \bibinfo
  {author} {\bibfnamefont {M.~A.}\ \bibnamefont {Tanatar}}, \bibinfo {author}
  {\bibfnamefont {L.}~\bibnamefont {Taillefer}}, \bibinfo {author}
  {\bibfnamefont {Y.}~\bibnamefont {Lee}}, \ and\ \bibinfo {author}
  {\bibfnamefont {C.}~\bibnamefont {Petrovic}},\ }\href@noop {} {\bibfield
  {journal} {\bibinfo  {journal} {Phys. Rev. B}\ }\textbf {\bibinfo {volume}
  {77}},\ \bibinfo {pages} {100505(R)} (\bibinfo {year} {2008})}\BibitemShut
  {NoStop}%
\bibitem [{\citenamefont {Chen}\ \emph {et~al.}(2006)\citenamefont {Chen},
  \citenamefont {Matsubayashi}, \citenamefont {Ban}, \citenamefont {Deguchi},\
  and\ \citenamefont {Sato}}]{Chen06:97}%
  \BibitemOpen
  \bibfield  {author} {\bibinfo {author} {\bibfnamefont {G.~F.}\ \bibnamefont
  {Chen}}, \bibinfo {author} {\bibfnamefont {K.}~\bibnamefont {Matsubayashi}},
  \bibinfo {author} {\bibfnamefont {S.}~\bibnamefont {Ban}}, \bibinfo {author}
  {\bibfnamefont {K.}~\bibnamefont {Deguchi}}, \ and\ \bibinfo {author}
  {\bibfnamefont {N.~K.}\ \bibnamefont {Sato}},\ }\href@noop {} {\bibfield
  {journal} {\bibinfo  {journal} {Phys. Rev. Lett.}\ }\textbf {\bibinfo
  {volume} {97}},\ \bibinfo {pages} {017005} (\bibinfo {year}
  {2006})}\BibitemShut {NoStop}%
\bibitem [{\citenamefont {Llobet}\ \emph {et~al.}(2005)\citenamefont {Llobet},
  \citenamefont {Christianson}, \citenamefont {Bao}, \citenamefont {Gardner},
  \citenamefont {Swainson}, \citenamefont {Lynn}, \citenamefont {Mignot},
  \citenamefont {Prokes}, \citenamefont {Pagliuso}, \citenamefont {Moreno},
  \citenamefont {Sarrao}, \citenamefont {Thompson},\ and\ \citenamefont
  {Lacerda}}]{Llobet05:95}%
  \BibitemOpen
  \bibfield  {author} {\bibinfo {author} {\bibfnamefont {A.}~\bibnamefont
  {Llobet}}, \bibinfo {author} {\bibfnamefont {A.~D.}\ \bibnamefont
  {Christianson}}, \bibinfo {author} {\bibfnamefont {W.}~\bibnamefont {Bao}},
  \bibinfo {author} {\bibfnamefont {J.~S.}\ \bibnamefont {Gardner}}, \bibinfo
  {author} {\bibfnamefont {I.~P.}\ \bibnamefont {Swainson}}, \bibinfo {author}
  {\bibfnamefont {J.~W.}\ \bibnamefont {Lynn}}, \bibinfo {author}
  {\bibfnamefont {J.~M.}\ \bibnamefont {Mignot}}, \bibinfo {author}
  {\bibfnamefont {K.}~\bibnamefont {Prokes}}, \bibinfo {author} {\bibfnamefont
  {P.~G.}\ \bibnamefont {Pagliuso}}, \bibinfo {author} {\bibfnamefont {N.~O.}\
  \bibnamefont {Moreno}}, \bibinfo {author} {\bibfnamefont {J.~L.}\
  \bibnamefont {Sarrao}}, \bibinfo {author} {\bibfnamefont {J.~D.}\
  \bibnamefont {Thompson}}, \ and\ \bibinfo {author} {\bibfnamefont {A.~H.}\
  \bibnamefont {Lacerda}},\ }\href@noop {} {\bibfield  {journal} {\bibinfo
  {journal} {Phys. Rev. Lett.}\ }\textbf {\bibinfo {volume} {95}},\ \bibinfo
  {pages} {217002} (\bibinfo {year} {2005})}\BibitemShut {NoStop}%
\bibitem [{\citenamefont {Fobes}\ \emph {et~al.}(2017)\citenamefont {Fobes},
  \citenamefont {Bauer}, \citenamefont {Thompson}, \citenamefont {Sazonov},
  \citenamefont {Hutanu}, \citenamefont {Zhang}, \citenamefont {Ronning},\ and\
  \citenamefont {Janoschek}}]{Fobes17:29}%
  \BibitemOpen
  \bibfield  {author} {\bibinfo {author} {\bibfnamefont {D.~M.}\ \bibnamefont
  {Fobes}}, \bibinfo {author} {\bibfnamefont {E.~D.}\ \bibnamefont {Bauer}},
  \bibinfo {author} {\bibfnamefont {J.~D.}\ \bibnamefont {Thompson}}, \bibinfo
  {author} {\bibfnamefont {A.}~\bibnamefont {Sazonov}}, \bibinfo {author}
  {\bibfnamefont {V.}~\bibnamefont {Hutanu}}, \bibinfo {author} {\bibfnamefont
  {S.}~\bibnamefont {Zhang}}, \bibinfo {author} {\bibfnamefont
  {F.}~\bibnamefont {Ronning}}, \ and\ \bibinfo {author} {\bibfnamefont
  {M.}~\bibnamefont {Janoschek}},\ }\href@noop {} {\bibfield  {journal}
  {\bibinfo  {journal} {J. Phys.: Condens. Matt.}\ }\textbf {\bibinfo {volume}
  {29}},\ \bibinfo {pages} {17LT01} (\bibinfo {year} {2017})}\BibitemShut
  {NoStop}%
\bibitem [{\citenamefont {Petrovic}\ \emph
  {et~al.}(2001{\natexlab{b}})\citenamefont {Petrovic}, \citenamefont
  {Movshovich}, \citenamefont {Jaime}, \citenamefont {Pagliuso}, \citenamefont
  {Hundley}, \citenamefont {Sarrao}, \citenamefont {Fisk},\ and\ \citenamefont
  {Thompson}}]{Petrovic01:53}%
  \BibitemOpen
  \bibfield  {author} {\bibinfo {author} {\bibfnamefont {C.}~\bibnamefont
  {Petrovic}}, \bibinfo {author} {\bibfnamefont {R.}~\bibnamefont
  {Movshovich}}, \bibinfo {author} {\bibfnamefont {M.}~\bibnamefont {Jaime}},
  \bibinfo {author} {\bibfnamefont {P.~G.}\ \bibnamefont {Pagliuso}}, \bibinfo
  {author} {\bibfnamefont {M.~F.}\ \bibnamefont {Hundley}}, \bibinfo {author}
  {\bibfnamefont {J.~L.}\ \bibnamefont {Sarrao}}, \bibinfo {author}
  {\bibfnamefont {Z.}~\bibnamefont {Fisk}}, \ and\ \bibinfo {author}
  {\bibfnamefont {J.~D.}\ \bibnamefont {Thompson}},\ }\href@noop {} {\bibfield
  {journal} {\bibinfo  {journal} {EPL}\ }\textbf {\bibinfo {volume} {53}},\
  \bibinfo {pages} {354} (\bibinfo {year} {2001}{\natexlab{b}})}\BibitemShut
  {NoStop}%
\bibitem [{\citenamefont {Tanatar}\ \emph {et~al.}(2005)\citenamefont
  {Tanatar}, \citenamefont {Paglione}, \citenamefont {Nakatsuji}, \citenamefont
  {Hawthorn}, \citenamefont {Boaknin}, \citenamefont {Hill}, \citenamefont
  {Ronning}, \citenamefont {Sutherland}, \citenamefont {Taillefer},
  \citenamefont {Petrovic}, \citenamefont {Canfield},\ and\ \citenamefont
  {Fisk}}]{Tanatar05:95}%
  \BibitemOpen
  \bibfield  {author} {\bibinfo {author} {\bibfnamefont {M.~A.}\ \bibnamefont
  {Tanatar}}, \bibinfo {author} {\bibfnamefont {J.}~\bibnamefont {Paglione}},
  \bibinfo {author} {\bibfnamefont {S.}~\bibnamefont {Nakatsuji}}, \bibinfo
  {author} {\bibfnamefont {D.~G.}\ \bibnamefont {Hawthorn}}, \bibinfo {author}
  {\bibfnamefont {E.}~\bibnamefont {Boaknin}}, \bibinfo {author} {\bibfnamefont
  {R.~W.}\ \bibnamefont {Hill}}, \bibinfo {author} {\bibfnamefont
  {F.}~\bibnamefont {Ronning}}, \bibinfo {author} {\bibfnamefont
  {M.}~\bibnamefont {Sutherland}}, \bibinfo {author} {\bibfnamefont
  {L.}~\bibnamefont {Taillefer}}, \bibinfo {author} {\bibfnamefont
  {C.}~\bibnamefont {Petrovic}}, \bibinfo {author} {\bibfnamefont {P.~C.}\
  \bibnamefont {Canfield}}, \ and\ \bibinfo {author} {\bibfnamefont
  {Z.}~\bibnamefont {Fisk}},\ }\href@noop {} {\bibfield  {journal} {\bibinfo
  {journal} {Phys. Rev. Lett.}\ }\textbf {\bibinfo {volume} {95}},\ \bibinfo
  {pages} {067002} (\bibinfo {year} {2005})}\BibitemShut {NoStop}%
\bibitem [{\citenamefont {Petrovic}\ \emph {et~al.}(2002)\citenamefont
  {Petrovic}, \citenamefont {Bud'ko}, \citenamefont {Kogan},\ and\
  \citenamefont {Canfield}}]{Petrovic02:66}%
  \BibitemOpen
  \bibfield  {author} {\bibinfo {author} {\bibfnamefont {C.}~\bibnamefont
  {Petrovic}}, \bibinfo {author} {\bibfnamefont {S.~L.}\ \bibnamefont
  {Bud'ko}}, \bibinfo {author} {\bibfnamefont {V.~G.}\ \bibnamefont {Kogan}}, \
  and\ \bibinfo {author} {\bibfnamefont {P.~C.}\ \bibnamefont {Canfield}},\
  }\href@noop {} {\bibfield  {journal} {\bibinfo  {journal} {Phys. Rev. B}\
  }\textbf {\bibinfo {volume} {66}},\ \bibinfo {pages} {054534} (\bibinfo
  {year} {2002})}\BibitemShut {NoStop}%
\bibitem [{\citenamefont {Raymond}\ \emph {et~al.}(2011)\citenamefont
  {Raymond}, \citenamefont {Panarin}, \citenamefont {Lapertot},\ and\
  \citenamefont {Flouquet}}]{Raymond11:80}%
  \BibitemOpen
  \bibfield  {author} {\bibinfo {author} {\bibfnamefont {S.}~\bibnamefont
  {Raymond}}, \bibinfo {author} {\bibfnamefont {J.}~\bibnamefont {Panarin}},
  \bibinfo {author} {\bibfnamefont {G.}~\bibnamefont {Lapertot}}, \ and\
  \bibinfo {author} {\bibfnamefont {J.}~\bibnamefont {Flouquet}},\ }\href@noop
  {} {\bibfield  {journal} {\bibinfo  {journal} {J. Phys. Soc. Jpn.}\ }\textbf
  {\bibinfo {volume} {80}},\ \bibinfo {pages} {SB023} (\bibinfo {year}
  {2011})}\BibitemShut {NoStop}%
\bibitem [{\citenamefont {Bauer}\ \emph {et~al.}(2005)\citenamefont {Bauer},
  \citenamefont {Capan}, \citenamefont {Ronning}, \citenamefont {Movshovich},
  \citenamefont {Thompson},\ and\ \citenamefont {Sarrao}}]{Bauer05:94}%
  \BibitemOpen
  \bibfield  {author} {\bibinfo {author} {\bibfnamefont {E.~D.}\ \bibnamefont
  {Bauer}}, \bibinfo {author} {\bibfnamefont {C.}~\bibnamefont {Capan}},
  \bibinfo {author} {\bibfnamefont {F.}~\bibnamefont {Ronning}}, \bibinfo
  {author} {\bibfnamefont {R.}~\bibnamefont {Movshovich}}, \bibinfo {author}
  {\bibfnamefont {J.~D.}\ \bibnamefont {Thompson}}, \ and\ \bibinfo {author}
  {\bibfnamefont {J.~L.}\ \bibnamefont {Sarrao}},\ }\href@noop {} {\bibfield
  {journal} {\bibinfo  {journal} {Phys. Rev. Lett.}\ }\textbf {\bibinfo
  {volume} {94}},\ \bibinfo {pages} {047001} (\bibinfo {year}
  {2005})}\BibitemShut {NoStop}%
\bibitem [{\citenamefont {Gofryk}\ \emph {et~al.}(2012)\citenamefont {Gofryk},
  \citenamefont {Ronning}, \citenamefont {Zhu}, \citenamefont {Ou},
  \citenamefont {Tobash}, \citenamefont {Stoyko}, \citenamefont {Lu},
  \citenamefont {Mar}, \citenamefont {Park}, \citenamefont {Bauer},
  \citenamefont {Thompson},\ and\ \citenamefont {Fisk}}]{Gofryk12:109}%
  \BibitemOpen
  \bibfield  {author} {\bibinfo {author} {\bibfnamefont {K.}~\bibnamefont
  {Gofryk}}, \bibinfo {author} {\bibfnamefont {F.}~\bibnamefont {Ronning}},
  \bibinfo {author} {\bibfnamefont {J.~X.}\ \bibnamefont {Zhu}}, \bibinfo
  {author} {\bibfnamefont {M.~N.}\ \bibnamefont {Ou}}, \bibinfo {author}
  {\bibfnamefont {P.~H.}\ \bibnamefont {Tobash}}, \bibinfo {author}
  {\bibfnamefont {S.~S.}\ \bibnamefont {Stoyko}}, \bibinfo {author}
  {\bibfnamefont {X.}~\bibnamefont {Lu}}, \bibinfo {author} {\bibfnamefont
  {A.}~\bibnamefont {Mar}}, \bibinfo {author} {\bibfnamefont {T.}~\bibnamefont
  {Park}}, \bibinfo {author} {\bibfnamefont {E.~D.}\ \bibnamefont {Bauer}},
  \bibinfo {author} {\bibfnamefont {J.~D.}\ \bibnamefont {Thompson}}, \ and\
  \bibinfo {author} {\bibfnamefont {Z.}~\bibnamefont {Fisk}},\ }\href@noop {}
  {\bibfield  {journal} {\bibinfo  {journal} {Phys. Rev. Lett.}\ }\textbf
  {\bibinfo {volume} {109}},\ \bibinfo {pages} {186402} (\bibinfo {year}
  {2012})}\BibitemShut {NoStop}%
\bibitem [{\citenamefont {Sakai}\ \emph {et~al.}(2015)\citenamefont {Sakai},
  \citenamefont {Ronning}, \citenamefont {Zhu}, \citenamefont {Wakeham},
  \citenamefont {Yasuoka}, \citenamefont {Tokunaga}, \citenamefont {Kambe},
  \citenamefont {Bauer},\ and\ \citenamefont {Thompson}}]{Sakai15:92}%
  \BibitemOpen
  \bibfield  {author} {\bibinfo {author} {\bibfnamefont {H.}~\bibnamefont
  {Sakai}}, \bibinfo {author} {\bibfnamefont {F.}~\bibnamefont {Ronning}},
  \bibinfo {author} {\bibfnamefont {J.~X.}\ \bibnamefont {Zhu}}, \bibinfo
  {author} {\bibfnamefont {N.}~\bibnamefont {Wakeham}}, \bibinfo {author}
  {\bibfnamefont {H.}~\bibnamefont {Yasuoka}}, \bibinfo {author} {\bibfnamefont
  {Y.}~\bibnamefont {Tokunaga}}, \bibinfo {author} {\bibfnamefont
  {S.}~\bibnamefont {Kambe}}, \bibinfo {author} {\bibfnamefont {E.~D.}\
  \bibnamefont {Bauer}}, \ and\ \bibinfo {author} {\bibfnamefont {J.~D.}\
  \bibnamefont {Thompson}},\ }\href@noop {} {\bibfield  {journal} {\bibinfo
  {journal} {Phys. Rev. B}\ }\textbf {\bibinfo {volume} {92}},\ \bibinfo
  {pages} {121105(R)} (\bibinfo {year} {2015})}\BibitemShut {NoStop}%
\bibitem [{\citenamefont {Capan}\ \emph {et~al.}(2010)\citenamefont {Capan},
  \citenamefont {Jo}, \citenamefont {Balicas}, \citenamefont {Goodrich},
  \citenamefont {DiTusa}, \citenamefont {Vekhter}, \citenamefont {Murphy},
  \citenamefont {Bianchi}, \citenamefont {Pham}, \citenamefont {Cho},
  \citenamefont {Chan}, \citenamefont {Young},\ and\ \citenamefont
  {Fisk}}]{Capan10:82}%
  \BibitemOpen
  \bibfield  {author} {\bibinfo {author} {\bibfnamefont {C.}~\bibnamefont
  {Capan}}, \bibinfo {author} {\bibfnamefont {Y.~J.}\ \bibnamefont {Jo}},
  \bibinfo {author} {\bibfnamefont {L.}~\bibnamefont {Balicas}}, \bibinfo
  {author} {\bibfnamefont {R.~G.}\ \bibnamefont {Goodrich}}, \bibinfo {author}
  {\bibfnamefont {J.~F.}\ \bibnamefont {DiTusa}}, \bibinfo {author}
  {\bibfnamefont {I.}~\bibnamefont {Vekhter}}, \bibinfo {author} {\bibfnamefont
  {T.~P.}\ \bibnamefont {Murphy}}, \bibinfo {author} {\bibfnamefont {A.~D.}\
  \bibnamefont {Bianchi}}, \bibinfo {author} {\bibfnamefont {L.~D.}\
  \bibnamefont {Pham}}, \bibinfo {author} {\bibfnamefont {J.~Y.}\ \bibnamefont
  {Cho}}, \bibinfo {author} {\bibfnamefont {J.~Y.}\ \bibnamefont {Chan}},
  \bibinfo {author} {\bibfnamefont {D.~P.}\ \bibnamefont {Young}}, \ and\
  \bibinfo {author} {\bibfnamefont {Z.}~\bibnamefont {Fisk}},\ }\href@noop {}
  {\bibfield  {journal} {\bibinfo  {journal} {Phys. Rev. B}\ }\textbf {\bibinfo
  {volume} {82}},\ \bibinfo {pages} {035112} (\bibinfo {year}
  {2010})}\BibitemShut {NoStop}%
\bibitem [{\citenamefont {Ou}\ \emph {et~al.}(2013)\citenamefont {Ou},
  \citenamefont {Gofryk}, \citenamefont {Baumbach}, \citenamefont {Stoyko},
  \citenamefont {Thompson}, \citenamefont {Lawrence}, \citenamefont {Bauer},
  \citenamefont {Ronning}, \citenamefont {Mar},\ and\ \citenamefont
  {Chen}}]{Ou13:88}%
  \BibitemOpen
  \bibfield  {author} {\bibinfo {author} {\bibfnamefont {M.~N.}\ \bibnamefont
  {Ou}}, \bibinfo {author} {\bibfnamefont {K.}~\bibnamefont {Gofryk}}, \bibinfo
  {author} {\bibfnamefont {R.~E.}\ \bibnamefont {Baumbach}}, \bibinfo {author}
  {\bibfnamefont {S.~S.}\ \bibnamefont {Stoyko}}, \bibinfo {author}
  {\bibfnamefont {J.~D.}\ \bibnamefont {Thompson}}, \bibinfo {author}
  {\bibfnamefont {J.~M.}\ \bibnamefont {Lawrence}}, \bibinfo {author}
  {\bibfnamefont {E.~D.}\ \bibnamefont {Bauer}}, \bibinfo {author}
  {\bibfnamefont {F.}~\bibnamefont {Ronning}}, \bibinfo {author} {\bibfnamefont
  {A.}~\bibnamefont {Mar}}, \ and\ \bibinfo {author} {\bibfnamefont {Y.~Y.}\
  \bibnamefont {Chen}},\ }\href@noop {} {\bibfield  {journal} {\bibinfo
  {journal} {Phys. Rev. B}\ }\textbf {\bibinfo {volume} {88}},\ \bibinfo
  {pages} {195134} (\bibinfo {year} {2013})}\BibitemShut {NoStop}%
\bibitem [{\citenamefont {Xu}\ \emph {et~al.}(2016)\citenamefont {Xu},
  \citenamefont {Dong}, \citenamefont {Lum}, \citenamefont {Zhang},
  \citenamefont {Hong}, \citenamefont {He}, \citenamefont {Wang}, \citenamefont
  {Ma}, \citenamefont {Petrovic}, \citenamefont {Maple}, \citenamefont {Shu},\
  and\ \citenamefont {Li}}]{Xu16:93}%
  \BibitemOpen
  \bibfield  {author} {\bibinfo {author} {\bibfnamefont {Y.}~\bibnamefont
  {Xu}}, \bibinfo {author} {\bibfnamefont {J.~K.}\ \bibnamefont {Dong}},
  \bibinfo {author} {\bibfnamefont {I.~K.}\ \bibnamefont {Lum}}, \bibinfo
  {author} {\bibfnamefont {J.}~\bibnamefont {Zhang}}, \bibinfo {author}
  {\bibfnamefont {X.~C.}\ \bibnamefont {Hong}}, \bibinfo {author}
  {\bibfnamefont {L.~P.}\ \bibnamefont {He}}, \bibinfo {author} {\bibfnamefont
  {K.~F.}\ \bibnamefont {Wang}}, \bibinfo {author} {\bibfnamefont {Y.~C.}\
  \bibnamefont {Ma}}, \bibinfo {author} {\bibfnamefont {C.}~\bibnamefont
  {Petrovic}}, \bibinfo {author} {\bibfnamefont {M.~B.}\ \bibnamefont {Maple}},
  \bibinfo {author} {\bibfnamefont {L.}~\bibnamefont {Shu}}, \ and\ \bibinfo
  {author} {\bibfnamefont {S.~Y.}\ \bibnamefont {Li}},\ }\href@noop {}
  {\bibfield  {journal} {\bibinfo  {journal} {Phys. Rev. B}\ }\textbf {\bibinfo
  {volume} {93}},\ \bibinfo {pages} {064502} (\bibinfo {year}
  {2016})}\BibitemShut {NoStop}%
\bibitem [{\citenamefont {Shu}\ \emph {et~al.}(2011)\citenamefont {Shu},
  \citenamefont {Baumbach}, \citenamefont {Janoschek}, \citenamefont
  {Gonzales}, \citenamefont {Huang}, \citenamefont {Sayles}, \citenamefont
  {Paglione}, \citenamefont {O'Brien}, \citenamefont {Hamlin}, \citenamefont
  {Zocco}, \citenamefont {Ho}, \citenamefont {McElroy},\ and\ \citenamefont
  {Maple}}]{Shu11:106}%
  \BibitemOpen
  \bibfield  {author} {\bibinfo {author} {\bibfnamefont {L.}~\bibnamefont
  {Shu}}, \bibinfo {author} {\bibfnamefont {R.~E.}\ \bibnamefont {Baumbach}},
  \bibinfo {author} {\bibfnamefont {M.}~\bibnamefont {Janoschek}}, \bibinfo
  {author} {\bibfnamefont {E.}~\bibnamefont {Gonzales}}, \bibinfo {author}
  {\bibfnamefont {K.}~\bibnamefont {Huang}}, \bibinfo {author} {\bibfnamefont
  {T.~A.}\ \bibnamefont {Sayles}}, \bibinfo {author} {\bibfnamefont
  {J.}~\bibnamefont {Paglione}}, \bibinfo {author} {\bibfnamefont
  {J.}~\bibnamefont {O'Brien}}, \bibinfo {author} {\bibfnamefont {J.~J.}\
  \bibnamefont {Hamlin}}, \bibinfo {author} {\bibfnamefont {D.~A.}\
  \bibnamefont {Zocco}}, \bibinfo {author} {\bibfnamefont {P.~C.}\ \bibnamefont
  {Ho}}, \bibinfo {author} {\bibfnamefont {C.~A.}\ \bibnamefont {McElroy}}, \
  and\ \bibinfo {author} {\bibfnamefont {M.~B.}\ \bibnamefont {Maple}},\
  }\href@noop {} {\bibfield  {journal} {\bibinfo  {journal} {Phys. Rev. Lett.}\
  }\textbf {\bibinfo {volume} {106}},\ \bibinfo {pages} {156403} (\bibinfo
  {year} {2011})}\BibitemShut {NoStop}%
\bibitem [{\citenamefont {Kim}\ \emph {et~al.}(2015)\citenamefont {Kim},
  \citenamefont {Tanatar}, \citenamefont {Flint}, \citenamefont {Petrovic},
  \citenamefont {Hu}, \citenamefont {White}, \citenamefont {Lum}, \citenamefont
  {Maple},\ and\ \citenamefont {Prozorov}}]{Kim15:114}%
  \BibitemOpen
  \bibfield  {author} {\bibinfo {author} {\bibfnamefont {H.}~\bibnamefont
  {Kim}}, \bibinfo {author} {\bibfnamefont {M.~A.}\ \bibnamefont {Tanatar}},
  \bibinfo {author} {\bibfnamefont {R.}~\bibnamefont {Flint}}, \bibinfo
  {author} {\bibfnamefont {C.}~\bibnamefont {Petrovic}}, \bibinfo {author}
  {\bibfnamefont {R.}~\bibnamefont {Hu}}, \bibinfo {author} {\bibfnamefont
  {B.~D.}\ \bibnamefont {White}}, \bibinfo {author} {\bibfnamefont {I.~K.}\
  \bibnamefont {Lum}}, \bibinfo {author} {\bibfnamefont {M.~B.}\ \bibnamefont
  {Maple}}, \ and\ \bibinfo {author} {\bibfnamefont {R.}~\bibnamefont
  {Prozorov}},\ }\href@noop {} {\bibfield  {journal} {\bibinfo  {journal}
  {Phys. Rev. Lett.}\ }\textbf {\bibinfo {volume} {114}},\ \bibinfo {pages}
  {027003} (\bibinfo {year} {2015})}\BibitemShut {NoStop}%
\bibitem [{\citenamefont {Zhong}\ \emph {et~al.}(2017)\citenamefont {Zhong},
  \citenamefont {Zhang}, \citenamefont {Shao},\ and\ \citenamefont
  {Luo}}]{Zhong17:12}%
  \BibitemOpen
  \bibfield  {author} {\bibinfo {author} {\bibfnamefont {Y.}~\bibnamefont
  {Zhong}}, \bibinfo {author} {\bibfnamefont {L.}~\bibnamefont {Zhang}},
  \bibinfo {author} {\bibfnamefont {C.}~\bibnamefont {Shao}}, \ and\ \bibinfo
  {author} {\bibfnamefont {H.~G.}\ \bibnamefont {Luo}},\ }\href@noop {}
  {\bibfield  {journal} {\bibinfo  {journal} {Front. Phys.}\ }\textbf {\bibinfo
  {volume} {12}},\ \bibinfo {pages} {127101} (\bibinfo {year}
  {2017})}\BibitemShut {NoStop}%
\bibitem [{\citenamefont {Pham}\ \emph {et~al.}(2006)\citenamefont {Pham},
  \citenamefont {Park}, \citenamefont {Maquilon}, \citenamefont {Thompson},\
  and\ \citenamefont {Fisk}}]{Pham06:97}%
  \BibitemOpen
  \bibfield  {author} {\bibinfo {author} {\bibfnamefont {L.~D.}\ \bibnamefont
  {Pham}}, \bibinfo {author} {\bibfnamefont {T.}~\bibnamefont {Park}}, \bibinfo
  {author} {\bibfnamefont {S.}~\bibnamefont {Maquilon}}, \bibinfo {author}
  {\bibfnamefont {J.~D.}\ \bibnamefont {Thompson}}, \ and\ \bibinfo {author}
  {\bibfnamefont {Z.}~\bibnamefont {Fisk}},\ }\href@noop {} {\bibfield
  {journal} {\bibinfo  {journal} {Phys. Rev. Lett.}\ }\textbf {\bibinfo
  {volume} {97}},\ \bibinfo {pages} {056404} (\bibinfo {year}
  {2006})}\BibitemShut {NoStop}%
\bibitem [{\citenamefont {Bauer}\ \emph {et~al.}(2011)\citenamefont {Bauer},
  \citenamefont {Yang}, \citenamefont {Capan}, \citenamefont {Urbano},
  \citenamefont {Miclea}, \citenamefont {Sakai}, \citenamefont {Ronning},
  \citenamefont {Graf}, \citenamefont {Balatsky}, \citenamefont {Movshovich},
  \citenamefont {Bianchi}, \citenamefont {Reyes}, \citenamefont {Kuhns},
  \citenamefont {Thompson},\ and\ \citenamefont {Fisk}}]{Bauer11:108}%
  \BibitemOpen
  \bibfield  {author} {\bibinfo {author} {\bibfnamefont {E.~D.}\ \bibnamefont
  {Bauer}}, \bibinfo {author} {\bibfnamefont {Y.}~\bibnamefont {Yang}},
  \bibinfo {author} {\bibfnamefont {C.}~\bibnamefont {Capan}}, \bibinfo
  {author} {\bibfnamefont {R.~R.}\ \bibnamefont {Urbano}}, \bibinfo {author}
  {\bibfnamefont {C.~F.}\ \bibnamefont {Miclea}}, \bibinfo {author}
  {\bibfnamefont {H.}~\bibnamefont {Sakai}}, \bibinfo {author} {\bibfnamefont
  {F.}~\bibnamefont {Ronning}}, \bibinfo {author} {\bibfnamefont {M.~J.}\
  \bibnamefont {Graf}}, \bibinfo {author} {\bibfnamefont {A.~V.}\ \bibnamefont
  {Balatsky}}, \bibinfo {author} {\bibfnamefont {R.}~\bibnamefont
  {Movshovich}}, \bibinfo {author} {\bibfnamefont {A.~D.}\ \bibnamefont
  {Bianchi}}, \bibinfo {author} {\bibfnamefont {A.~P.}\ \bibnamefont {Reyes}},
  \bibinfo {author} {\bibfnamefont {P.~L.}\ \bibnamefont {Kuhns}}, \bibinfo
  {author} {\bibfnamefont {J.~D.}\ \bibnamefont {Thompson}}, \ and\ \bibinfo
  {author} {\bibfnamefont {Z.}~\bibnamefont {Fisk}},\ }\href@noop {} {\bibfield
   {journal} {\bibinfo  {journal} {PNAS}\ }\textbf {\bibinfo {volume} {108}},\
  \bibinfo {pages} {6857} (\bibinfo {year} {2011})}\BibitemShut {NoStop}%
\bibitem [{\citenamefont {Seo}\ \emph {et~al.}(2014)\citenamefont {Seo},
  \citenamefont {Lu}, \citenamefont {Zhu}, \citenamefont {Urbano},
  \citenamefont {Curro}, \citenamefont {Bauer}, \citenamefont {Sidorov},
  \citenamefont {Pham}, \citenamefont {Park}, \citenamefont {Fisk},\ and\
  \citenamefont {Thompson}}]{Sheo13:10}%
  \BibitemOpen
  \bibfield  {author} {\bibinfo {author} {\bibfnamefont {S.}~\bibnamefont
  {Seo}}, \bibinfo {author} {\bibfnamefont {X.}~\bibnamefont {Lu}}, \bibinfo
  {author} {\bibfnamefont {J.~X.}\ \bibnamefont {Zhu}}, \bibinfo {author}
  {\bibfnamefont {R.~R.}\ \bibnamefont {Urbano}}, \bibinfo {author}
  {\bibfnamefont {N.}~\bibnamefont {Curro}}, \bibinfo {author} {\bibfnamefont
  {E.~D.}\ \bibnamefont {Bauer}}, \bibinfo {author} {\bibfnamefont {V.~A.}\
  \bibnamefont {Sidorov}}, \bibinfo {author} {\bibfnamefont {L.~D.}\
  \bibnamefont {Pham}}, \bibinfo {author} {\bibfnamefont {T.}~\bibnamefont
  {Park}}, \bibinfo {author} {\bibfnamefont {Z.}~\bibnamefont {Fisk}}, \ and\
  \bibinfo {author} {\bibfnamefont {J.~D.}\ \bibnamefont {Thompson}},\
  }\href@noop {} {\bibfield  {journal} {\bibinfo  {journal} {Nat. Phys.}\
  }\textbf {\bibinfo {volume} {10}},\ \bibinfo {pages} {120} (\bibinfo {year}
  {2014})}\BibitemShut {NoStop}%
\bibitem [{\citenamefont {Nicklas}\ \emph {et~al.}(2007)\citenamefont
  {Nicklas}, \citenamefont {Stockert}, \citenamefont {Park}, \citenamefont
  {Habicht}, \citenamefont {Kiefer}, \citenamefont {Pham}, \citenamefont
  {Thompson}, \citenamefont {Fisk},\ and\ \citenamefont
  {Steglich}}]{Nicklas07:76}%
  \BibitemOpen
  \bibfield  {author} {\bibinfo {author} {\bibfnamefont {M.}~\bibnamefont
  {Nicklas}}, \bibinfo {author} {\bibfnamefont {O.}~\bibnamefont {Stockert}},
  \bibinfo {author} {\bibfnamefont {T.}~\bibnamefont {Park}}, \bibinfo {author}
  {\bibfnamefont {K.}~\bibnamefont {Habicht}}, \bibinfo {author} {\bibfnamefont
  {K.}~\bibnamefont {Kiefer}}, \bibinfo {author} {\bibfnamefont {L.~D.}\
  \bibnamefont {Pham}}, \bibinfo {author} {\bibfnamefont {J.~D.}\ \bibnamefont
  {Thompson}}, \bibinfo {author} {\bibfnamefont {Z.}~\bibnamefont {Fisk}}, \
  and\ \bibinfo {author} {\bibfnamefont {F.}~\bibnamefont {Steglich}},\
  }\href@noop {} {\bibfield  {journal} {\bibinfo  {journal} {Phys. Rev. B}\
  }\textbf {\bibinfo {volume} {76}},\ \bibinfo {pages} {052401} (\bibinfo
  {year} {2007})}\BibitemShut {NoStop}%
\bibitem [{\citenamefont {Bao}\ \emph {et~al.}(2009)\citenamefont {Bao},
  \citenamefont {Gasparovic}, \citenamefont {Lynn}, \citenamefont {Ronning},
  \citenamefont {Bauer}, \citenamefont {Thompson},\ and\ \citenamefont
  {Fisk}}]{Bao09:79}%
  \BibitemOpen
  \bibfield  {author} {\bibinfo {author} {\bibfnamefont {W.}~\bibnamefont
  {Bao}}, \bibinfo {author} {\bibfnamefont {Y.~C.}\ \bibnamefont {Gasparovic}},
  \bibinfo {author} {\bibfnamefont {J.~W.}\ \bibnamefont {Lynn}}, \bibinfo
  {author} {\bibfnamefont {F.}~\bibnamefont {Ronning}}, \bibinfo {author}
  {\bibfnamefont {E.~D.}\ \bibnamefont {Bauer}}, \bibinfo {author}
  {\bibfnamefont {J.~D.}\ \bibnamefont {Thompson}}, \ and\ \bibinfo {author}
  {\bibfnamefont {Z.}~\bibnamefont {Fisk}},\ }\href@noop {} {\bibfield
  {journal} {\bibinfo  {journal} {Phys. Rev. B}\ }\textbf {\bibinfo {volume}
  {79}},\ \bibinfo {pages} {092415} (\bibinfo {year} {2009})}\BibitemShut
  {NoStop}%
\bibitem [{\citenamefont {Kenzelmann}\ \emph {et~al.}(2008)\citenamefont
  {Kenzelmann}, \citenamefont {Straessle}, \citenamefont {Niedermayer},
  \citenamefont {Sigrist}, \citenamefont {Padmanabhan}, \citenamefont
  {Zolliker}, \citenamefont {Bianchi}, \citenamefont {Movshovich},
  \citenamefont {Bauer}, \citenamefont {Sarrao},\ and\ \citenamefont
  {Thompson}}]{Kenzelmann08:321}%
  \BibitemOpen
  \bibfield  {author} {\bibinfo {author} {\bibfnamefont {M.}~\bibnamefont
  {Kenzelmann}}, \bibinfo {author} {\bibfnamefont {T.}~\bibnamefont
  {Straessle}}, \bibinfo {author} {\bibfnamefont {C.}~\bibnamefont
  {Niedermayer}}, \bibinfo {author} {\bibfnamefont {M.}~\bibnamefont
  {Sigrist}}, \bibinfo {author} {\bibfnamefont {B.}~\bibnamefont
  {Padmanabhan}}, \bibinfo {author} {\bibfnamefont {M.}~\bibnamefont
  {Zolliker}}, \bibinfo {author} {\bibfnamefont {A.~D.}\ \bibnamefont
  {Bianchi}}, \bibinfo {author} {\bibfnamefont {R.}~\bibnamefont {Movshovich}},
  \bibinfo {author} {\bibfnamefont {E.~D.}\ \bibnamefont {Bauer}}, \bibinfo
  {author} {\bibfnamefont {J.~L.}\ \bibnamefont {Sarrao}}, \ and\ \bibinfo
  {author} {\bibfnamefont {J.~D.}\ \bibnamefont {Thompson}},\ }\href@noop {}
  {\bibfield  {journal} {\bibinfo  {journal} {{Science}}\ }\textbf {\bibinfo
  {volume} {{321}}},\ \bibinfo {pages} {{1652}} (\bibinfo {year}
  {{2008}})}\BibitemShut {NoStop}%
\bibitem [{\citenamefont {Blackburn}\ \emph {et~al.}(2010)\citenamefont
  {Blackburn}, \citenamefont {Das}, \citenamefont {Eskildsen}, \citenamefont
  {Forgan}, \citenamefont {Laver}, \citenamefont {Niedermayer}, \citenamefont
  {Petrovic},\ and\ \citenamefont {White}}]{Blackburn10:105}%
  \BibitemOpen
  \bibfield  {author} {\bibinfo {author} {\bibfnamefont {E.}~\bibnamefont
  {Blackburn}}, \bibinfo {author} {\bibfnamefont {P.}~\bibnamefont {Das}},
  \bibinfo {author} {\bibfnamefont {M.~R.}\ \bibnamefont {Eskildsen}}, \bibinfo
  {author} {\bibfnamefont {E.~M.}\ \bibnamefont {Forgan}}, \bibinfo {author}
  {\bibfnamefont {M.}~\bibnamefont {Laver}}, \bibinfo {author} {\bibfnamefont
  {C.}~\bibnamefont {Niedermayer}}, \bibinfo {author} {\bibfnamefont
  {C.}~\bibnamefont {Petrovic}}, \ and\ \bibinfo {author} {\bibfnamefont
  {J.~S.}\ \bibnamefont {White}},\ }\href@noop {} {\bibfield  {journal}
  {\bibinfo  {journal} {Phys. Rev. Lett.}\ }\textbf {\bibinfo {volume} {105}},\
  \bibinfo {pages} {187001} (\bibinfo {year} {2010})}\BibitemShut {NoStop}%
\bibitem [{\citenamefont {Kim}\ \emph {et~al.}(2017)\citenamefont {Kim},
  \citenamefont {Lin}, \citenamefont {Bauer}, \citenamefont {Ronning},
  \citenamefont {Thompson},\ and\ \citenamefont {Movshovich}}]{Kim17:95}%
  \BibitemOpen
  \bibfield  {author} {\bibinfo {author} {\bibfnamefont {D.~Y.}\ \bibnamefont
  {Kim}}, \bibinfo {author} {\bibfnamefont {S.-Z.}\ \bibnamefont {Lin}},
  \bibinfo {author} {\bibfnamefont {E.~D.}\ \bibnamefont {Bauer}}, \bibinfo
  {author} {\bibfnamefont {F.}~\bibnamefont {Ronning}}, \bibinfo {author}
  {\bibfnamefont {J.~D.}\ \bibnamefont {Thompson}}, \ and\ \bibinfo {author}
  {\bibfnamefont {R.}~\bibnamefont {Movshovich}},\ }\href@noop {} {\bibfield
  {journal} {\bibinfo  {journal} {{Phys. Rev. B}}\ }\textbf {\bibinfo {volume}
  {{95}}},\ \bibinfo {pages} {{241110}} (\bibinfo {year} {{2017}})}\BibitemShut
  {NoStop}%
\bibitem [{\citenamefont {Michal}\ and\ \citenamefont
  {Mineev}(2011)}]{Michal11:84}%
  \BibitemOpen
  \bibfield  {author} {\bibinfo {author} {\bibfnamefont {V.~P.}\ \bibnamefont
  {Michal}}\ and\ \bibinfo {author} {\bibfnamefont {V.~P.}\ \bibnamefont
  {Mineev}},\ }\href@noop {} {\bibfield  {journal} {\bibinfo  {journal} {Phys.
  Rev. B}\ }\textbf {\bibinfo {volume} {84}},\ \bibinfo {pages} {052508}
  (\bibinfo {year} {2011})}\BibitemShut {NoStop}%
\bibitem [{\citenamefont {Koutroulakis}\ \emph {et~al.}(2010)\citenamefont
  {Koutroulakis}, \citenamefont {Stewart}, \citenamefont
  {Mitrovi\ifmmode~\acute{c}\else \'{c}\fi{}}, \citenamefont
  {Horvati\ifmmode~\acute{c}\else \'{c}\fi{}}, \citenamefont {Berthier},
  \citenamefont {Lapertot},\ and\ \citenamefont
  {Flouquet}}]{Koutroulakis10:104}%
  \BibitemOpen
  \bibfield  {author} {\bibinfo {author} {\bibfnamefont {G.}~\bibnamefont
  {Koutroulakis}}, \bibinfo {author} {\bibfnamefont {M.~D.}\ \bibnamefont
  {Stewart}}, \bibinfo {author} {\bibfnamefont {V.~F.}\ \bibnamefont
  {Mitrovi\ifmmode~\acute{c}\else \'{c}\fi{}}}, \bibinfo {author}
  {\bibfnamefont {M.}~\bibnamefont {Horvati\ifmmode~\acute{c}\else
  \'{c}\fi{}}}, \bibinfo {author} {\bibfnamefont {C.}~\bibnamefont {Berthier}},
  \bibinfo {author} {\bibfnamefont {G.}~\bibnamefont {Lapertot}}, \ and\
  \bibinfo {author} {\bibfnamefont {J.}~\bibnamefont {Flouquet}},\ }\href@noop
  {} {\bibfield  {journal} {\bibinfo  {journal} {Phys. Rev. Lett.}\ }\textbf
  {\bibinfo {volume} {104}},\ \bibinfo {pages} {087001} (\bibinfo {year}
  {2010})}\BibitemShut {NoStop}%
\bibitem [{\citenamefont {Raymond}\ \emph {et~al.}(2014)\citenamefont
  {Raymond}, \citenamefont {Ramos}, \citenamefont {Aoki}, \citenamefont
  {Knebel}, \citenamefont {Mineev},\ and\ \citenamefont
  {Lapertot}}]{Raymond14:83}%
  \BibitemOpen
  \bibfield  {author} {\bibinfo {author} {\bibfnamefont {S.}~\bibnamefont
  {Raymond}}, \bibinfo {author} {\bibfnamefont {S.~M.}\ \bibnamefont {Ramos}},
  \bibinfo {author} {\bibfnamefont {D.}~\bibnamefont {Aoki}}, \bibinfo {author}
  {\bibfnamefont {G.}~\bibnamefont {Knebel}}, \bibinfo {author} {\bibfnamefont
  {V.~P.}\ \bibnamefont {Mineev}}, \ and\ \bibinfo {author} {\bibfnamefont
  {G.}~\bibnamefont {Lapertot}},\ }\href@noop {} {\bibfield  {journal}
  {\bibinfo  {journal} {J. Phys. Soc. Jpn.}\ }\textbf {\bibinfo {volume}
  {83}},\ \bibinfo {pages} {013707} (\bibinfo {year} {2014})}\BibitemShut
  {NoStop}%
\bibitem [{\citenamefont {Collins}(1989)}]{Collins}%
  \BibitemOpen
  \bibfield  {author} {\bibinfo {author} {\bibfnamefont {M.~F.}\ \bibnamefont
  {Collins}},\ }\href@noop {} {\emph {\bibinfo {title} {Magnetic Critical
  Scattering}}}\ (\bibinfo  {publisher} {Oxford University Press},\ \bibinfo
  {year} {1989})\BibitemShut {NoStop}%
\bibitem [{\citenamefont {Das}\ \emph {et~al.}(2014)\citenamefont {Das},
  \citenamefont {Lin}, \citenamefont {Ghimire}, \citenamefont {Huang},
  \citenamefont {Ronning}, \citenamefont {Bauer}, \citenamefont {Thompson},
  \citenamefont {Batista}, \citenamefont {Ehlers},\ and\ \citenamefont
  {Janoschek}}]{Das14:113}%
  \BibitemOpen
  \bibfield  {author} {\bibinfo {author} {\bibfnamefont {P.}~\bibnamefont
  {Das}}, \bibinfo {author} {\bibfnamefont {S.~Z.}\ \bibnamefont {Lin}},
  \bibinfo {author} {\bibfnamefont {N.~J.}\ \bibnamefont {Ghimire}}, \bibinfo
  {author} {\bibfnamefont {K.}~\bibnamefont {Huang}}, \bibinfo {author}
  {\bibfnamefont {F.}~\bibnamefont {Ronning}}, \bibinfo {author} {\bibfnamefont
  {E.~D.}\ \bibnamefont {Bauer}}, \bibinfo {author} {\bibfnamefont {J.~D.}\
  \bibnamefont {Thompson}}, \bibinfo {author} {\bibfnamefont {C.~D.}\
  \bibnamefont {Batista}}, \bibinfo {author} {\bibfnamefont {G.}~\bibnamefont
  {Ehlers}}, \ and\ \bibinfo {author} {\bibfnamefont {M.}~\bibnamefont
  {Janoschek}},\ }\href@noop {} {\bibfield  {journal} {\bibinfo  {journal}
  {Phys. Rev. Lett.}\ }\textbf {\bibinfo {volume} {113}},\ \bibinfo {pages}
  {246403} (\bibinfo {year} {2014})}\BibitemShut {NoStop}%
\bibitem [{\citenamefont {Bao}\ \emph {et~al.}(2002)\citenamefont {Bao},
  \citenamefont {Aeppli}, \citenamefont {Lynn}, \citenamefont {Pagliuso},
  \citenamefont {Sarrao}, \citenamefont {Hundley}, \citenamefont {Thompson},\
  and\ \citenamefont {Fisk}}]{Bao02:65}%
  \BibitemOpen
  \bibfield  {author} {\bibinfo {author} {\bibfnamefont {W.}~\bibnamefont
  {Bao}}, \bibinfo {author} {\bibfnamefont {G.}~\bibnamefont {Aeppli}},
  \bibinfo {author} {\bibfnamefont {J.~W.}\ \bibnamefont {Lynn}}, \bibinfo
  {author} {\bibfnamefont {P.~G.}\ \bibnamefont {Pagliuso}}, \bibinfo {author}
  {\bibfnamefont {J.~L.}\ \bibnamefont {Sarrao}}, \bibinfo {author}
  {\bibfnamefont {M.~F.}\ \bibnamefont {Hundley}}, \bibinfo {author}
  {\bibfnamefont {J.~D.}\ \bibnamefont {Thompson}}, \ and\ \bibinfo {author}
  {\bibfnamefont {Z.}~\bibnamefont {Fisk}},\ }\href@noop {} {\bibfield
  {journal} {\bibinfo  {journal} {Phys. Rev. B}\ }\textbf {\bibinfo {volume}
  {65}},\ \bibinfo {pages} {100505(R)} (\bibinfo {year} {2002})}\BibitemShut
  {NoStop}%
\bibitem [{\citenamefont {Wilson}\ \emph
  {et~al.}(2010{\natexlab{a}})\citenamefont {Wilson}, \citenamefont {Rotundu},
  \citenamefont {Yamani}, \citenamefont {Valdivia}, \citenamefont {Freelon},
  \citenamefont {Bourret-Courchesne},\ and\ \citenamefont
  {Birgeneau}}]{Wilson10:81}%
  \BibitemOpen
  \bibfield  {author} {\bibinfo {author} {\bibfnamefont {S.~D.}\ \bibnamefont
  {Wilson}}, \bibinfo {author} {\bibfnamefont {C.~R.}\ \bibnamefont {Rotundu}},
  \bibinfo {author} {\bibfnamefont {Z.}~\bibnamefont {Yamani}}, \bibinfo
  {author} {\bibfnamefont {P.~N.}\ \bibnamefont {Valdivia}}, \bibinfo {author}
  {\bibfnamefont {B.}~\bibnamefont {Freelon}}, \bibinfo {author} {\bibfnamefont
  {E.}~\bibnamefont {Bourret-Courchesne}}, \ and\ \bibinfo {author}
  {\bibfnamefont {R.~J.}\ \bibnamefont {Birgeneau}},\ }\href@noop {} {\bibfield
   {journal} {\bibinfo  {journal} {Phys. Rev. B}\ }\textbf {\bibinfo {volume}
  {81}},\ \bibinfo {pages} {014501} (\bibinfo {year}
  {2010}{\natexlab{a}})}\BibitemShut {NoStop}%
\bibitem [{\citenamefont {Wilson}\ \emph
  {et~al.}(2010{\natexlab{b}})\citenamefont {Wilson}, \citenamefont {Yamani},
  \citenamefont {Rotundu}, \citenamefont {Freelon}, \citenamefont {Valdivia},
  \citenamefont {Bourret-Courchesne}, \citenamefont {Lynn}, \citenamefont
  {Chi}, \citenamefont {Hong},\ and\ \citenamefont {Birgeneau}}]{Wilson10:82}%
  \BibitemOpen
  \bibfield  {author} {\bibinfo {author} {\bibfnamefont {S.~D.}\ \bibnamefont
  {Wilson}}, \bibinfo {author} {\bibfnamefont {Z.}~\bibnamefont {Yamani}},
  \bibinfo {author} {\bibfnamefont {C.~R.}\ \bibnamefont {Rotundu}}, \bibinfo
  {author} {\bibfnamefont {B.}~\bibnamefont {Freelon}}, \bibinfo {author}
  {\bibfnamefont {P.~N.}\ \bibnamefont {Valdivia}}, \bibinfo {author}
  {\bibfnamefont {E.}~\bibnamefont {Bourret-Courchesne}}, \bibinfo {author}
  {\bibfnamefont {J.~W.}\ \bibnamefont {Lynn}}, \bibinfo {author}
  {\bibfnamefont {S.}~\bibnamefont {Chi}}, \bibinfo {author} {\bibfnamefont
  {T.}~\bibnamefont {Hong}}, \ and\ \bibinfo {author} {\bibfnamefont {R.~J.}\
  \bibnamefont {Birgeneau}},\ }\href@noop {} {\bibfield  {journal} {\bibinfo
  {journal} {Phys. Rev. B}\ }\textbf {\bibinfo {volume} {82}},\ \bibinfo
  {pages} {144502} (\bibinfo {year} {2010}{\natexlab{b}})}\BibitemShut
  {NoStop}%
\bibitem [{\citenamefont {Pajerowski}\ \emph
  {et~al.}(2013{\natexlab{a}})\citenamefont {Pajerowski}, \citenamefont
  {Rotundu}, \citenamefont {Lynn},\ and\ \citenamefont {Birgeneau}}]{Paj13:87}%
  \BibitemOpen
  \bibfield  {author} {\bibinfo {author} {\bibfnamefont {D.~M.}\ \bibnamefont
  {Pajerowski}}, \bibinfo {author} {\bibfnamefont {C.~R.}\ \bibnamefont
  {Rotundu}}, \bibinfo {author} {\bibfnamefont {J.~W.}\ \bibnamefont {Lynn}}, \
  and\ \bibinfo {author} {\bibfnamefont {R.~J.}\ \bibnamefont {Birgeneau}},\
  }\href@noop {} {\bibfield  {journal} {\bibinfo  {journal} {Phys. Rev. B}\
  }\textbf {\bibinfo {volume} {87}},\ \bibinfo {pages} {134507} (\bibinfo
  {year} {2013}{\natexlab{a}})}\BibitemShut {NoStop}%
\bibitem [{\citenamefont {Urbano}\ \emph {et~al.}(2007)\citenamefont {Urbano},
  \citenamefont {Young}, \citenamefont {Curro}, \citenamefont {Thompson},
  \citenamefont {Pham},\ and\ \citenamefont {Fisk}}]{Urbano07:99}%
  \BibitemOpen
  \bibfield  {author} {\bibinfo {author} {\bibfnamefont {R.~R.}\ \bibnamefont
  {Urbano}}, \bibinfo {author} {\bibfnamefont {B.~L.}\ \bibnamefont {Young}},
  \bibinfo {author} {\bibfnamefont {N.~J.}\ \bibnamefont {Curro}}, \bibinfo
  {author} {\bibfnamefont {J.~D.}\ \bibnamefont {Thompson}}, \bibinfo {author}
  {\bibfnamefont {L.~D.}\ \bibnamefont {Pham}}, \ and\ \bibinfo {author}
  {\bibfnamefont {Z.}~\bibnamefont {Fisk}},\ }\href@noop {} {\bibfield
  {journal} {\bibinfo  {journal} {Phys. Rev. Lett.}\ }\textbf {\bibinfo
  {volume} {99}},\ \bibinfo {pages} {146402} (\bibinfo {year}
  {2007})}\BibitemShut {NoStop}%
\bibitem [{\citenamefont {Blume}\ \emph {et~al.}(1962)\citenamefont {Blume},
  \citenamefont {Freeman},\ and\ \citenamefont {Watson}}]{Blume62:37}%
  \BibitemOpen
  \bibfield  {author} {\bibinfo {author} {\bibfnamefont {M.}~\bibnamefont
  {Blume}}, \bibinfo {author} {\bibfnamefont {A.~J.}\ \bibnamefont {Freeman}},
  \ and\ \bibinfo {author} {\bibfnamefont {R.~E.}\ \bibnamefont {Watson}},\
  }\href@noop {} {\bibfield  {journal} {\bibinfo  {journal} {J. Chem. Phys.}\
  }\textbf {\bibinfo {volume} {37}},\ \bibinfo {pages} {1245} (\bibinfo {year}
  {1962})}\BibitemShut {NoStop}%
\bibitem [{\citenamefont {Raymond}\ and\ \citenamefont
  {Lapertot}(2015)}]{Raymond15:115}%
  \BibitemOpen
  \bibfield  {author} {\bibinfo {author} {\bibfnamefont {S.}~\bibnamefont
  {Raymond}}\ and\ \bibinfo {author} {\bibfnamefont {G.}~\bibnamefont
  {Lapertot}},\ }\href@noop {} {\bibfield  {journal} {\bibinfo  {journal}
  {Phys. Rev. Lett.}\ }\textbf {\bibinfo {volume} {115}},\ \bibinfo {pages}
  {037001} (\bibinfo {year} {2015})}\BibitemShut {NoStop}%
\bibitem [{\citenamefont {Mazzone}\ \emph {et~al.}(2017)\citenamefont
  {Mazzone}, \citenamefont {Raymond}, \citenamefont {Gavilano}, \citenamefont
  {Steffens}, \citenamefont {Schneidewind}, \citenamefont {Lapertot},\ and\
  \citenamefont {Kenzelmann}}]{Mazzone17:119}%
  \BibitemOpen
  \bibfield  {author} {\bibinfo {author} {\bibfnamefont {D.~G.}\ \bibnamefont
  {Mazzone}}, \bibinfo {author} {\bibfnamefont {S.}~\bibnamefont {Raymond}},
  \bibinfo {author} {\bibfnamefont {J.~L.}\ \bibnamefont {Gavilano}}, \bibinfo
  {author} {\bibfnamefont {P.}~\bibnamefont {Steffens}}, \bibinfo {author}
  {\bibfnamefont {A.}~\bibnamefont {Schneidewind}}, \bibinfo {author}
  {\bibfnamefont {G.}~\bibnamefont {Lapertot}}, \ and\ \bibinfo {author}
  {\bibfnamefont {M.}~\bibnamefont {Kenzelmann}},\ }\href@noop {} {\bibfield
  {journal} {\bibinfo  {journal} {Phys. Rev. Lett.}\ }\textbf {\bibinfo
  {volume} {119}},\ \bibinfo {pages} {187002} (\bibinfo {year}
  {2017})}\BibitemShut {NoStop}%
\bibitem [{\citenamefont {Movshovich}\ \emph {et~al.}(2001)\citenamefont
  {Movshovich}, \citenamefont {Jaime}, \citenamefont {Thompson}, \citenamefont
  {Petrovic}, \citenamefont {Fisk}, \citenamefont {Pagliuso},\ and\
  \citenamefont {Sarrao}}]{Mov01:86}%
  \BibitemOpen
  \bibfield  {author} {\bibinfo {author} {\bibfnamefont {R.}~\bibnamefont
  {Movshovich}}, \bibinfo {author} {\bibfnamefont {M.}~\bibnamefont {Jaime}},
  \bibinfo {author} {\bibfnamefont {J.~D.}\ \bibnamefont {Thompson}}, \bibinfo
  {author} {\bibfnamefont {C.}~\bibnamefont {Petrovic}}, \bibinfo {author}
  {\bibfnamefont {Z.}~\bibnamefont {Fisk}}, \bibinfo {author} {\bibfnamefont
  {P.~G.}\ \bibnamefont {Pagliuso}}, \ and\ \bibinfo {author} {\bibfnamefont
  {J.~L.}\ \bibnamefont {Sarrao}},\ }\href@noop {} {\bibfield  {journal}
  {\bibinfo  {journal} {Phys. Rev. Lett.}\ }\textbf {\bibinfo {volume} {86}},\
  \bibinfo {pages} {5152} (\bibinfo {year} {2001})}\BibitemShut {NoStop}%
\bibitem [{\citenamefont {Fujita}\ \emph {et~al.}(2002)\citenamefont {Fujita},
  \citenamefont {Yamada}, \citenamefont {Hiraka}, \citenamefont {Gehring},
  \citenamefont {Lee}, \citenamefont {Wakimoto},\ and\ \citenamefont
  {Shirane}}]{Fujita02:65}%
  \BibitemOpen
  \bibfield  {author} {\bibinfo {author} {\bibfnamefont {M.}~\bibnamefont
  {Fujita}}, \bibinfo {author} {\bibfnamefont {K.}~\bibnamefont {Yamada}},
  \bibinfo {author} {\bibfnamefont {H.}~\bibnamefont {Hiraka}}, \bibinfo
  {author} {\bibfnamefont {P.~M.}\ \bibnamefont {Gehring}}, \bibinfo {author}
  {\bibfnamefont {S.~H.}\ \bibnamefont {Lee}}, \bibinfo {author} {\bibfnamefont
  {S.}~\bibnamefont {Wakimoto}}, \ and\ \bibinfo {author} {\bibfnamefont
  {G.}~\bibnamefont {Shirane}},\ }\href@noop {} {\bibfield  {journal} {\bibinfo
   {journal} {Phys. Rev. B}\ }\textbf {\bibinfo {volume} {65}},\ \bibinfo
  {pages} {064505} (\bibinfo {year} {2002})}\BibitemShut {NoStop}%
\bibitem [{\citenamefont {Stock}\ \emph {et~al.}(2006)\citenamefont {Stock},
  \citenamefont {Buyers}, \citenamefont {Yamani}, \citenamefont {Broholm},
  \citenamefont {Chung}, \citenamefont {Tun}, \citenamefont {Liang},
  \citenamefont {Bonn}, \citenamefont {Hardy},\ and\ \citenamefont
  {Birgeneau}}]{Stock06:73}%
  \BibitemOpen
  \bibfield  {author} {\bibinfo {author} {\bibfnamefont {C.}~\bibnamefont
  {Stock}}, \bibinfo {author} {\bibfnamefont {W.~J.~L.}\ \bibnamefont
  {Buyers}}, \bibinfo {author} {\bibfnamefont {Z.}~\bibnamefont {Yamani}},
  \bibinfo {author} {\bibfnamefont {C.~L.}\ \bibnamefont {Broholm}}, \bibinfo
  {author} {\bibfnamefont {J.~H.}\ \bibnamefont {Chung}}, \bibinfo {author}
  {\bibfnamefont {Z.}~\bibnamefont {Tun}}, \bibinfo {author} {\bibfnamefont
  {R.}~\bibnamefont {Liang}}, \bibinfo {author} {\bibfnamefont
  {D.}~\bibnamefont {Bonn}}, \bibinfo {author} {\bibfnamefont {W.~N.}\
  \bibnamefont {Hardy}}, \ and\ \bibinfo {author} {\bibfnamefont {R.~J.}\
  \bibnamefont {Birgeneau}},\ }\href@noop {} {\bibfield  {journal} {\bibinfo
  {journal} {Phys. Rev. B}\ }\textbf {\bibinfo {volume} {73}},\ \bibinfo
  {pages} {100504(R)} (\bibinfo {year} {2006})}\BibitemShut {NoStop}%
\bibitem [{\citenamefont {Stock}\ \emph
  {et~al.}(2008{\natexlab{b}})\citenamefont {Stock}, \citenamefont {Buyers},
  \citenamefont {Yamani}, \citenamefont {Tun}, \citenamefont {Birgeneau},
  \citenamefont {Liang}, \citenamefont {Bonn},\ and\ \citenamefont
  {Hardy}}]{Stock08:77}%
  \BibitemOpen
  \bibfield  {author} {\bibinfo {author} {\bibfnamefont {C.}~\bibnamefont
  {Stock}}, \bibinfo {author} {\bibfnamefont {W.~J.~L.}\ \bibnamefont
  {Buyers}}, \bibinfo {author} {\bibfnamefont {Z.}~\bibnamefont {Yamani}},
  \bibinfo {author} {\bibfnamefont {Z.}~\bibnamefont {Tun}}, \bibinfo {author}
  {\bibfnamefont {R.~J.}\ \bibnamefont {Birgeneau}}, \bibinfo {author}
  {\bibfnamefont {R.}~\bibnamefont {Liang}}, \bibinfo {author} {\bibfnamefont
  {D.}~\bibnamefont {Bonn}}, \ and\ \bibinfo {author} {\bibfnamefont {W.~N.}\
  \bibnamefont {Hardy}},\ }\href@noop {} {\bibfield  {journal} {\bibinfo
  {journal} {Phys. Rev. B}\ }\textbf {\bibinfo {volume} {77}},\ \bibinfo
  {pages} {104513} (\bibinfo {year} {2008}{\natexlab{b}})}\BibitemShut
  {NoStop}%
\bibitem [{\citenamefont {Yamani}\ \emph {et~al.}(2015)\citenamefont {Yamani},
  \citenamefont {Buyers}, \citenamefont {Wang}, \citenamefont {Kim},
  \citenamefont {Chung}, \citenamefont {Chang}, \citenamefont {Gehring},
  \citenamefont {Gasparovic}, \citenamefont {Stock}, \citenamefont {Broholm},
  \citenamefont {Baglo}, \citenamefont {Liang}, \citenamefont {Bonn},\ and\
  \citenamefont {Hardy}}]{Yamani15:91}%
  \BibitemOpen
  \bibfield  {author} {\bibinfo {author} {\bibfnamefont {Z.}~\bibnamefont
  {Yamani}}, \bibinfo {author} {\bibfnamefont {W.~J.~L.}\ \bibnamefont
  {Buyers}}, \bibinfo {author} {\bibfnamefont {F.}~\bibnamefont {Wang}},
  \bibinfo {author} {\bibfnamefont {Y.~J.}\ \bibnamefont {Kim}}, \bibinfo
  {author} {\bibfnamefont {J.~H.}\ \bibnamefont {Chung}}, \bibinfo {author}
  {\bibfnamefont {S.}~\bibnamefont {Chang}}, \bibinfo {author} {\bibfnamefont
  {P.~M.}\ \bibnamefont {Gehring}}, \bibinfo {author} {\bibfnamefont
  {G.}~\bibnamefont {Gasparovic}}, \bibinfo {author} {\bibfnamefont
  {C.}~\bibnamefont {Stock}}, \bibinfo {author} {\bibfnamefont {C.~L.}\
  \bibnamefont {Broholm}}, \bibinfo {author} {\bibfnamefont {J.~C.}\
  \bibnamefont {Baglo}}, \bibinfo {author} {\bibfnamefont {R.}~\bibnamefont
  {Liang}}, \bibinfo {author} {\bibfnamefont {D.~A.}\ \bibnamefont {Bonn}}, \
  and\ \bibinfo {author} {\bibfnamefont {W.~N.}\ \bibnamefont {Hardy}},\
  }\href@noop {} {\bibfield  {journal} {\bibinfo  {journal} {Phys. Rev. B}\
  }\textbf {\bibinfo {volume} {91}},\ \bibinfo {pages} {134427} (\bibinfo
  {year} {2015})}\BibitemShut {NoStop}%
\bibitem [{\citenamefont {Pajerowski}\ \emph
  {et~al.}(2013{\natexlab{b}})\citenamefont {Pajerowski}, \citenamefont
  {Rotundu}, \citenamefont {Lynn},\ and\ \citenamefont
  {Birgeneau}}]{Pajerowski13:87}%
  \BibitemOpen
  \bibfield  {author} {\bibinfo {author} {\bibfnamefont {D.~M.}\ \bibnamefont
  {Pajerowski}}, \bibinfo {author} {\bibfnamefont {C.~R.}\ \bibnamefont
  {Rotundu}}, \bibinfo {author} {\bibfnamefont {J.~W.}\ \bibnamefont {Lynn}}, \
  and\ \bibinfo {author} {\bibfnamefont {R.~J.}\ \bibnamefont {Birgeneau}},\
  }\href@noop {} {\bibfield  {journal} {\bibinfo  {journal} {Phys. Rev. B}\
  }\textbf {\bibinfo {volume} {87}},\ \bibinfo {pages} {134507} (\bibinfo
  {year} {2013}{\natexlab{b}})}\BibitemShut {NoStop}%
\bibitem [{\citenamefont {Headings}\ \emph {et~al.}(2011)\citenamefont
  {Headings}, \citenamefont {Hayden}, \citenamefont {Kulda}, \citenamefont
  {Babu},\ and\ \citenamefont {Cardwell}}]{Headings10:105}%
  \BibitemOpen
  \bibfield  {author} {\bibinfo {author} {\bibfnamefont {N.~S.}\ \bibnamefont
  {Headings}}, \bibinfo {author} {\bibfnamefont {S.~M.}\ \bibnamefont
  {Hayden}}, \bibinfo {author} {\bibfnamefont {J.}~\bibnamefont {Kulda}},
  \bibinfo {author} {\bibfnamefont {N.~H.}\ \bibnamefont {Babu}}, \ and\
  \bibinfo {author} {\bibfnamefont {D.~A.}\ \bibnamefont {Cardwell}},\
  }\href@noop {} {\bibfield  {journal} {\bibinfo  {journal} {Phys. Rev. B}\
  }\textbf {\bibinfo {volume} {84}},\ \bibinfo {pages} {104513} (\bibinfo
  {year} {2011})}\BibitemShut {NoStop}%
\bibitem [{\citenamefont {Stock}\ \emph {et~al.}(2004)\citenamefont {Stock},
  \citenamefont {Buyers}, \citenamefont {Liang}, \citenamefont {Peets},
  \citenamefont {Tun}, \citenamefont {Bonn}, \citenamefont {Hardy},\ and\
  \citenamefont {Birgeneau}}]{Stock04:69}%
  \BibitemOpen
  \bibfield  {author} {\bibinfo {author} {\bibfnamefont {C.}~\bibnamefont
  {Stock}}, \bibinfo {author} {\bibfnamefont {W.~J.~L.}\ \bibnamefont
  {Buyers}}, \bibinfo {author} {\bibfnamefont {R.}~\bibnamefont {Liang}},
  \bibinfo {author} {\bibfnamefont {D.}~\bibnamefont {Peets}}, \bibinfo
  {author} {\bibfnamefont {Z.}~\bibnamefont {Tun}}, \bibinfo {author}
  {\bibfnamefont {D.}~\bibnamefont {Bonn}}, \bibinfo {author} {\bibfnamefont
  {W.~N.}\ \bibnamefont {Hardy}}, \ and\ \bibinfo {author} {\bibfnamefont
  {R.~J.}\ \bibnamefont {Birgeneau}},\ }\href@noop {} {\bibfield  {journal}
  {\bibinfo  {journal} {Phys. Rev. B}\ }\textbf {\bibinfo {volume} {69}},\
  \bibinfo {pages} {014502} (\bibinfo {year} {2004})}\BibitemShut {NoStop}%
\bibitem [{\citenamefont {Akbari}\ and\ \citenamefont
  {Thalmeier}(2012)}]{Akbari12:86}%
  \BibitemOpen
  \bibfield  {author} {\bibinfo {author} {\bibfnamefont {A.}~\bibnamefont
  {Akbari}}\ and\ \bibinfo {author} {\bibfnamefont {P.}~\bibnamefont
  {Thalmeier}},\ }\href@noop {} {\bibfield  {journal} {\bibinfo  {journal}
  {Phys. Rev. B}\ }\textbf {\bibinfo {volume} {86}},\ \bibinfo {pages} {134516}
  (\bibinfo {year} {2012})}\BibitemShut {NoStop}%
\bibitem [{\citenamefont {Ismer}\ \emph {et~al.}(2007)\citenamefont {Ismer},
  \citenamefont {Eremin}, \citenamefont {Rossi},\ and\ \citenamefont
  {Morr}}]{Ismer07:99}%
  \BibitemOpen
  \bibfield  {author} {\bibinfo {author} {\bibfnamefont {J.~P.}\ \bibnamefont
  {Ismer}}, \bibinfo {author} {\bibfnamefont {I.}~\bibnamefont {Eremin}},
  \bibinfo {author} {\bibfnamefont {E.}~\bibnamefont {Rossi}}, \ and\ \bibinfo
  {author} {\bibfnamefont {D.~K.}\ \bibnamefont {Morr}},\ }\href@noop {}
  {\bibfield  {journal} {\bibinfo  {journal} {Phys. Rev. Lett.}\ }\textbf
  {\bibinfo {volume} {99}},\ \bibinfo {pages} {047005} (\bibinfo {year}
  {2007})}\BibitemShut {NoStop}%
\bibitem [{\citenamefont {Stock}\ \emph {et~al.}(2007)\citenamefont {Stock},
  \citenamefont {Cowley}, \citenamefont {Buyers}, \citenamefont {Coldea},
  \citenamefont {Broholm}, \citenamefont {Frost}, \citenamefont {Birgeneau},
  \citenamefont {Liang}, \citenamefont {Bonn},\ and\ \citenamefont
  {Hardy}}]{Stock07:75}%
  \BibitemOpen
  \bibfield  {author} {\bibinfo {author} {\bibfnamefont {C.}~\bibnamefont
  {Stock}}, \bibinfo {author} {\bibfnamefont {R.~A.}\ \bibnamefont {Cowley}},
  \bibinfo {author} {\bibfnamefont {W.~J.~L.}\ \bibnamefont {Buyers}}, \bibinfo
  {author} {\bibfnamefont {R.}~\bibnamefont {Coldea}}, \bibinfo {author}
  {\bibfnamefont {C.~L.}\ \bibnamefont {Broholm}}, \bibinfo {author}
  {\bibfnamefont {C.~D.}\ \bibnamefont {Frost}}, \bibinfo {author}
  {\bibfnamefont {R.~J.}\ \bibnamefont {Birgeneau}}, \bibinfo {author}
  {\bibfnamefont {R.}~\bibnamefont {Liang}}, \bibinfo {author} {\bibfnamefont
  {D.}~\bibnamefont {Bonn}}, \ and\ \bibinfo {author} {\bibfnamefont {W.~N.}\
  \bibnamefont {Hardy}},\ }\href@noop {} {\bibfield  {journal} {\bibinfo
  {journal} {Phys. Rev. B}\ }\textbf {\bibinfo {volume} {75}},\ \bibinfo
  {pages} {172510} (\bibinfo {year} {2007})}\BibitemShut {NoStop}%
\bibitem [{\citenamefont {Stock}\ \emph {et~al.}(2010)\citenamefont {Stock},
  \citenamefont {Cowley}, \citenamefont {Buyers}, \citenamefont {Frost},
  \citenamefont {Taylor}, \citenamefont {Peets}, \citenamefont {Liang},
  \citenamefont {Bonn},\ and\ \citenamefont {Hardy}}]{Stock10:82}%
  \BibitemOpen
  \bibfield  {author} {\bibinfo {author} {\bibfnamefont {C.}~\bibnamefont
  {Stock}}, \bibinfo {author} {\bibfnamefont {R.~A.}\ \bibnamefont {Cowley}},
  \bibinfo {author} {\bibfnamefont {W.~J.~L.}\ \bibnamefont {Buyers}}, \bibinfo
  {author} {\bibfnamefont {C.~D.}\ \bibnamefont {Frost}}, \bibinfo {author}
  {\bibfnamefont {J.~W.}\ \bibnamefont {Taylor}}, \bibinfo {author}
  {\bibfnamefont {D.}~\bibnamefont {Peets}}, \bibinfo {author} {\bibfnamefont
  {R.}~\bibnamefont {Liang}}, \bibinfo {author} {\bibfnamefont
  {D.}~\bibnamefont {Bonn}}, \ and\ \bibinfo {author} {\bibfnamefont {W.~N.}\
  \bibnamefont {Hardy}},\ }\href@noop {} {\bibfield  {journal} {\bibinfo
  {journal} {Phys. Rev. B}\ }\textbf {\bibinfo {volume} {82}},\ \bibinfo
  {pages} {174505} (\bibinfo {year} {2010})}\BibitemShut {NoStop}%
\bibitem [{\citenamefont {Stock}\ \emph {et~al.}(2014)\citenamefont {Stock},
  \citenamefont {Rodriguez}, \citenamefont {Sobolev}, \citenamefont
  {Rodriguez-Rivera}, \citenamefont {Ewings}, \citenamefont {Taylor},
  \citenamefont {Christianson},\ and\ \citenamefont {Green}}]{Stock14:90}%
  \BibitemOpen
  \bibfield  {author} {\bibinfo {author} {\bibfnamefont {C.}~\bibnamefont
  {Stock}}, \bibinfo {author} {\bibfnamefont {E.~E.}\ \bibnamefont
  {Rodriguez}}, \bibinfo {author} {\bibfnamefont {O.}~\bibnamefont {Sobolev}},
  \bibinfo {author} {\bibfnamefont {J.~A.}\ \bibnamefont {Rodriguez-Rivera}},
  \bibinfo {author} {\bibfnamefont {R.~A.}\ \bibnamefont {Ewings}}, \bibinfo
  {author} {\bibfnamefont {J.~W.}\ \bibnamefont {Taylor}}, \bibinfo {author}
  {\bibfnamefont {A.~D.}\ \bibnamefont {Christianson}}, \ and\ \bibinfo
  {author} {\bibfnamefont {M.~A.}\ \bibnamefont {Green}},\ }\href@noop {}
  {\bibfield  {journal} {\bibinfo  {journal} {Phys. Rev. B}\ }\textbf {\bibinfo
  {volume} {90}},\ \bibinfo {pages} {121113} (\bibinfo {year}
  {2014})}\BibitemShut {NoStop}%
\bibitem [{\citenamefont {Stock}\ \emph {et~al.}(2017)\citenamefont {Stock},
  \citenamefont {Rodriguez}, \citenamefont {Bourges}, \citenamefont {Ewings},
  \citenamefont {Cao}, \citenamefont {Chi}, \citenamefont {Rodriguez-Rivera},\
  and\ \citenamefont {Green}}]{Stock17:95}%
  \BibitemOpen
  \bibfield  {author} {\bibinfo {author} {\bibfnamefont {C.}~\bibnamefont
  {Stock}}, \bibinfo {author} {\bibfnamefont {E.~E.}\ \bibnamefont
  {Rodriguez}}, \bibinfo {author} {\bibfnamefont {P.}~\bibnamefont {Bourges}},
  \bibinfo {author} {\bibfnamefont {R.~A.}\ \bibnamefont {Ewings}}, \bibinfo
  {author} {\bibfnamefont {H.}~\bibnamefont {Cao}}, \bibinfo {author}
  {\bibfnamefont {S.}~\bibnamefont {Chi}}, \bibinfo {author} {\bibfnamefont
  {J.~A.}\ \bibnamefont {Rodriguez-Rivera}}, \ and\ \bibinfo {author}
  {\bibfnamefont {M.~A.}\ \bibnamefont {Green}},\ }\href@noop {} {\bibfield
  {journal} {\bibinfo  {journal} {Phys. Rev. B}\ }\textbf {\bibinfo {volume}
  {95}},\ \bibinfo {pages} {144407} (\bibinfo {year} {2017})}\BibitemShut
  {NoStop}%
\bibitem [{\citenamefont {Song}\ \emph {et~al.}(2016)\citenamefont {Song},
  \citenamefont {Dyke}, \citenamefont {Lum}, \citenamefont {White},
  \citenamefont {Jang}, \citenamefont {Yazici}, \citenamefont {Shu},
  \citenamefont {Schneidewind}, \citenamefont {Cermak}, \citenamefont {Qiu},
  \citenamefont {Maple}, \citenamefont {Morr},\ and\ \citenamefont
  {Dai}}]{Song16:7}%
  \BibitemOpen
  \bibfield  {author} {\bibinfo {author} {\bibfnamefont {Y.}~\bibnamefont
  {Song}}, \bibinfo {author} {\bibfnamefont {J.~V.}\ \bibnamefont {Dyke}},
  \bibinfo {author} {\bibfnamefont {I.~K.}\ \bibnamefont {Lum}}, \bibinfo
  {author} {\bibfnamefont {B.~D.}\ \bibnamefont {White}}, \bibinfo {author}
  {\bibfnamefont {S.}~\bibnamefont {Jang}}, \bibinfo {author} {\bibfnamefont
  {D.}~\bibnamefont {Yazici}}, \bibinfo {author} {\bibfnamefont
  {L.}~\bibnamefont {Shu}}, \bibinfo {author} {\bibfnamefont {A.}~\bibnamefont
  {Schneidewind}}, \bibinfo {author} {\bibfnamefont {P.}~\bibnamefont
  {Cermak}}, \bibinfo {author} {\bibfnamefont {Y.}~\bibnamefont {Qiu}},
  \bibinfo {author} {\bibfnamefont {M.~B.}\ \bibnamefont {Maple}}, \bibinfo
  {author} {\bibfnamefont {D.~K.}\ \bibnamefont {Morr}}, \ and\ \bibinfo
  {author} {\bibfnamefont {P.}~\bibnamefont {Dai}},\ }\href@noop {} {\bibfield
  {journal} {\bibinfo  {journal} {Nat. Commun.}\ }\textbf {\bibinfo {volume}
  {7}},\ \bibinfo {pages} {12774} (\bibinfo {year} {2016})}\BibitemShut
  {NoStop}%
\bibitem [{\citenamefont {Chubukov}\ and\ \citenamefont
  {Gor'kov}(2008)}]{Chubukov08:101}%
  \BibitemOpen
  \bibfield  {author} {\bibinfo {author} {\bibfnamefont {A.~V.}\ \bibnamefont
  {Chubukov}}\ and\ \bibinfo {author} {\bibfnamefont {L.~P.}\ \bibnamefont
  {Gor'kov}},\ }\href@noop {} {\bibfield  {journal} {\bibinfo  {journal} {Phys.
  Rev. Lett.}\ }\textbf {\bibinfo {volume} {101}},\ \bibinfo {pages} {147004}
  (\bibinfo {year} {2008})}\BibitemShut {NoStop}%
\end{thebibliography}

%

\end{document}


\title{Supplementary Information: ``From Ising resonant fluctuations to static uniaxial order in antiferromagnetic and weakly superconducting CeCo(In$_{1-x}$Hg$_{x}$)$_{5}$ ($x$=0.01)"}

\author{C. Stock}
\affiliation{School of Physics and Astronomy, University of Edinburgh, Edinburgh EH9 3JZ, UK}
\author{J. A. Rodriguez-Rivera}
\affiliation{NIST Center for Neutron Research, National Institute of Standards and Technology, 100 Bureau Dr., Gaithersburg, MD 20899}
\affiliation{Department of Materials Science, University of Maryland, College Park, MD  20742}
\author{K. Schmalzl}
\affiliation{Forschungszentrum Juelich GmbH, Juelich Centre for Neutron Science at ILL, 71 avenue des Martyrs, 38000 Grenoble, France}
\author{F. Demmel}
\affiliation{ISIS Facility, Rutherford Appleton Labs, Chilton, Didcot, OX11 0QX}
\author{D. K. Singh}
\affiliation{Department of Physics and Astronomy, University of Missouri, Missouri 65211, USA}
\author{F. Ronning}
\affiliation{Los Alamos National Laboratory, Los Alamos, New Mexico 87545, USA}
\author{J. D. Thompson}
\affiliation{Los Alamos National Laboratory, Los Alamos, New Mexico 87545, USA}
\author{E.D. Bauer}
\affiliation{Los Alamos National Laboratory, Los Alamos, New Mexico 87545, USA}
\date{\today}

\begin{abstract}

Supplementary information is provided in support of the main text discussing the magnetic structure and excitations in CeCo(In$_{1-x}$Hg$_{x}$)$_{5}$ ($x$=0.01).  We present details of sample characterization and also experimental configurations in each of the neutron experiments.  Further neutron diffraction data is also shown illustrating the $c$-axis low temperature magnetic structure.  Finally, information is presented on the background subtraction used to separate the magnetic inelastic scattering and a discussion of the origins of the observed background.

\end{abstract}

\maketitle

\section{Materials preparation and characterization:}

Single crystals of CeCo(In$_{1-x}$Hg$_{x}$)$_{5}${} were grown from In/Hg flux.  The elements were placed in an alumina crucible in the nominal molar ratio Ce:Co:In:Hg = 1:1:20(1-$x_{nom}$):20$x_{nom}$ and sealed under vacuum in a silica ampoule.  The ampoule was heated to 1150$^{\circ}$C, held there for 6 hours, cooled quickly to 800 $^{\circ}$C at 150$^{\circ}$C/hr., then slowly cooled to 400 $^{\circ}$C at 2$^{\circ}$C/hr., at which point the In/Hg flux was removed by centrifugation.  The  CeCo(In$_{1-x}$Hg$_{x}$)$_{5}${} plate-like, single crystals were separated from the crucible by etching in dilute HCl.    The actual concentration of Hg present in samples was determined via Energy Dispersive Spectroscopy (EDS) to be $x_{actual}$ = 0.154 $x_{nom}$.  The values of $x$ listed in the paper refer to the actual Hg concentration $x_{actual}$.  A similar ratio of   $x_{actual}$ / $x_{nom}$ $\simeq$ 0.1 was found for CeCo(In$_{1-x}$Cd$_{x}$)$_{5}$.

The specific heat, plotted as $C/T$, versus temperature $T$ of  CeCo(In$_{1-x}$Hg$_{x}$)$_{5}${} for $0\leq x \leq 2.6 \%$  is shown in Fig. \ref{Cp}.  A large anomaly in $C/T$ is observed for $x\leq 0.74\%$ around 2 K, which corresponds to the superconducting transition.  For $x = 1.0\%$ ($1.3\%$), two somewhat broadened transitions are observed in $C/T$ at $T_N = 2.8$ K and $T_c = 1.75$ K ($T_N = 3.4$ K and $T_c = 1.4$ K), associated with coexistent antiferromagnetic order and unconventional superconductivity as displayed in the inset of Fig. \ref{Cp}.  The superconducting anomalies observed in $C/T$ for samples provide evidence for bulk superconductivity  in CeCo(In$_{1-x}$Hg$_{x}$)$_{5}${} crystals with $x \leq 1.3\%$. These value for the superconducting and antiferromagnetic transitions are in reasonable agreement with the neutron scattering results, considering the neutron samples consisted of approximately 150 single crystals. The superconducting transition temperature was determined by an entropy-conserving construction assuming an ideal BCS-like jump in $C/T$ at $T_c$, while the $T_N$ was determined as the temperature of the peak of the N{\'{e}}el transition. For $x> 1.3\%$, only antiferromagnetic order was observed with $T_N>$ 4 K.  A phase diagram based upon these specific heat results is shown in Fig. 1(d) of the manuscript. 
  
\renewcommand{\thefigure}{S1}
\begin{figure}[t]
\includegraphics[angle=0,width=8.5cm] {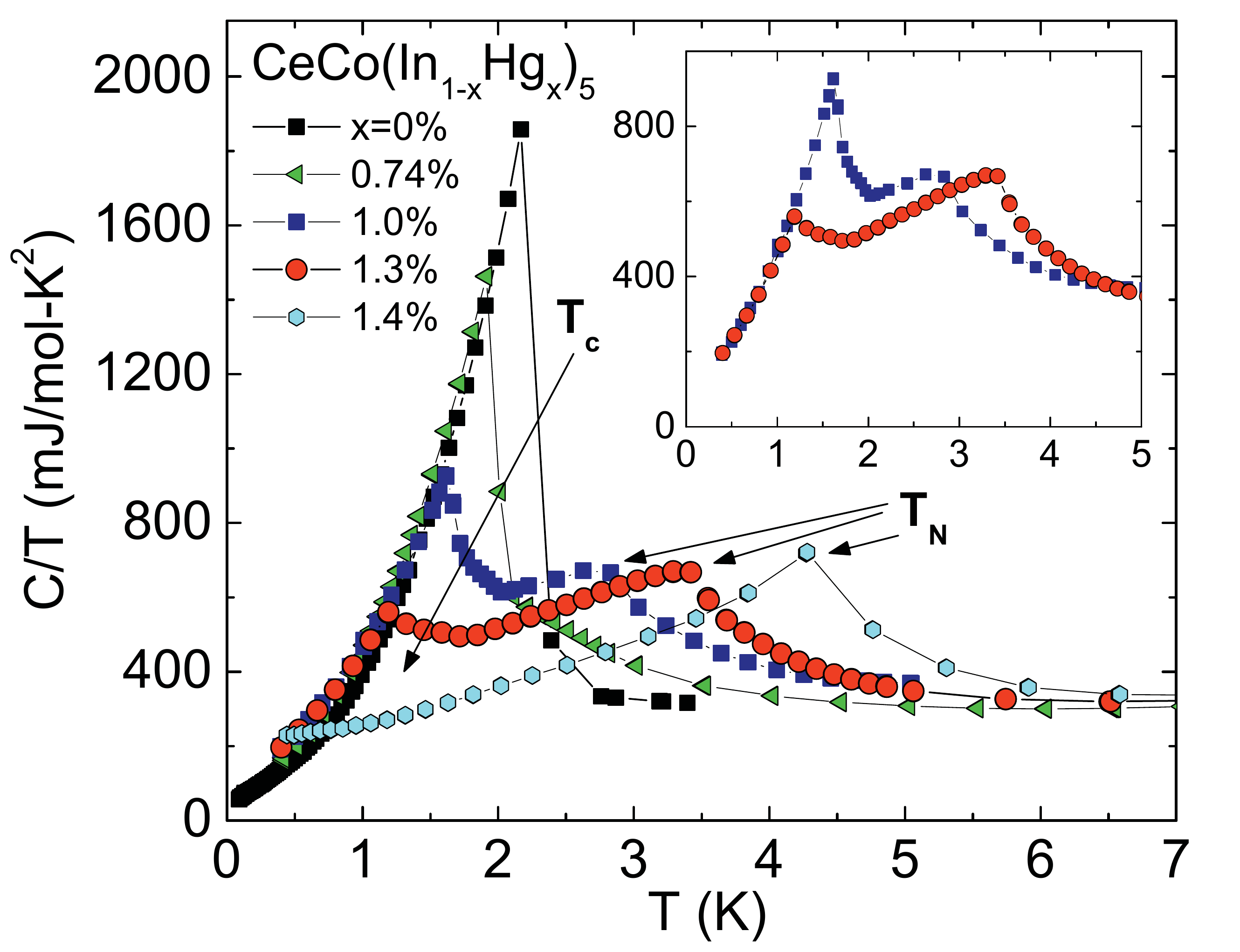}
\caption{\label{Cp}  The specific heat, $C/T$,  of  CeCo(In$_{1-x}$Hg$_{x}$)$_{5}$ for $0\leq x \leq 1.4 \%$.  Inset: $C/T$ vs T of the $x = 1.0\%$ and $1.3\%$ samples showing the coexistence of superconductivity and antiferromagnetism.}
\end{figure}

\renewcommand{\thefigure}{S2}
\begin{figure}[t]
\includegraphics[angle=0,width=8.5cm] {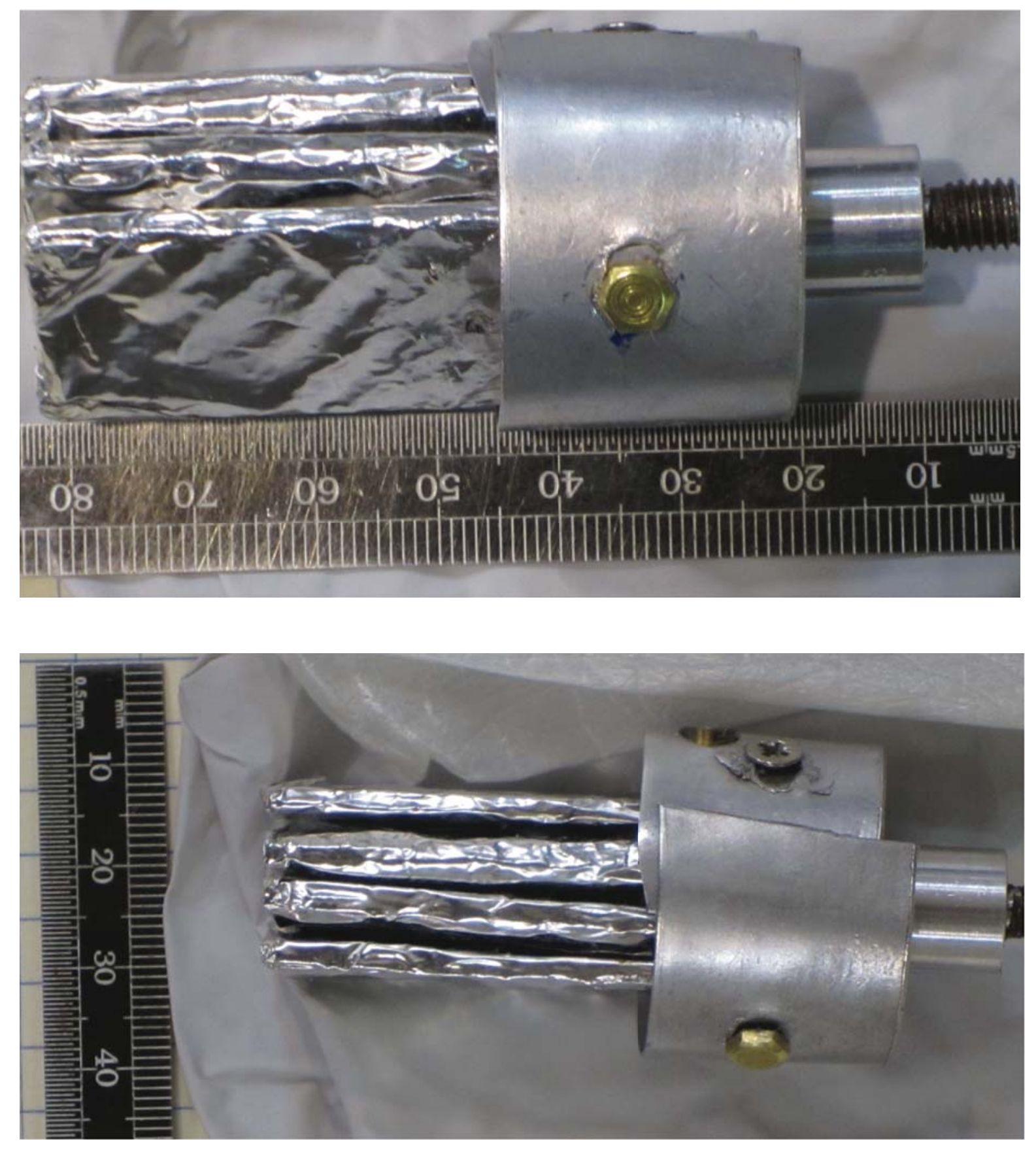}
\caption{\label{picture_sample}  The CeCo(In$_{1-x}$Hg$_{x}$)$_{5}$ ($x$=0.01) sample wrapped in Aluminum foil mounted on 4 individual plates in the (HHL) scattering plane.  The samples were fixed to the Aluminum plates using hydrogen free Fomblin oil.}
\end{figure}

\section{Neutron scattering:}

\textit{Elastic scattering using D23:}  Unpolarized diffraction studies were performed on the D23 diffractometer (incident $\lambda$=2.38 \AA).  D23 is a two-axis diffractometer with no energy analyzing crystal on the scattered side.  The experiment was used to confirm the low temperature $c$-axis oriented magnetic structure in the Neel ordered phase.  Measurements were performed on a single crystal of CeCo(In$_{0.987}$Hg$_{0.013}$)$_{5}$ with a mass of 28 mg.

The sample mount used in the neutron inelastic scattering experiments is displayed in Fig. \ref{picture_sample}.  The plate like samples of CeCo(In$_{1-x}$Hg$_{x}$)$_{5}$ ($x$=0.010) were  aligned such that reflections of the form (HHL) lay within the horizontal scattering plane.  The samples were fixed to the plates using high viscosity Fomblin oil which we confirmed to be hydrogen free using prompt gamma analysis (NCNR, NIST).  The overall mosaic (full-width at half maximum) was 3.0 degrees determined from rocking scans of the (004) nuclear Bragg peak.

\textit{Inelastic scattering using IN12}: On IN12 the final neutron energy was fixed at E$_{f}$=3.3 meV using PG(002) crystals.  Higher order harmonics were filtered with an incident beam velocity selector removing the requirement for cooled filters either before or after the sample.  For all measurements, both the monochromator and analyzers were horizontally and vertically focussed.  

\renewcommand{\thefigure}{S3}
\begin{figure}[t]
\includegraphics[angle=0,width=9.5cm] {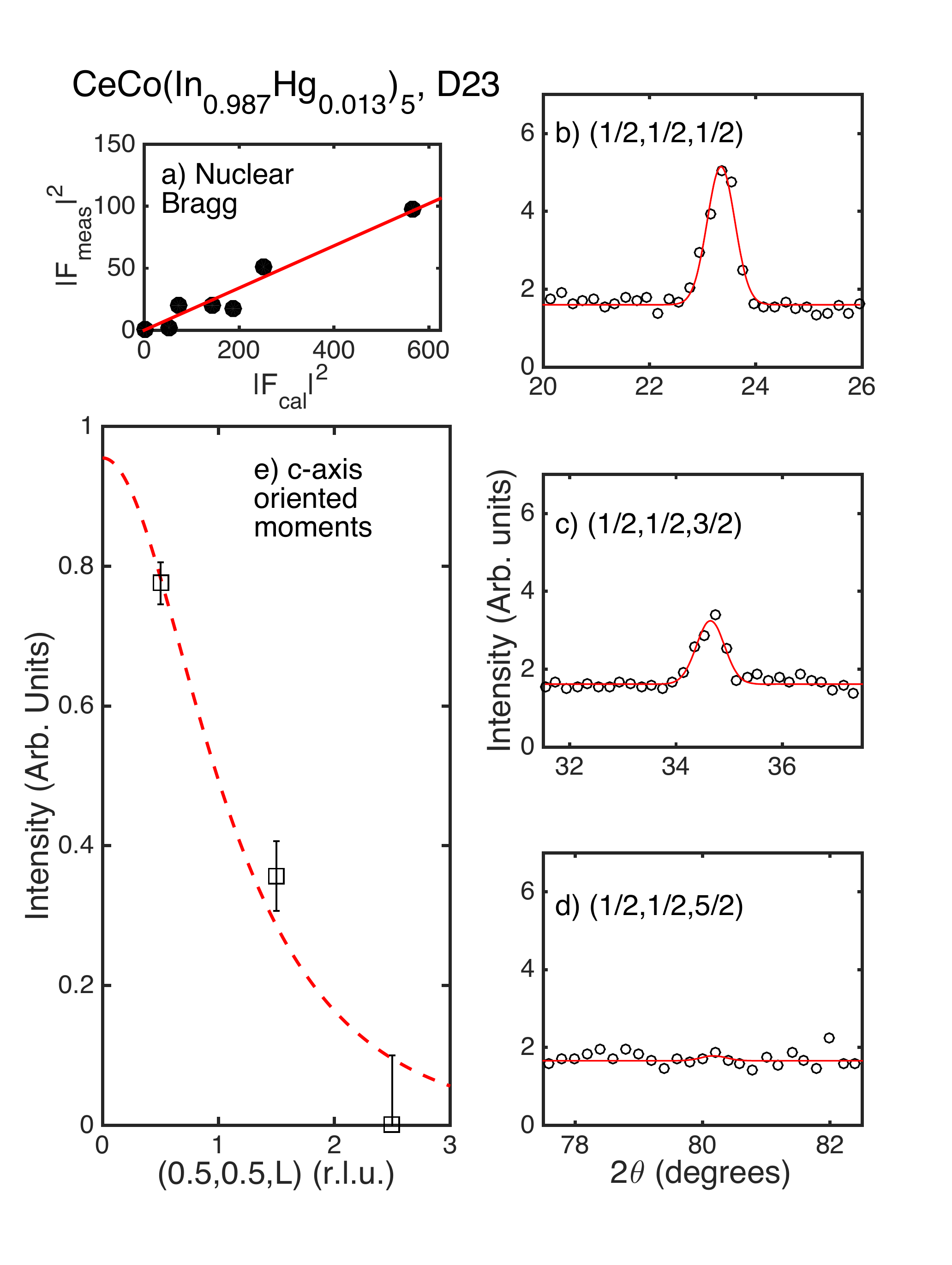}
\caption{\label{diffraction}  Diffraction data for CeCo(In$_{0.987}$Hg$_{0.013}$)$_{5}$ taken on D23 (ILL, France) at T=2 K in the antiferromagnetic phase.  $(a)$ The plot used to extract a calibration constant for an absolute static moment.  $(b-d)$ $\theta-2\theta$ scans at (1/2,1/2,L=1/2, 3/2, 5/2). $(e)$ shows the intensities compared against a model with $c$-axis oriented moments.}
\end{figure}

\textit{Inelastic scattering using MACS:}  MACS is a cold triple-axis spectrometer consisting of 20 double-bounce PG(002) analyzing crystals and detectors.  Each channel on the scattered side is collimated using 90$'$ Soller slits before the energy analyzing crystal.  The final energy was fixed at E$_{f}$=3.7 meV and cooled Be and BeO filters were placed before the monochromator and after the sample.  This configuration allowed energy transfers up to 1.3 meV to be measured given the differing cutoffs for Be and BeO filters of 5 and 3.7 meV, respectively.  The advantage of this configuration over the use of a single filter is that it removes any higher order scattering from the monochromator from reaching the sample which contributes to the background from high energy phonon scattering.  Given the presence of a significant amount of Aluminum and Fomblin oil, for mounting purposes, this double filter configuration resulted in a substantial decrease in background.  For all measurements, the monochromator was both horizontally and vertically focussed.  

\textit{Inelastic scattering using SPINS:}   To search for any resonance peaks reminiscent of the excitation in CeCoIn$_{5}$, we used the SPINS cold triple-axis spectrometer at NIST.  The same instrument settings as used to observe the resonance peak in pure CeCoIn$_{5}$ at T=0.1 K were used, albeit at T=0.5 K for CeCo(In$_{0.990}$Hg$_{0.010}$)$_{5}$.  The final energy was fixed to E$_{f}$=3.7 meV using an 11 blade graphite analyzer giving an approximately 11$^{\circ}$ horizontal acceptance.  The incident energy was varied using a vertically focussing graphite monochromator such that the energy transfer was defined by $\hbar \omega$ = E$_{i}$-E$_{f}$.  A Be filter was used in the incident beam (with a cutoff of 5 meV) and a BeO filter was used on the scattered side (with a cutoff of 3.7 meV) providing a range in energy transfer of $\sim$ 1.3 meV.  The sample was cooled in a He$^{3}$ cryostat to reach temperatures of $\sim$ 0.5 K.

\renewcommand{\thefigure}{S4}
\begin{figure}[t]
\includegraphics[angle=0,width=8.5cm] {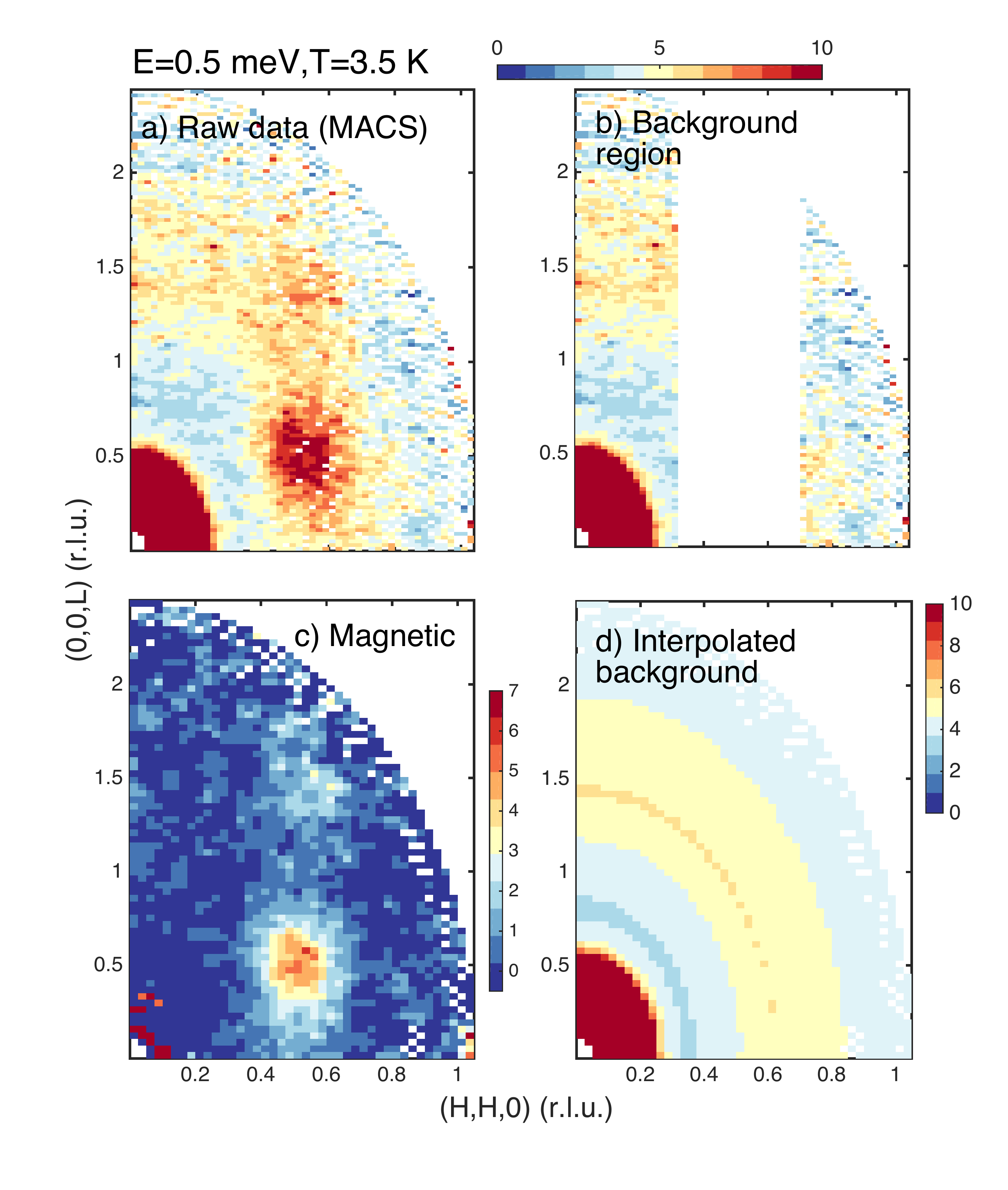}
\caption{\label{MACS_background}  The figure illustrates the methodology used to extract a background at each temperature for CeCo(In$_{0.990}$Hg$_{0.010}$)$_{5}$.  $(a)$ illustrates the raw data.  $(b)$ the region in reciprocal space used to extract the background. $(c)$ the background subtracted magnetic signal. $(d)$ the interpolated background derived assuming an angular independent background.}
\end{figure}

\renewcommand{\thefigure}{S5}
\begin{figure}[t]
\includegraphics[angle=0,width=8.5cm] {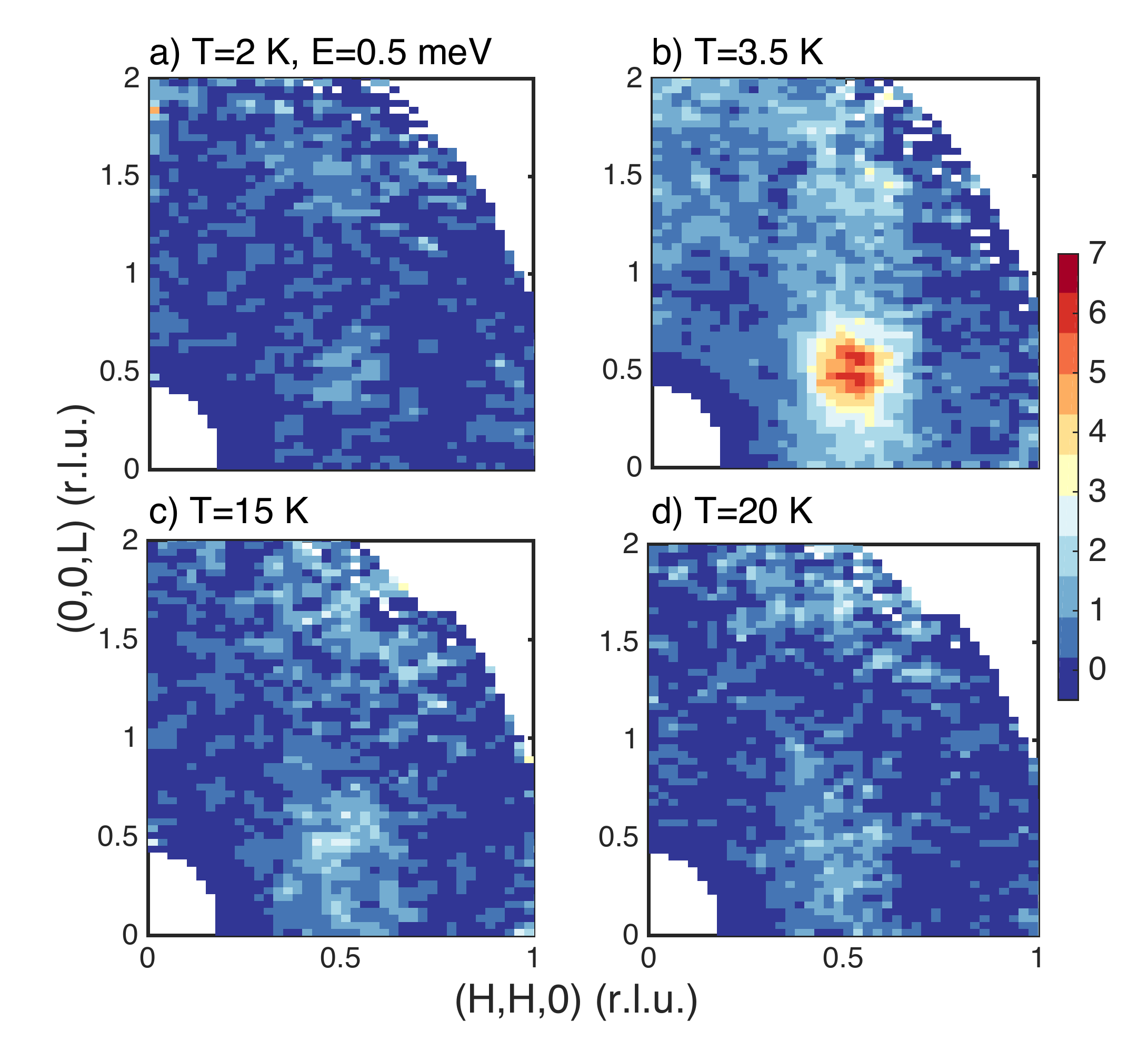}
\caption{\label{T_all}  The figure shows examples of background-corrected constant E=0.5 meV slices at various temperatures for CeCo(In$_{0.990}$Hg$_{0.010}$)$_{5}$.  The method used to subtract the background is discussed in the main text and the figures above.}
\end{figure}

\textit{Inelastic scattering using OSIRIS:}  High energy resolution data was obtained using the OSIRIS backscattering spectrometer located at ISIS.  A whitebeam of neutrons was incident on the sample and the final energy was fixed at E$_{f}$=1.84 meV using cooled graphite analyzers.  The default time configuration is set for a dynamic range of $\pm$ 0.5 meV.  An elastic energy resolution (full-width) of 2$\delta E$=0.025 meV was obtained for these experiments.

\section{Absolute Calibration and Magnetic Structure:}

The absolute moment quoted in the main text was extracted from diffraction data on a single crystal of  CeCo(In$_{0.987}$Hg$_{0.013}$)$_{5}$ is summarized in Fig. \ref{diffraction}.   Fig. \ref{diffraction}$(a)$ illustrates a plot of the measured nuclear Bragg peak structure factor squared ($|F_{meas}|^{2}$) against the calculated structure factor squared ($|F_{cal}|^{2}$).  The linear curve was used to extract the calibration constant for the absolute moment.  The large structure factor nuclear Bragg peaks (220) and (112) were found to deviate from this curve substantially owing to extinction and therefore were omitted from the analysis.

Figs. \ref{diffraction}$(b-d)$ display $\theta-2\theta$ scans through the magnetic Bragg peaks of CeCo(In$_{0.987}$Hg$_{0.013}$)$_{5}$  at 2 K.  The peaks show a strong decrease in intensity with increasing L and, as shown in Fig. \ref{diffraction}$(e)$, are consistent with magnetic moments oriented along the crystallographic $c$-axis.  The dashed curve in Fig. \ref{diffraction}$(e)$ is a plot of the orientation factor for magnetic neutron scattering multiplied by the magnetic form factor squared ($I(\vec{Q}) \propto f(Q)^{2}  [1-(\hat{Q} \cdot \hat{c})^{2}] $).  We note that measurements of magnetic peaks of the form (3/2, 3/2, L) failed to observe any magnetic signal.  

\section{Background Subtraction for Inelastic Scattering:}

We outline here how the background was corrected for from the Aluminum sample mount and Fomblin grease used to fixed the CeCo(In$_{1-x}$Hg$_{x}$)$_{5}$ samples (Fig. \ref{picture_sample}).  For the IN12 data, a combination of data at 20 K and 0.5 K was used as an estimate of the background.  Similarly, for OSIRIS a background was measured at 15 K and directly subtracted from the data.  

For MACS, the large detector coverage afforded a simultaneous measure of the background and also the magnetic inelastic signal of CeCo(In$_{0.990}$Hg$_{0.010}$)$_{5}$.  The methodology used to estimate the background at each temperature is illustrated in Fig. \ref{MACS_background}.  Fig. \ref{MACS_background}$(a)$ illustrates the uncorrected (raw) intensity including both magnetic and background scattering at 3.5 K at an energy transfer of 0.5 meV.   The intensity consists of a component which is ``powder-like" appearing as a ring of scattering independent of the angle $\omega$ or A3.  Further background measurements found this component to originate from the Fomblin oil used to fix the sample and also measured this background component to be temperature dependent.  Therefore, the background needed to be estimated independently at each temperature.  

To subtract this background due to the Fomblin oil, we have removed a strip in (HHL) near the H=0.5 position as shown in Fig. \ref{MACS_background}$(b)$.  This masked region was found to have a large temperature dependence indicative of a magnetic contribution.    The remaining data was then used to estimate an angular independent background shown in Fig. \ref{MACS_background}$(d)$.  Subtracting Fig. \ref{MACS_background}$(d)$ from $(a)$ results in the purely magnetic signal displayed in Fig. \ref{MACS_background}$(c)$ and discussed in the main text.

Further examples of this subtraction method are illustrated in Fig. \ref{T_all} which shows constant E=0.5 meV of CeCo(In$_{0.990}$Hg$_{0.010}$)$_{5}$ at T=2, 3.5, 15, and 20 K.  The data shows little magnetic scattering at 2 K and a monotonic decrease at 15 and 20 K as discussed in the main text.  Strong magnetic scattering is seen at 3.5 K associated with critical Ising fluctuations near the antiferromagnetic transition.  The consistency of this method with the temperature subtraction employed on IN12 verifies the validity of this method.